\documentclass[ip,twocolumn,amsmath,amssymb]{jpsj3} 
\usepackage{bm} 
\usepackage{enumerate}
\usepackage{graphicx,color}%  
\usepackage{amsmath,amsfonts,amsthm,amssymb,slashbox} 
\usepackage{dcolumn}
\usepackage{ulem}
\def\etal{{\itshape et al.}}
\def\journal#1#2#3#4{#1 {\bf #2} (#4) #3}
\def\JPSJ{J.\ Phys.\ Soc.\ Jpn.}
\def\PR{Phys.\ Rev.}
\def\PRL{Phys.\ Rev.\ Lett.}
\def\PRB{Phys.\ Rev.\ B}

\def\JCP{J.\ Chem.\ Phys.}

\newcommand{\bk}{\mbox{\boldmath $k$}}

\newcommand{\bdot}{\bm{\cdot}}
\newcommand{\Ha}{\mathcal{H}}
\newcommand{\mh}{\mathsf{h}}

\newcommand{\mS}{\mathsf{S}}

\newcommand{\mX}{\mathsf{X}}
\newcommand{\sP}{\mathcal{P}}
\newcommand{\sL}{\mathcal{L}}
\newcommand{\sO}{\mathcal{O}}
\newcommand{\sH}{\mathcal{H}}
\newcommand{\la}{\langle}
\newcommand{\ra}{\rangle}
\newcommand{\ga}{\alpha}
\newcommand{\gb}{\beta}
\newcommand{\gc}{\gamma}

\newcommand{\vk}{{\bm{k}}}

\newcommand{\vr}{{\bm{r}}}
\newcommand{\vR}{{\bm{R}}}
\newcommand{\vQ}{{\bm{Q}}}
\newcommand{\vga}{{\bm{\alpha}}}
\newcommand{\vgc}{{\bm{\gamma}}}
\newcommand{\Ns}{N_{\text{s}}}
\def\runauthor{Imada and Miyake}

\title{
Electronic Structure Calculation by First Principles for Strongly Correlated Electron Systems
} 
 
\author{  
Masatoshi \textsc{Imada}$^{1,3}$ and Takashi \textsc{Miyake}$^{2,3,}$}
\inst{
$^{1}$ Department of Applied Physics,  
 University of Tokyo, 7-3-1 Hongo, Bunkyo-ku, Tokyo 113-8656 \\  
$^{2}$ Nanosystem Research Institute, AIST, Tsukuba 305-8568, Japan \\ 
$^{3}$ Japan Science and Technology Agency, CREST, Honcho, Kawaguchi, Saitama 332-0012, Japan 
}

\recdate{today} 
 
\abst{ Recent trends of {\it ab initio} studies and progress in methodologies for electronic structure calculations of strongly correlated electron systems are discussed.  The interest for developing efficient methods is motivated by recent discoveries and characterizations of strongly correlated electron materials and by requirements for understanding mechanisms of intriguing phenomena beyond a single-particle picture.  A three-stage scheme is developed as renormalized multi-scale solvers (RMS) utilizing the hierarchical electronic structure in the energy space. It provides us with an {\it ab initio} downfolding of the global band structure into low-energy effective models followed by low-energy solvers for the models.  The RMS method is illustrated with examples of several materials. In particular, we overview cases such as dynamics of semiconductors, transition metals and its compounds including iron-based superconductors and perovskite oxides, and organic conductors of $\kappa$-ET type.
 } 
\kword{% 
first-principles calculation, effective Hamiltonian, downfolding, constrained RPA method,
strongly correlated electron system} 
 
\begin{document} 
\maketitle 
{\large Contents}
\begin{description}
{\bf \item{1.} Introduction}
{\bf \item{2.} Global Electronic Structure}
\item{2.1} Density functional theory (DFT)
\item{2.2} Basis functions for DFT
\item{2.2.1} Plane wave
\item{2.2.2} Argumented plane wave and muffin-tin orbital
\item{2.3} GW approximation
\item{2.4} LDA+U method
{\bf \item{3.} Downfolding}
\item{3.1} General framework
\item{3.2} Wannier functions
\item{3.3} Screened interaction
\item{3.4} Self-energy correction
\item{3.5} Low-energy Hamiltonian
\item{3.6} Vertex correction
\item{3.7} Disentanglement
\item{3.8} Dimensional downfolding
{\bf \item{4.}  Low-Energy Solver}
\item{4.1} Dynamical mean-field theory
\item{4.2} Variational Monte-Carlo method
\item{4.3} Path-integral renormalization group
{\bf \item{5.} Applications}
\item{5.1} Dynamics of semiconductors
\item{5.2} 3d transition metal and its oxides
\item{5.2.1} Transition metal
\item{5.2.2} SrVO$_3$
\item{5.2.3} VO$_2$
\item{5.2.4} Sr$_2$VO$_4$
\item{5.2.5} YVO$_3$
\item{5.2.6} Iron-based superconductors
\item{5.2.7} Organic conductors
{\bf \item{6.} Concluding Remark and Outlook}
\end{description}

\section{Introduction} \label{Introduction}  

Since the foundation of quantum mechanics, understanding and predicting properties of condensed matter from microscopic basis have continuously been a great challenge of modern science and technology. Behaviors of many electrons primarily determine the diversity and rich variety of materials in our environment with potential applications for future technology. At the same time, many electron systems have been a source of challenges of our intelligence on nature, because of their interacting and quantum mechanical nature. 

Among all, density functional theory (DFT)\cite{hohenberg64,kohn65} offers a standard method for calculating electronic structure of real materials.
Local density approximation (LDA)\cite{kohn65} and generalized gradient 
%\textcolor{red}{
approximation
%} 
(GGA) offer practical ways, and reasonably predict  physical properties in a wide range of materials if we consider cases with weak electron correlations such as semiconductors. 

However, in materials with strong correlation effects such as transition metal compounds, organic conductors and rare earth compounds, these methods not only lead to a quantitative inaccuracy but also to a qualitatively wrong answer.  The most famous example is the mother materials of copper oxide superconductors such as La$_2$CuO$_4$, where DFT predicts a good metal with a half-filled band while La$_2$CuO$_4$ is a typical and good Mott insulator with a gap amplitude of about 2 eV.\cite{Mattheiss,Hybertsen}

In LDA, the many-body Schr\"{o}dinger equation is replaced by the Kohn-Sham equation 
%\textcolor{red}{
\begin{eqnarray}
%[-\nabla^2 + V_{\rm ext} +V_H + V_{\rm XC}] \psi_{k,j}&=& \epsilon_{k,j} \psi_{k,j}, 
\left[-\frac{1}{2} \nabla^2 + V_{\rm ext} +V_H + V_{\rm XC}\right] \psi_{k,j}&=& \epsilon_{k,j} \psi_{k,j}, 
\label{eqn:l1}
\end{eqnarray}
%他の部分に合わせてファクター1/2をつけました}
which contains the external potential $V_{\rm ext}$ coming from nuclei and the Hartree term of the electron-electron Coulomb interaction $V_H$. The eigenfunction and the eigenvalue with the momentum $k$ and other quantum number $j$ such as orbital indices are denoted by $\psi_{k,j}$ and $\epsilon_{k,j}$, respectively. By solving this single-particle Schr\"{o}dinger equation, the ground state energy and the charge density are obtained.  Here, the electron correlation effect is accounted by the exchange correlation potential $V_{\rm XC}$. From the Hohenberg-Kohn theorem,\cite{hohenberg64} in principle, the solution of the Kohn-Sham equation gives the exact ground-state energy of the many-body system by a functional of the electron charge density.\cite{kohn65}  However, since we do not know how to treat $V_{\rm XC}$ exactly, we resort to approximations, for instance by LDA.

In LDA, the exchange correlation potential is replaced by
results of a uniform electron gas such as quantum Monte Carlo calculations etc.~\cite{ceperley80} Therefore, it becomes a good approximation when the electron density does not have large spatial variations, which is justified in a good metal with electron wavefunctions extended uniformly in space. However, in strongly correlated electron systems, where electrons become nearly localized and electron density fluctuations are large with large spatial dependence, the approximation becomes poor.  

Typical strongly correlated electron systems are found in transition metal compounds, where the Fermi level $E_F$ crosses $d$ bands. Rare earth compounds with $f$-electron bands crossing $E_F$ and organic conductors with $p$ bands at $E_F$ are also well known correlated electron systems. A characteristic and common feature of strongly correlated electron systems is that their bands crossing the Fermi level have narrow bandwidths. The origin of the narrow bands is that the spreads of $d$, $p$ and $f$ orbitals are relatively small as compared to lattice constants, which makes the overlap of two orbitals each on the neighboring atoms small. The relatively small spreads also make the local electron interactions large.  Experimentally, these compounds are often insulators and "bad metals".\cite{ImadaRMP}  

In addition, competitions of tendencies for various orders and fluctuations such as magnetic, charge and superconducting orders invalidate mean-field treatments including LDA. Electron correlations have to be treated at much higher level of accuracies. This is a grand challenge of first-principles calculations for electronic structure.
Strongly correlated electron systems have attracted interest as platforms of possible innovative devices and realizing functions and efficiencies beyond the semiconductor applications in the 20th century.  {\it Ab initio} methods hold a key of clarifying basic properties from the scientific points of view.     

Strong correlation effects appear not only in typical correlated electron materials, but also show up even in weakly correlated systems such as semiconductors, if excitations and dynamics are involved.\cite{rohlfing-louie98,albrecht,benedict98}
This typically emerges in excitonic effects, where an electron and a hole interact strongly with attractive interactions.  Dynamical fluctuations also generate effective interaction in a small energy scale such as van-der-Waals interaction and dispersive forces, where the force is mediated by dynamical electronic polarizations due to electron correlation effects.  Such dynamical fluctuations are beyond the tractability of LDA. However, these weak forces play essential roles in solutions, complex systems and biological systems.  For example, they are crucial in determining structures of proteins and DNA. In this article, these dynamical effects on dispersive forces are not discussed in detail, though it remains a challenge. 
 
%電子相関を取り入れる方法としては、主に量子化学の分野で配置相互作用（configuration interaction, CI)法が知られているが、この方法のみで固体中の電子のような自由度の大きな系を扱うことはできない。

%\section{Historical Overview}
%There exist a number of attempts to overcome the difficulty of LDA. 
%Actually, 
Methods of electronic structure calculations can be classified into two categories.\cite{kohn99}  
The first one is the density functional theory described above, where the ground state is obtained only from the charge density.
The other is the wavefunction method that explicitly seeks for solutions of many-body wavefunctions. One of the simplest wavefunction methods is, though not sufficient, the well known Hartree-Fock theory.
Variational wavefunction method and Monte Carlo method based on the path integral may also be regarded as wavefunction approaches. 
  
The density functional theory reduces the problem to a single-particle one through solving the Kohn-Sham equation. Here, the self-consistent 
equation is reduced to obtaining the charge density and the computational load is much smaller. On the other hand, the wavefunction method allows more flexibility of treating the electron correlation effects, giving us more information on the ground state while it is in general more time consuming even for the Hartree-Fock level. Within the limited computer power, the density functional method has thus been used more widely.  However, from the incentive for treating the electron correlation effects more accurately, the serious limitation of the density functional theory revealed recently has urged reexaminations of standard methods.  In fact, the standard density functional theory so far offers no ways of systematic improvements on this electron correlation problem.    

Electron correlation effects can be taken into account by considering the standard many-body perturbation theory, 
if the correlation effects are not too large.  When they interact each other, one can view that each electron is dressed by other electrons from the single-particle picture and moves in the cloud of other electrons.  The dressed electron is called a quasiparticle, where the particle-like identity is retained in a gedanken experiment of switching on the interaction gradually in an adiabatic fashion. The quasiparticle may, however, have a mass and a dispersion different from a bare electron.  This is manifested by the self-energy of electrons $\Sigma$, where the pole of the quasiparticle (dispersion) is renormalized as $\omega=\epsilon^*(k,\omega)$ with $\epsilon^*(k,\omega) =\epsilon(k) + \Sigma (k,\omega)$, depending on momentum $k$ and frequency $\omega$.  In the lowest order perturbation, $\Sigma$ is given by $GW$, where $G$ is the the electron Green's function $G({\bf k},\omega)$ obtained from the Kohn-Sham Hamiltonian and $W$ is the screened Coulomb interaction. The screened interaction is calculated based on the random phase approximation (RPA). This is called GW approximation as we describe details in \S \ref{GW approximation}.\cite{hedin,aryasetiawan98,aulbur99} This theory has succeeded in improving the gap of semiconductors and insulators as we see in \S \ref{Dynamics of semiconductors}.

To circumvent the failure of reproducing band gaps in correlated insulators, 
the LDA+U method has been developed.\cite{Anisimov91,Anisimov1,Anisimov2}
%\textcolor{red}{
%（後の部分も含めてLDA+Uの文献に1991年のAnisimovの論文を加えました。）}
It combines LDA with an artificially introduced ``$U$" term which raise the energy of electrons only for the unoccupied part to take into account the onsite Coulomb interaction in the same spirit as the Hartree Fock approximation.  

Recently, more thorough efforts have been made to overcome the difficulty of 
DFT by combining and utilizing the flexibility of the wavefunction method for a more accurate description of electron correlation effects. This hybrid approach allows reducing the heavy computational task of the wavefunction methods and simultaneously allows accurate solutions by improving wavefunction methods.  In this review, we figure out recent studies along this line and discuss achievements as well as future perspectives.       
%多体波動関数を直接計算する方法には
%配置相互作用法（CI法）やハートレーフォック法などが知られているが、これらはLDAなどを用いる
%密度汎関数法に比べて多大な計算時間を要する。また、CI法は固体中の電子のように自由度の大きな系全体に用いると計算時間が指数関数的に増大するため実用的ではなく、ハートレーフォック近似は電子相関に伴うゆらぎを考慮できないため、精度に問題がある。
%We first briefly outline the history of first-principles methodology for strongly correlated electron systems.

The density functional theory is formulated to give ground state energies as a functional of the electron density only. 
By extending this, several attempts have been made to represent not by the electron density functional but by functionals of more information. For example, one attempt is to represent by a functional of the whole electron Green's function $G({\bf k},\omega)$ depending on the momentum $k$ and the frequency $\omega$.
The density functional theory can be formulated as a theory to minimize a functional $\Gamma(n,V_{\rm XC})$ of the electron density $n(r)$ and the exchange correlation potential $V_{\rm XC}$. By extending it and replacing by an extremum problem for a functional of the Green's function $G(k,\omega)$ and its conjugate field $X$, one can have a formalism equivalent to that by the Luttinger-Ward or Baym-Kadanoff functional.\cite{kotliar06} 
As we discuss later, the dynamical mean field theory can be regarded as one of these attempts. The GW method can also be formulated by the Luttinger-Ward formalism. 
In addition, an attempt for a formalism including the two-body correlation functions in the functional has also been made.\cite{ziesche}

The origin of the failure of LDA in treating strongly correlated electrons is ascribed to reconstructions of electronic states near the Fermi level $E_{\rm F}$ taking place beyond the 
expectation by LDA. The electrons whose energies are far away from the Fermi level are either fully occupied or empty and do not have a polarizability even in the strongly correlated systems.  Since the electronic polarizability is large near the Fermi level, electron correlation effects appear in the energy window around the Fermi level in the order of the effective electron interaction (, which is typically several eV). Therefore, when the widths of bands near the Fermi level become comparable or even smaller than the effective interaction, the whole band structure of this band may be seriously reconstructed. Since this happens in the whole Brillouin zone, the reconstruction may happen locally, namely in a spatially inhomogeneous fashion.  This invalidates the applicability of LDA around the Fermi level. 
%that caused by the electron correlation effect.  
%フェルミレベル周辺の電子のエネルギーウィンドウを設定し、この範囲の電子についてのみクーロン相互作用の効果を考えればよい。
In other words, the LDA may give an adequate band structure in the global energy scale, while it is seriously reconstructed near the Fermi level in the range of the effective interaction ($\sim$ several eV), which constitutes a hierarchy structure in energy. 
Meanwhile physical properties of materials around or below the room temperature are determined in this low-energy part of the hierarchy. 

By considering this hierarchy structure, one can develop a first-principles method that starts from the density functional theory, and then eliminates the degrees of freedom far away from $E_{\rm F}$ by following the spirit of the renormalization group. This method of eliminating degrees of freedom and restricting the Hilbert space is called the downfolding method.\cite{Andersen,aryasetiawan04,Solovyev1,Solovyev2} The downfolding leaves an effective model represented only by the degrees of freedom near $E_{\rm F}$. Since the effective model contains only a small number of bands near the Fermi level, for which we call ``target band", it is constructed on the lattice in real space with this number of retained orbitals in the unit cell, as in the multi-band Hubbard-type models in Lagrangian forms in general or in Hamiltonian forms if the retardation effects caused by the downfolded (eliminated) bands are small. Recently, effective low-energy models obtained after the downfolding have extensively been employed and solved by accurate low-energy solvers to discuss strong correlation effects. 

\begin{figure*}[htb] 
\begin{center} 
\includegraphics[width=0.6\textwidth]{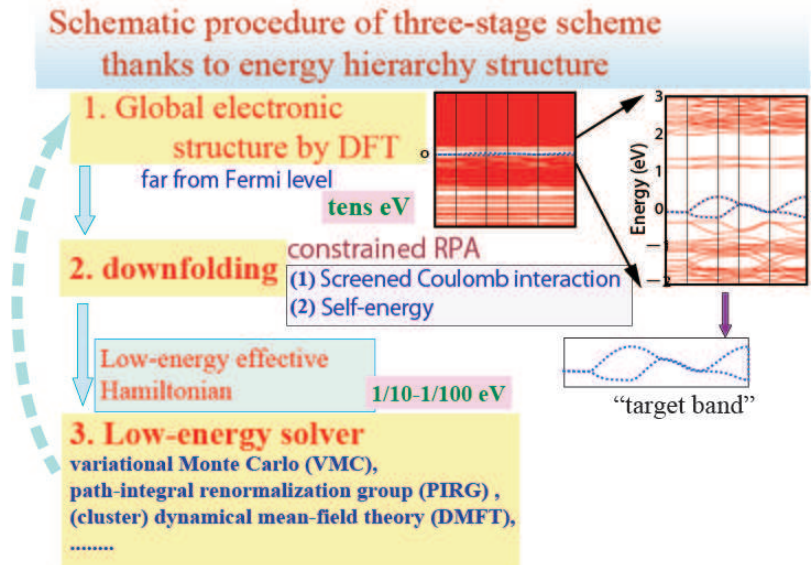}
\end{center} 
\caption{(Color online) Schematic procedure of three-stage scheme for hierarchical electronic structure reviewed in this article.} 
\label{fig:schematics} 
\end{figure*}

The whole procedures 
%of 1) calculating global band structure by DFT 2) downfolding 3) solving by low-energy solvers,
constitute the three-stage scheme of the renormalized multi-scale solvers (RMS).
We here summarize the present RMS method for the electronic structure calculation as the hybrid-type three-stage scheme as we illustrate in Fig.\ref{fig:schematics}:
\begin{enumerate}
\item Calculate the global band structure including bands far from the Fermi level by relying on DFT such as LDA, GGA or GW. 
\item Perform the renormalization procedure to downfold the higher energy degrees of freedom.  This yields effective models for low-energy degrees of freedom near the Fermi level.
\item Solve the low-energy effective model by an accurate low-energy solver.
\end{enumerate}
Although we do not describe, a possible iterative procedure to feed back the solution of the low-energy solver into the global structure in the first step taken until the self-consistent solution is a future issue when the low-energy solution seriously modifies the original global structure.

This article is organized as follows:
In \S \ref{Global Electronic Structure}, several points useful for the first stage of the three-stage RMS scheme in calculating global electronic structures are summarized. After an elementary remark on DFT in \S \ref{Density Functional Theory (DFT)}, we introduce basically three basis functions developed for solving Kohn-Sham equation. The first is the plane-wave basis suited for $sp$ electron systems, and the second and the third are the augmented wave and the muffin-tin orbital, respectively, suited for $d$ and $f$ electron systems.
In \S \ref{GW approximation}, we review GW method as a method to take into account electron correlation effects for the global band structure within a perturbative approach.
In \S \ref{LDA+U}, the LDA+U method proposed to implement Hartree-Fock level corrections to LDA is reviewed.  
In \S \ref{Downfolding}, the second stage of the three-stage scheme of RMS is introduced for the purpose of downfolding and eliminating the degrees of freedom far from the Fermi level. Effective low-energy models are derived from this downfolding procedure.  Tools for solving the derived effective models are listed in  \S \ref{Low-Energy Solver}, where we review, dynamical mean-field theory in \S \ref{Dynamical mean-field theory}, many-variable variational Monte Carlo method in \S \ref{Variational Monte Carlo method}, and path-integral renormalization group method in \S \ref{Path-integral renormalization group}. Section \ref{Applications} describes some applications to various materials as dynamics of semiconductors, transition metal compounds, and organic conductors. Section \ref{Concluding Remarks} is devoted to summary and future scope.

%%%%%%%%%%%%%%%%%%%%%%%%%%
\section{Global Electronic Structure} \label{Global Electronic Structure}
\subsection{Density Functional Theory (DFT)} \label{Density Functional Theory (DFT)} 
Understanding properties of matter from first principles 
is a central problem in condensed matter physics. 
The properties are, in principle,  described by the many-body Hamiltonian, 
\begin{eqnarray}
H &=& 
- \sum_i  \frac{\hbar^2}{2m} \nabla_i^2 
+ \frac{1}{2} \sum_{i \neq j} \frac{e^2}{| {\mathbf r}_i - {\mathbf r}_j | } 
\nonumber \\
&-& 
 \sum_{i,I} \frac{Z_I e^2}{| {\mathbf r}_i - {\mathbf R}_I | }
+ \frac{1}{2} \sum_{I \neq J} \frac{Z_I Z_J e^2}{| {\mathbf R}_I - {\mathbf R}_J | }
\; ,
\label{Schroedinger}
\end{eqnarray}
where the first term represents the kinetic energy of electrons. 
The second, third and fourth terms are interactions between 
electrons, electron-nucleus, and nuclei, respectively.
Electrons are labeled by real space coordinate ${\mathbf r}$ with suffices with the lower case as $i$ and $j$, while nuclei are denoted by coordinate ${\mathbf R}$ with the upper-case suffices as $I$ and $J$. The electronic bare mass and charge are $m$ and $e$, while the atomic number is denoted by $Z$.
The spin degrees of freedom, relativistic effects and quantum effects 
of nuclei are neglected for simplicity. 
The Hamiltonian is solved exactly only in very limited cases, 
hence developing a practical procedure for treating many-electron systems 
has long been an important issue. 

DFT gives an approximate but reasonably accurate and 
practical method for this problem. 
DFT is based on the Hohenberg-Kohn (HK) theorem \cite{hohenberg64}, 
that asserts: 

\begin{description}
\item{Theorem 1)} For any many electron systems under the influence of an external potential 
$V_{\rm ext}({\bf r})$, the potential is, apart from a trivial additive constant, 
a unique functional of the one electron density of the ground state.

\item{Theorem 2)} For any external potential, 
there exists a total energy functional of one electron density $n({\bf r})$, 
\begin{equation}
E_{\rm tot}[n] = F[n] + \int V_{\rm ext}({\bf r}) n({\bf r}) d{\bf r} \;, 
\label{eq:hk}
\end{equation}
where $F[n]$ is a {\it universal} functional of  $n({\bf r})$. 
The ground state energy of the many electron system is the minimum of $E_{\rm tot}[n]$, 
and associated $n({\bf r})$ is the electron density of the ground state.
\end{description}
The HK theorem was originally proved for systems having non-degenerate ground state. 
Later on it was extended to degenerate cases by Levy \cite{levy79}.
The theorem is an exact theory of interacting many electron systems. 
Since the Hamiltonian is determined by the ground state electron density, 
all properties of matter are implicitly determined by the density. 
This gives a justification to take the electron density as a basic variable of the theory.

In DFT, the ground state total energy and density are obtained 
by minimizing the total energy functional with respect to $n({\bf r})$. 
The formulation may be regarded as a rigorous extension of the Thomas-Fermi (TF) 
theory \cite{thomas27,fermi27}, in which 
the total energy functional is given as 
\begin{eqnarray}
E_{\rm tot}^{\rm TF} [n] &=& T^{\rm TF} + \int  V_{\rm ext}({\bf r}) n({\bf r}) {\rm d}^3 r 
\nonumber  \\
& & + \frac{1}{2} \int \frac{n({\bf r}) n({\bf r}')}{| {\bf r} - {\bf r}' |} {\rm d}^3 r {\rm d}^3 r'
\; , \label{eq:tf} \\
T^{\rm TF} &=& \frac{3}{10}(3 \pi^2)^{2/3} \int n({\bf r})^{5/3} {\rm d}^3 r 
\;.
\end{eqnarray}
The kinetic energy in the TF theory is approximated as the integral of the local part over space, 
where the local part is the mean kinetic energy per electron multiplied by the electron density at the position. 
The TF theory was proposed in the 1920's and applied to real materials. 
However, the method turned out to be unsatisfactory not only 
quantitatively but also qualitatively: The theory cannot describe chemical bonds between atoms. 
It was clarified that the error comes mainly from the approximation for the kinetic energy. 

\begin{figure*}[htb]
\begin{center} 
\includegraphics[width=0.65\textwidth]{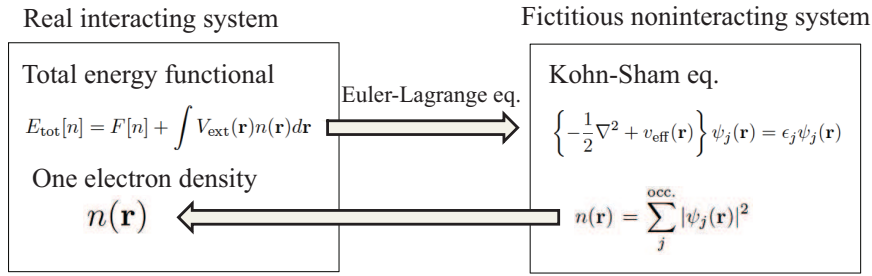}
\end{center} 
\caption{(Color online) Schematic representation of density functional theory} 
\label{fig:dft} 
\end{figure*}

Much better results are obtained by replacing the kinetic term 
with that of the noninteracting electron systems. 
This is nothing but the Hartree theory which was developed 
soon after the TF theory \cite{hartree}. 
Inspired by this observation, in 1965 Kohn and Sham \cite{kohn65} 
proposed a practical procedure for DFT \cite{kohn99}. 
They introduced an auxiliary noninteracting electron system 
that obeys the following single-particle equation (Fig.\ref{fig:dft})
\begin{equation}
\left\{ -\frac{1}{2} \nabla^2 + v_{\rm eff}({\bf r}) \right\} \psi_{j}({\bf r}) = \epsilon_{j} \psi_{j}({\bf r}) \;.
\label{eq:ks}
\end{equation}
The electron density and the kinetic energy of the system are
\begin{eqnarray}
n({\bf r}) &=& \sum_{j}^{\rm occ.} | \psi_j ({\bf r}) |^2 \;, \label{eq:nr} \\
T_{\rm s} &=& \sum_{j}^{\rm occ} \langle \psi_j | -\frac{1}{2} \nabla^2 | \psi_j \rangle \;.
\label{eq:ts}
\end{eqnarray}
Coming back to the original interacting system, 
the functional $F[n]$ in eq.(\ref{eq:hk}) can be divided as 
\begin{equation}
F[n] = T_s[n] + 
\frac{1}{2} \int \frac{n({\bf r}) n({\bf r}')}{| {\bf r} - {\bf r}' |} {\rm d}^3 r {\rm d}^3 r'
+ E_{\rm xc}[n] \;.
\label{eq:fn}
\end{equation}
The first term is the kinetic energy, but it is for the 
noninteracting system defined in eq.(\ref{eq:ts}), 
not the true kinetic energy of the interacting electrons. 
The second term is the electrostatic energy (Hartree energy). 
The last term, so-called exchange-correlation energy, contains all the remaining 
contributions including the difference between the noninteracting and 
interacting kinetic energies. 
%The total energy functional eq.(\ref{eq:hk})(\ref{eq:fn}) is minimized by solving the Euler-Lagrange equation. 
Now we assume that the ground state electron density of the interacting system 
can be represented as eq.(\ref{eq:nr}). 
Then, the stationary condition for the total energy functional eqs.(\ref{eq:hk}) and (\ref{eq:fn}) is satisfied when 
the self-consistent solution of eq.(\ref{eq:ks}), with the effective potential
\begin{equation}
v_{\rm eff}({\bf r}) = \int \frac{n({\bf r}')}{|{\bf r}-{\bf r}' |} {\rm d}^3 r' +
\frac{\delta E_{\rm xc}[n]}{\delta n({\bf r})} + V_{\rm ext}({\bf r}) \;. 
\label{eq:veff}
\end{equation}
is achieved.
%coupled with eq.(\ref{eq:nr}). 
The set of equations (\ref{eq:ks}), (\ref{eq:nr}) and (\ref{eq:veff}) is called Kohn-Sham equation. 

The remaining question is how to determine the exchange-correlation energy functional. 
First of all, the exact functional is not known, and trials to improve the functional is a hot topic even today. 
Formally the functional can be written as 
\begin{equation}
E_{\rm xc} [n] = \int e_{\rm xc}({\bf r}; [n]) n({\bf r}) {\rm d}^3 r \;,
\end{equation}
where $e_{\rm xc}({\bf r}; [n])$ is the exchange-correlation energy per electron at the position ${\bf r}$. 
In principle, full information of the density $n$, not only the value at ${\bf r}$ is necessary 
to determine $e_{\rm xc}$. 
A simple and most widely used approximation is the LDA
proposed in the Kohn-Sham work.\cite{kohn65} 
The LDA approximates $e_{\rm xc}$ to be that of a uniform electron gas of 
the density at the position. Namely, the energy functional is expressed as follows. 
\begin{equation}
E_{\rm xc}^{\rm LDA} [n] = \int e_{\rm xc}(n({\bf r})) n({\bf r}) {\rm d}^3 r \;.
\end{equation}
The explicit formula for $e_{\rm xc}$ has been proposed by several authors 
based on a perturbation theory \cite{wigner34}, the RPA~\cite{barth72}, 
or more accurately by the fit \cite{vosko80,perdew81} 
to the Ceperley-Alder quantum Monte Carlo simulation \cite{ceperley80}. 

The LDA is by construction exact in the limit of a uniform electron density, whereas 
the approximation gets worse as the spatial variation of 
the electron density becomes strong. 
%However, even in such cases of atoms and molecules, the LDA yields 
%reasonable the total energy differences. 
Typically the lattice constant of solids and bond lengths between atoms are computed to be 
within the 2-3 \% error to experiments. 
The accuracy of the ionization energy in molecules and 
cohesive energy in solids is with 10-20 \% errors.
The high accuracy is partially rationalized by the fact that the LDA satisfies 
a sum rule for the exchange-correlation hole \cite{gunnarsson76}.

An obvious modification of LDA is inclusion of density gradient effects
$\nabla n({\bf r})$. % It was mentioned already in the Kohn-Sham paper \cite{kohn65}. 
However, it turned out that the simple low-order gradient expansion 
does not improve but worsen the results in some cases. 
This suggests the spatial variation is so strong in real materials 
that simple gradient expansion 
does not work. Instead it may be a better idea to include the effect of 
gradient corrections while keeping various asymptotic behaviors and sum rules, 
such as the one for the exchange-correlation hole mentioned above. 
This is called the generalized gradient approximation (GGA), 
\begin{equation}
E_{\rm xc}^{\rm GGA}[n] = \int e_{\rm xc}^{\rm GGA}(n,\nabla n ) n({\bf r}) \;.
\end{equation}
There are many explicit formula of GGA proposed by today \cite{becke88,pw91,pbe96}.
The GGA tends to give more accurate results than the LDA 
in the atomization energy, cohesive energy, description of magnetism and so on.

While the total energy and the electron density are obtained from the total energy functional, 
the Kohn-Sham equation merely represents a fictitious system which is introduced 
to carry out the minimization. 
One may want to regard the eigenvalues $\{ \epsilon_j \}$ as the orbital energies. 
However, this interpretation is not justified rigorously. 
Physical meaning of the Kohn-Sham energy is known only for the highest occupied state. 
It is proved that $-\epsilon_j$ for the state is the ionization energy 
\cite{almbladh85}. 
For other states, a similar relation 
$\epsilon_j = \partial E_{\rm tot} / \partial f_j$ holds 
\cite{janak75}, but 
it is not the electron addition/removal energy. 
Here $f_j$ is the occupation number of the state $j$, namely, 
$n({\bf r}) = \sum_j f_j | \psi_j ({\bf r}) | ^2$.
Mathematically the Kohn-Sham eigenvalue is a Lagrange multiplier 
corresponding to the orthonormal condition $\langle \psi_i | \psi_j \rangle = \delta_{ij}$. 

In practice, the Kohn-Sham energy is useful information for understanding 
the electronic properties. 
The overall feature of the electronic structure is captured in LDA/GGA, and 
the Fermi surface is reasonably accurate in many materials. 
However, the band gap of semiconductors and insulators are underestimated significantly. 
This is the case for almost all materials including weakly correlated systems.
The low-energy electronic structure in strongly correlated materials are often very different 
from measurement, and sometimes qualitatively wrong. 

%%%%%%%%%%%%%%%%%%%%%%%%%%%%%%%%%%%%
\subsection{Basis Functions for DFT} \label{Basis Functions for DFT}
The first step of describing the low-energy properties of correlated materials is the global electronic structure 
by means of DFT in LDA, GGA or whatever.
Various numerical techniques have been developed to solve the Kohn-Sham equation 
accurately and efficiently. Below we will see a few key ingredients. 

\subsubsection{Plane wave} \label{Plane wave}
One of the most widely used basis functions is the plane wave basis set, where the wavefunction of an electron with the momentum ${\bf k}$ and quantum number $n$ are expanded as 
\begin{equation}
\psi_{{\bf k}n}({\bf r}) = \sum_{\bf G} C_{{\bf k}n}^{\bf G} e^{i ({\bf k + G}) \cdot {\bf r}}
\;.
\label{eq:planewave}
\end{equation}
A great advantage of the plane-wave basis is that 
numerical accuracy is improved systematically by increasing the number of plane waves 
({\bf G} points in eq.(\ref{eq:planewave})).
However, the calculation becomes tremendously heavy if a naive plane-wave expansion is adopted, 
because huge number of plane waves is required to express localized core electrons. 
To make calculations feasible, the interactions between core and valence electrons are
replaced with a pseudopotential. 
The pseudopotential eliminates explicit treatment of the core electrons from the Kohn-Sham equation. 
The valence orbitals are also modified to be smoother than the true (all-electron) ones near the core region, 
which reduces computational cost drastically. 
The concept of the pseudopotential dates back to the 1930's \cite{fermi34}. 
It has been developed continuously, and non-empirical pseudopotential appeared in the late 1970's. 

The pseudopotential is constructed in such a way that the scattering properties of valence electrons 
reproduce those of the all-electron calculation accurately. 
The pseudo wavefunction agrees with the all-electron wavefunction outside a certain radius $r_c$, 
whereas for $r < r_c$, the pseudo wavefunction is nodeless and much smoother than the all-electron one.
The pseudopotential can be written in the following form,
\begin{equation}
V_{\rm ps}({\bf r}) =  V_{\rm local}(r) + \sum_{lm} | Y_{lm} \rangle \delta V_l (r) \langle Y_{lm} | \;.
\end{equation}
The first term is independent of angular momentum $l$ and is called local part. 
The second term is dependent on $l$, and called non-local part. 
Since (i) the pseudo wavefunctions are equal to the 
all-electron ones at $r > r_c$ and (ii) the all-electron potential is independent of $l$, 
it follows that $\delta V_l = 0$ at $r > r_c$.
Various ways of pseudopotential construction have been proposed so far 
to increase accuracy, transferability  and computational efficiency  
\cite{hsc79,bhs82,troullier91,vanderbilt90}. 
For more technical details, see e.g. Refs.\citen{payne92,martin94}.

%%%%%%%%%%%%%%%%%%%%%%%%%%%%%%%%
\subsubsection{Augmented plane waves and muffin-tin orbital}
While the plane-wave basis is suitable for $sp$ electron systems, for systems containing $d$ and $f$ 
electrons, the computational cost becomes heavy because of localized nature of the electrons. 
All electron methods are powerful in such cases. 
The augmented plane waves (APW) and muffin-tin orbital (MTO) are commonly used basis functions. 

\begin{figure}[htbp] 
\begin{center} 
\includegraphics[width=0.4\textwidth]{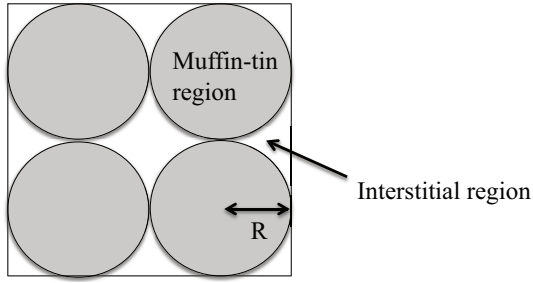}
\end{center} 
\caption{
%\textcolor{red}{
In both the APW and MTO methods, the space is divided into the 
muffin-tin region and the interstitial region. 
The basis function is constructed by approximating the potential to be 
spherically symmetric in the muffin-tin region, and constant in the 
interstitial region.}
%} 
\label{fig:muffin} 
\end{figure}

In the APW method \cite{slater37}, space is divided into two parts: the muffin-tin region and the interstitial region (Fig.\ref{fig:muffin}).
Inside the muffin-tin region, the basis function is constructed by solving the Schr\"odinger equation 
for the spherically symmetrized potential at a particular energy $\epsilon$. 
The solution $\phi$ is connected at the muffin-tin surface to the plane wave. 
Thus, the APW is expressed as 
\begin{equation}
\chi_{\bf k +G} ({\bf r};\epsilon) = 
\left \{
\begin{array}{l}
\sum_{lma} C_{lma}({\bf k+G}) \phi_{lma}({\bf r};\epsilon) \;, (r<R) \\
e^{i ({\bf k+G}) \cdot{\bf r}} \;, (r>R)
\end{array}
\right. 
\end{equation}
where $a$ is the index for a muffin-tin. 

The Kohn-Sham eigenvalue is obtained as a solution of the secular equation 
$| H - \epsilon S |  = 0$, where $S$ is the overlap matrix between the APW basis functions. 
Because the APW is implicitly energy dependent, the secular equation is a nonlinear equation. 
It is not a general eigenvalue problem, but one has to search for the  selfconsistency of $\epsilon$ numerically, 
which is computationally demanding. 
In 1975, Andersen proposed a linear method to solve this problem \cite{andersen75}. 
The energy dependent $\phi_{lma}({\bf r};\epsilon)$ is expanded at around a fixed energy 
$\epsilon_{la}$, and approximated as 
\begin{equation}
A_{lma} \phi_{lma}({\bf r};\epsilon_{la}) +
B_{lma} \dot{\phi}_{lma}({\bf r};\epsilon_{la}) \;,
\end{equation}
where $\dot{\phi}$ is the energy derivative of $\phi$. The coefficients $A$ and $B$ are 
determined from a matching condition at $r=R$ up to the first-derivative. 
The linearized function is called linear augmented plane wave (LAPW). 
It is the most accurate method among electronic structure methods available today.

Muffin-tin orbital (MTO) is another basis function for the all-electron method.
It is defined by 
\begin{equation}
\chi_{lma} ({\bf r};\epsilon) = 
\left \{
\begin{array}{l}
\phi_{lma}({\bf r};\epsilon) + \cot\delta_{la}(\kappa) j_l(\kappa r)Y_{lm}(\theta,\varphi), \;\\
                                                              \hfill              (r<R) \; \\
n_l(\kappa r)Y_{lm}(\theta,\varphi), \ \ (r>R) \;.
\end{array}
\right. 
\label{eq:mto}
\end{equation}
Here $j_l$ is the spherical Bessel function, and
$n_l$ is the spherical Hankel (Neumann) function for $\kappa^2 >0$  ($\kappa^2 <0$).
$\delta_{la}$ is determined by matching the logarithmic derivative at the muffin-tin boundary.
The MTO is not the eigenfunction of a single muffin-tin potential because of the second 
term for $r<R$ in eq.(\ref{eq:mto}), but it is useful for solving the many muffin-tin problem.  
The second term expresses approximately the tails of the MTO's centered at other sites. 
It reduces the number of required basis functions compared to other methods.

The linear muffin-tin orbital (LMTO) \cite{andersen75} is the linear method for the MTO, 
in which the following approximations are adopted.
Firstly the energy $\epsilon$ is fixed. It makes the basis function energy independent. 
Secondly, $j_l Y_{lm}$ in eq.(\ref{eq:mto}) is replaced with 
\begin{equation}
- \frac{\dot{\phi}_{lma}({\bf r}) }{\kappa \frac{d}{d\epsilon}\cot\delta_{la}} \;. 
\end{equation}
This form is chosen to satisfy the following condition at the fixed energy.
\begin{equation}
\frac{d}{d\epsilon} \chi_{lma}({\bf r}) = 0 \;.
\end{equation}
Thirdly, the $n_l Y_{lm}$ term is replaced with a linear combination of $\dot{\phi}_{lma}$ in other muffin-tins. 

Further efficiency is achieved by
approximating the whole space as a sum of muffin-tins. 
In this atomic sphere approximation (ASA), 
spherical anisotropy of the potential inside the muffin-tins is neglected.
The muffin tin radius is chosen so that the interstitial region becomes small, 
while the overlap between muffin-tins is small as well, since 
both the interstitial region and the overlap region is neglected in the ASA. 

Although the ASA is not accurate in materials with large empty space or strong anisotropy, 
the LMTO-ASA is a computationally cheap and powerful method for closed-pack and localized-electron 
systems. The ASA makes mapping onto lattice models easy. 
Therefore, the LMTO-ASA has played an important role in the 
development of electronic structure techniques for strongly correlated materials. 
For example, both the LDA+U and the LDA+DMFT methods were developed on top of 
the LMTO-ASA in the beginning and extended to other basis sets later. 

To go beyond the linear approximation and improve the accuracy, NMTO was 
developed recently \cite{andersen00}.  
In the LMTO, $\phi$ is computed at a fixed energy, 
whereas the NMTO basis is a linear combination of $N$ such functions evaluated at 
$N$ different energies.

%%%%%%%%%%%%%%%%%%%%%%%%%%%%%%%%%%%
\subsection{GW approximation} \label{GW approximation} 
Many-body perturbation expansion is a traditional theory for interacting electron systems. 
Expansion in the Coulomb interaction, $v({\bf r}) = 1 / | {\bf r} | $, 
gives the Hartree-Fock (HF) approximation \cite{fock30} in the lowest order but in a self-consistent  manner. 
The HF self-energy is a sum of two terms: 
the static Coulomb interaction (Hartree term) and the exchange interaction (Fock term). 
Assuming that the one-electron wavefunction is not modified by electron addition or removal, 
it can be shown that the eigenvalues of the Hartree-Fock equation are electron addition or removal energies  
(Koopman's theorem) \cite{koopman33}.
In other words, the eigenvalue is equal to the total energy difference between 
the $N$ and $N\pm1$ electron systems. 

The HF theory is a good approximation in finite systems, but 
the accuracy goes down in extended systems. 
In fact, the result is often even worse than the Hartree theory in solids. 
%Looking at the band gap of insulators, 
The Hartree theory yields 
too small band gap of insulators. In the HF theory, the exchange term pushes down 
occupied energy levels and widen the band gap. 
The correction is, however, too large, consequently the HF overestimates the band gap. 
Another well-known drawback in the HF theory is the anomaly at 
the Fermi level. The Fermi velocity in metals diverges, 
and the density of states vanishes at the Fermi level. 
(Comparing to the DFT-LDA, the HF approximation is computationally 
more demanding because of non-local Fock term, and 
the results are worse in extended systems.)
These facts suggest that proper treatment of screening effects is 
crucial in solids. A sensible way would be the expansion 
in the series of the {\it screened} Coulomb interaction. 
This is the basic idea of the GW approximation.\cite{hedin,aryasetiawan98,aulbur99}

Before the GW method is established, there are a few works reported in the 1950's. 
Quinn and Ferrel studied the electron gas, and attempted to include correlation effects in the form
of the GW approximation, with several other approximations \cite{quinn58}. 
DuBois did a related work on the electron gas in high density region \cite{dubois59}.  
Baym and Kadanoff also mentioned a GW form of the self-energy 
in their paper on the conserving approximation \cite{baym61}. 
In 1965, Hedin derived an exact closed set of equations for the self-energy 
in which the self-energy was expanded in powers of the screened Coulomb interaction.
In particular, the first term in the expansion yields the GW approximation, 
which can be viewed as the time-dependent Hartree approximation 
for the self-energy.

In his 1965 paper, Hedin presented a full self-energy calculation for 
the electron gas. Shortly after that, Lundqvist did extensive calculations of the electron
gas self-energy and spectral functions \cite{lundqvist}. 
The calculation is so heavy that it took 20 years before 
the applications to real materials were reported. 
Hybertsen and Louie performed the GW calculation of semiconductors 
and found that the GW approximation cures the band gap problem of DFT \cite{hybertsen85}. 
Their seminal work used the plasmon-pole approximation, which is 
a simplification for the frequency dependence of the dielectric function. 
Soon after this, Godby, Schlueter and Sham carried out the GW calculation 
without the plasmon-pole approximation, and obtained similar results \cite{godby}. 
These calculations were performed using the pseudopotential methods 
based on the plane-wave basis. However, pseudopotential GW calculations 
for materials containing localized electrons are computationally demanding. 
This difficulty motivated all-electron GW calculations. 
The all-electron GW calculations were done by Aryasetiawan in 1990's 
\cite{aryasetiawan92,aryasetiawan95}. 
With the rapid increase in computer performance, GW calculations can now be
performed for systems containing more than 50 atoms. 
%\textcolor{red}{ここまでがGWのhistorical view.}

The basic quantity of the GW approximation is the one-particle Green's function 
defined by 
\begin{equation}
G(1,2) = i \langle N | T[ \hat{\psi}(1) \hat{\psi}^{\dagger}(2)] | N\rangle \;,
\label{G12}
\end{equation}
where $| N \rangle$ is the ground state of the $N$ electron system, 
$T$ is the time-ordered product, 
$\hat{\psi}$ and $\hat{\psi}^{\dagger}$ are field operators, 
and $1 = ({\bf r}_1,t_1)$ is the composite variable. 
Starting from the equation of motion of the Green's function, 
Hedin derived a set of equations between
the Green's function $G$, self-energy $\Sigma$, screened Coulomb interaction $W$, 
polarization function $P$, and vertex function $\Gamma$: 

\begin{equation}
\Sigma(1,2) = i \int G(1,3^+)W(1 4) \Gamma(3,2,4) {\rm d} (34) \; , \label{eq:sigma}
\end{equation}
\begin{equation}
W(1,2) = v(1,2) + i \int v(1,3) P(3,4) W(4,2) {\rm d} (34) \; , \label{eq:w}
\end{equation}
\begin{equation}
P(1,2) = -i \int G(1,3)\Gamma(3,4,2)G(4,1^+) {\rm d}(34) \;, \label{eq:p}
\end{equation}
\begin{eqnarray}
\Gamma(1,2,3) &=& \delta(1-2)\delta(1-3) \nonumber \\
&+&  \int \frac{\delta \Sigma(1,2)}{\delta G(4,5)}G(4,6) 
G(7,5) \Gamma(6,7,3) {\rm d}(4567) \;, \nonumber \\
&& \label{eq:gamma}
\end{eqnarray}
\begin{equation}
G(1,2) = G_0(1,2) + \int G_0(1,3)\Sigma(3,4)G(4,2) {\rm d}(34) \,. \label{eq:dyson}
\end{equation}

\begin{figure}[htbp] 
\begin{center} 
\includegraphics[width=0.4\textwidth]{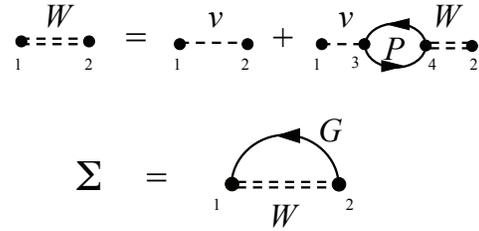}
\end{center} 
\caption{Feynman diagrams included in the GW approximation. Solid lines with arrows represent Green's function represented by eigenstates of Kohn-Sham equation.  Dashed lines represent the bare Coulomb interaction $v$, while double dashed lines are for the screened interaction $W$. } 
\label{fig:gw} 
\end{figure}

A key to solving the equations is the vertex function. To solve the
integral equation (\ref{eq:gamma}), we must know $\delta\Sigma/\delta G$, 
which requires an explicit expression for the self-energy in terms of the Green
function. But the self-energy depends in turn on the vertex as can be seen
in (\ref{eq:sigma}). 
In the GW approximation, the vertex function is approximated as 
\begin{equation}
\Gamma(1,2,3)= \delta (1-2)\delta(2-3) \;,
\end{equation}
which leads to the following form of the self-energy 
(, to which the name of the approximation owes).
\begin{equation}
\Sigma^{\rm GW}(1,2) = i G(1,2)W(1,2) \; .
\label{eq:sigmagw}
\end{equation}
The diagrams for eqs.(\ref{eq:w}) and (\ref{eq:sigmagw}) are illustrated in Fig.~\ref{fig:gw}.
We note that the Hartree contribution in the self-energy shown in the first line of Fig.~\ref{fig:SigmaDiagram}
is not considered, because it is already considered in the Green's function of Kohn-Sham equation in the LDA level.
The polarization function is expressed as 
\begin{equation}
P(1,2) = -i G(1,2)G(2,1^+) \;. \label{eq:rpa}
\end{equation}
The GW approximation is the lowest order expansion of the 
self-energy in the screened Coulomb interaction. 
Considering that the Fock term is $\Sigma_{\rm x}(1,2) = i G(1, 2)v(1-2)$, 
GW may be regarded as the screened Hartree-Fock approximation 
using the screened Coulomb interaction in the RPA. 

Most applications to real materials are carried out non self-consistently 
on top of the LDA solution. The starting Green's function equivalent to eq.(\ref{G12}) is 
\begin{equation}
G_{0}({\bf r},{\bf r}';\omega) = 
\sum_v^{\rm occ} \frac{\psi_v({\bf r})\psi_v^*({\bf r}')}{\omega - \epsilon_v - i \delta} +
\sum_c^{\rm unocc} \frac{\psi_c({\bf r})\psi_c^*({\bf r}')}{\omega - \epsilon_c + i \delta}
\label{eq:g0}
\; ,
\end{equation}
where $\psi_v$ and $\psi_c$ are the eigenstates of Kohn-Sham equation. 
The polarizability eq.(\ref{eq:rpa}) is then expressed as 
\begin{eqnarray}
P_0({\bf r},{\bf r}';\omega) &=& \sum_{v}^{\rm occ}\sum_{c}^{\rm unocc}
\left\{
\frac{\psi_v({\bf r}) \psi_c^*({\bf r}) \psi_c({\bf r}') \psi_v^*({\bf r}')}{\omega - (\epsilon_c - \epsilon_v) + i\delta} 
\right.
\nonumber \\
&& 
\left.
- \frac{\psi_v^*({\bf r}) \psi_c({\bf r}) \psi_c^*({\bf r}') \psi_v({\bf r}')}{\omega + (\epsilon_c - \epsilon_v) - i\delta}
\right\}
\;,
\label{eq:p0}
\end{eqnarray}
from which $W$ is computed. 
Putting the obtained $W$ into eq.(\ref{eq:sigmagw}), 
$\Sigma$ is computed. 
Representing eq.(\ref{eq:sigmagw}) in the frequency domain, 
the self-energy can be divided into two terms. 
One is the contribution from the poles of the Green's function
\begin{equation}
\Sigma_{\rm SEX}({\bf r},{\bf r}';\omega) = - \sum_{v}^{\rm occ}
\psi_v({\bf r}) \psi_v^*({\bf r}')W({\bf r},{\bf r}'; \epsilon_v - \omega) \;. \label{eq:sex} 
\end{equation}
This is the same form as the Fock potential in the HF approximation, 
but the Coulomb interaction is replaced with the screened one. 
It is called screened exchange term.
The other term comes from the poles of $W$, 
called Coulomb hole term. 
\begin{equation}
\Sigma_{\rm COH}({\bf r},{\bf r}';\omega) = \sum_{i}^{\rm all} 
\psi_i({\bf r}) \psi_i^*({\bf r}'){\rm P} \int_{0}^{\infty}
\frac{B({\bf r},{\bf r}';\omega')}{\omega - \epsilon_i - \omega'} d\omega'
\;, \label{eq:coh}
\end{equation}
where $B$ is the spectral function of $W$. 
The meaning of this term is understood by taking the limit, 
\begin{equation} 
\omega-\epsilon_i \rightarrow 0 \;. 
\label{eq:cohsex}
\end{equation}
In this static Coulomb hole + static exchange (COHSEX) approximation, the Coulomb hole term is 
reduced to 
\begin{equation}
\Sigma_{\rm COH}({\bf r},{\bf r}') \rightarrow \frac{1}{2}
\delta({\bf r}-{\bf r}') 
\{v({\bf r},{\bf r}') - W({\bf r},{\bf r}';\omega=0) \} \;.
\label{eq:scoh}
\end{equation}
This can be interpreted as the change in the potential 
associated with redistribution of the electron density, 
which is induced by an added electron at the position ${\bf r}$.
The factor 1/2 comes from the adiabatic growth of the screened interaction. 

Once the self-energy is computed, the Green's function is 
obtained by solving the Dyson equation (\ref{eq:dyson}).
Its spectral function, $A({\bf k},\omega)$, 
\begin{equation}
A({\bf k},\omega) = \frac{1}{\pi} \sum_{n} 
| \langle \psi_{{\bf k}n} | {\rm Im} G(\omega) 
| \psi_{{\bf k}n} \rangle | 
\; ,
\end{equation}
is the one electron excitation spectrum for electron addition or removal. 
This is the quantity to be compared to (inverse) photoemission measurements.
The quasiparticle energy, which is the peak position of the spectral function, is given by
\begin{equation}
E_{{\bf k}n}^{\rm QP} = \epsilon_{{\bf k}n} + {\rm Re} \langle \psi_{{\bf k}n} | \Sigma(E_{{\bf k}n}^{\rm QP}) - v_{\rm xc} | \psi_{{\bf k}n} \rangle
\;.
\label{eq:egw}
\end{equation}
It is often assumed that the self-energy is weakly frequency dependent. 
Then the self-energy is safely expanded around the Kohn-Sham eigenvalue. Taking up to the linear order, eq.(\ref{eq:egw}) is 
reduced to 
\begin{equation}
E_{{\bf k}n}^{\rm QP} = \epsilon_{{\bf k}n} + Z_{{\bf k}n} {\rm Re} \langle \psi_{{\bf k}n} | \Sigma(\epsilon_{{\bf k}n}) - v_{\rm xc} | \psi_{{\bf k}n} \rangle
\;,
\label{eq:egw2}
\end{equation}
where the renormalization factor $Z$ is defined by 
\begin{equation}
Z_{{\bf k}n} = \left(1 - \partial {\rm Re} \Sigma_{{\bf k}n} / \partial \omega |_{\omega=\epsilon_{{\bf k}n}} \right)^{-1} 
\; .
\end{equation}
%The linear expansion is valid in weakly correlated materials, where it should be checked carefully 
%in strongly correlated materials. 

\subsection{LDA+U method} \label{LDA+U}
For the Mott insulator, the LDA and GGA often give qualitatively wrong answer. 
Because of the strong repulsion in short-ranged part of the Coulomb interaction,
electrons cannot come close each other and the consequential segregated localization is the origin of the Mott insulator.
However, LDA treats the distribution of the interacting electrons as that averaged over the space. Since it does not 
well take into account the electron configuration avoiding each other, it predicts a metal erroneously.  
On the other hand, if a symmetry breaking such as antiferromagnetic or orbital order takes place, 
the electron distribution of each spin or orbital component loses its uniformity and electrons with
different spin or orbital avoid each other.  If these symmetry breakings 
are taken into account by a mean field, such a segregated localization can be described.  
In the Hartree-Fock theory of symmetry broken states, though very simple, insulators then emerge. 
 
To fill up and correct the deficiency of LDA, Anisimov {\it et al.} proposed the LDA+U method for the orbitals of strongly correlated electrons 
(typically, $d$ orbital for transition metal elements and $f$ orbital for rare earth elements),
following the spirit of the Hartree-Fock theory.\cite{Anisimov91,Anisimov1,Anisimov2}
In this method, we start from the Kohn-Sham Hamiltonian $H_{\rm LDA}$ based on DFT and add a correction term
$\Delta H$ as
%\begin{eqnarray}
%H&=&\sum_{k\sigma}\left( \xi_1 + \xi_2\right)c_{k\sigma}^{\dagger}c_{k\sigma}\nonumber \\
%& + & U\sum_i n_{i\uparrow}n_{i\downarrow} + V\sum_{\langle ij\rangle}n_in_j-\mu\sum_j n_i \label{eqn:2}
%\end{eqnarray}
\begin{eqnarray}
H&=&H_{\rm LDA}+\Delta H 
\label{eqn:1}
\end{eqnarray}
and then we solve $H$.
Here, the correction is 
\begin{eqnarray}
\Delta H &=& H_{\rm HF}-H_{\rm DC} \;,\\
H_{\rm HF}&=&\frac{1}{2}U\sum_{i, \mu\ne \nu \in d} n_{i,\mu}n_{i,\nu},
\label{eqn:2}
\end{eqnarray}
where the density operator $n_{i,\nu}\equiv \sum_{\sigma} c_{i,\mu,\sigma}^{\dagger}c_{i,\mu,\sigma}$ at site $i$, orbital $\mu$ and spin $\sigma$ is given by the creation (annihilation) operator $c^{\dagger}$ ($c$).
The Hartree-Fock term $H_{\rm HF}$ represents the short-ranged Coulomb interaction 
on specified electron orbitals $\mu$ and $\nu$ (3$d$ orbital for the transition metal elements, for instance and it is denoted as $d$ for
simplicity) within a unit cell $i$. 
(Here we described a simplified case where the interaction $U$ within the unit cell does not depend on the orbitals $\mu$ and $\nu$. For the moment, the exchange interaction is also ignored for simplicity.)
Since the electron density $n_{i,\mu}$ at the unit cell $i$ and orbital $\mu$ (including spin degrees of freedom) is determined selfconsistently with
the solution of Kohn-Sham equation after decoupling, this turns out to be equivalent to the level of the Hartree-Fock approximation.  

When we add $H_{HF}$, the Coulomb interaction already to some extent counted in the exchange correlation potential $V_{XC}$ in LDA is doubly counted.  The term $H_{\rm DC}$ is subtracted to remove this
%the interaction already considered in LDA as the Hartree and exchange correlation potential in order to avoid the 
double counting.
For this purpose,
\begin{eqnarray}
H_{\rm DC}&=&\frac{1}{2}\sum_i Un_{di}(n_{di}-1) \;,\\
n_{di}&=&\sum_{\nu\in d}n_{i\nu}
\label{eqn:H_DC}
\end{eqnarray}
is frequently employed.

The single-particle energy $\epsilon_{i,\nu}$ of an electron at the site $i$ and orbital $\nu$ is given from the total energy $E= \langle H \rangle$ as $\epsilon_{i,\nu} =d \langle H\rangle/dn_{i,\nu}$. This is rewritten as $E_{\rm LDA}+U(\frac{1}{2}-n_{i})$. Then the occupied level at the $i$th site gives $n_i=1$ and its energy gets $U/2$ lower than the LDA energy, while the unoccupied level is as large as $U/2$ higher than the LDA level. This makes the energy difference of $U$ between the occupied and unoccupied levels generating a energy gap for the electron excitation.  This allows a description of an insulator.  

The framework can be extended to include the exchange interaction, where eqs. (\ref{eqn:2}) and (\ref{eqn:H_DC}) are modified to include the exchange interaction as 
\begin{eqnarray}
H_{\rm HF}&=&\frac{1}{2}\sum_{i, \mu,\mu',\mu'',\mu''' \sigma} [\langle \mu,\mu''|U|\mu'\mu'''\rangle n_{i,\mu\mu'}^{\sigma}n_{i,\mu''.\mu'''}^{-\sigma} \nonumber \\
&-&( \langle \mu,\mu''|U|\mu'\mu'''\rangle -  \langle \mu,\mu''|U|\mu'''\mu'\rangle ) n_{i,\mu\mu'}^{\sigma}n_{i,\mu''.\mu'''}^{\sigma} \nonumber \\ 
\label{eqn:2ex}
\end{eqnarray}
and the double counting term 
\begin{eqnarray}
H_{\rm DC}&=& \sum_i \frac{1}{2}(Un_{di}(n_{di}-1) \nonumber \\
&-& \frac{1}{2}J(n_{di}^{\uparrow}(n_{di}^{\uparrow}-1)+n_{di}^{\downarrow}(n_{di}^{\downarrow}-1)), 
\label{eqn:H_DCex}
\end{eqnarray}
where spin $\sigma$ is separated from the orbital indices $\mu$ and $J$ specifies the exchange parameter. We introduced the notation $n_{i,\mu\mu'}^{\sigma}\equiv c_{i,\mu,\sigma}^{\dagger}c_{i,\mu',\sigma}$.

Several different choices of the double counting correction $H_{DC}$ such as fully localized limit and that around mean field have been proposed.  For details readers are referred to Refs.\citen{Anisimov2} and \citen{Ylvisaker}.  

%\textcolor{red}{LDA+Uといえば通常LSDA+Uのことを指すので、この段落は必要ないと思います。
%原理的にはスピン分極なしのものも考えられますが、ほとんどのプログラムでは
%実装上そうなっていないと思います。
%Instead of LDA, one can extend Kohn-Sham equation by including the spin degrees of freedom
%as the functional of spin density, where the extension of LDA called LSDA（local spin density %approximation)
%allow the possible magnetic symmetry breaking. By using LSDA, LSDA$+U$ method is formulated.
%\textcolor{red}{\cite{Anisimov91,Anisimov1}}
%}
%
%ではLSDAに対する相殺項はスピン自由度$\sigma$を用いて
%\begin{eqnarray}
%H_{\rm HF}&=&\frac{1}{2}Un_d(n_d-1)-\frac{1}{2}J[n_{d\uparrow}(n_{d\uparrow}-1)+ n_{d\downarrow}(n_{d\downarrow}-1)] \\
%n_{d,\sigma}&=&\sum_{i,\nu\in d}n_{i,\sigma}, \\
%n_{d}&=&\sum_{i,\sigma,\nu\in d}n_{i,\sigma}
%\label{eqn:2}
%\end{eqnarray}
%と書ける。

In the LDA+U method, one has to determine the value of $U$ from the first principles point of view.
For this purpose, constrained LDA (c-LDA) method is often employed.\cite{Dederichs,Dederichs-1,Gunnarsson,Solovyev1}
Since $U$ is the quadratic coefficient of the $n$ expansion in the Hamiltonian, 
$U$ is calculated from $\partial^2 E/ \partial n^2$ by changing $n$ slightly.\cite{Herring}
To change the electron density on a specified orbital by keeping the density on other orbitals to calculate $U$ on that specified orbital, one has to switch off the transfer of electrons between the specified and the other orbitals.
Then it needs to keep off the hybridization contained in the original Kohn-Sham equation by hand. This disturbs the original electronic structure and introduces an ambiguity in the way of cutting the transfer. However, if it is done carefully, a reasonable value of the interaction is obtained.  

In fact, the LDA$+U$ method has been applied widely to transition metal oxides and has shown reasonable correction to the underestimate of the gap known as the deficiency of LDA.\cite{Anisimov91,Anisimov1,Anisimov2,Terakura-Solovyev}　
%$\Delta H$による補正には任意性があることに注意する。
%短距離クーロン斥力を$U$と表わしているが、これは電子軌道$i$や$j$のとり方、すなわちこの問題での基底の取り方に依存する。
%基底としてより拡がったワニエ軌道を取れば、クーロン斥力は小さくなり、より局在した軌道に属する電子同士の斥力は大きくなる。
%つまり、ワニエ軌道のとり方によって、$U$の値が変わってくることになる。
%しかし、このワニエ軌道が広がりすぎていると、$U$の大きさが小さすぎて、対称性の破れによる棲み分けが起きなくなってしまう。
A problem with the LDA$+U$ method is that the result may depend on the choice of the basis function. In the principle of quantum mechanics,
the ground state should not depend on the choice of the basis function.  However, since LDA$+U$ normally takes into account only the short-ranged part of the repulsion, the final result may depend on the choice of the basis function. 
In addition, the Hartree-Fock approximation overestimates the ``order parameter" $n_{i,\mu}$ as usual as one of the mean-field approximation, and ignores quantum (dynamical) fluctuations to reduce the localization. 
Then the band gap is usually overestimated. On the contrary, the correlation effect is ignored when the 
symmetry breaking (or localization) is absent.
Another drawback is the ignorance of spatial fluctuations. A part of the limitations, namely the missing dynamical fluctuation has been examined in a comparison of the static mean field theory with the dynamical mean field theory described in the later section.\cite{Sangiovanni06}

%%%%%%%%%%%%%%%%%%%%%%%%%%%%%%%%%%%%%%%%%%%%%%%%
%\section{Three-Stage Scheme} \label{Three-Stage} 
%%%%%%%%%%%%%%%%%%%%%%%%%%%%%%%%%%%%%%%%%%%%%%
\section{Downfolding} \label{Downfolding}
\subsection{General Framework}  \label{General Framework}  
A way to derive an effective low-energy model from first principles can be performed by calculating the renormalization effect to the low-energy degrees of freedom caused by the elimination of the high-energy one, based on the idea of the Wilson renormalization group.  This is the basis of the downfolding method.    
Choice of the low-energy degrees near the Fermi level retained in the effective model is not unique.  However, if the renormalization procedure is adequate, finally obtained properties do not depend on the choice.  In many of strongly correlated materials, there exists a well defined group of bands near the Fermi level and isolated from the bands away from the Fermi level as we will see examples in Figs.~\ref{Fig.SVOLDADispersion} and \ref{fig:kappa-band}. This is not accidental because strongly correlated electrons at high density as in solid can be realized only when the mutual screening of electrons is poor, which is ideally seen for isolated small number of bands.  This isolated group is the natural target of the low-energy degrees of freedom. In transition metal oxides, this group consists of the bands whose main component is $3d$ orbitals at the transition metal atoms.  If the crystal field splitting is large as in many cases with the perovskite structure, the low-energy degrees of freedom may further be restricted either to $t_{2g}$ or to $e_g$ orbitals only.  In the organic conductors, the group of HOMO (highest occupied molecular orbital) and LUMO (lowest unoccupied molecular orbital) are frequently isolated from others as we see later. 

In general, a complete set of basis can be constructed from the basis functions of a Hamiltonian obtained by ignoring the electron-electron interaction. In this basis, the second-quantized total Hamiltonian of electrons equivalent to the electronic part of eq.(\ref{Schroedinger}) is written as
\begin{eqnarray}
H[c^{\dagger},c]&=&\sum_{\mu,\nu}h_K(\mu,\nu)c_{\mu}^{\dagger}c_{\nu} \nonumber \\
&+&\sum_{\mu.\mu',\nu,\nu'}h_V(\mu,\mu',\nu,\nu')c_{\mu}^{\dagger}c_{\mu'}^{\dagger}c_{\nu}c_{\nu'}.
\label{eqn:d1}
\end{eqnarray}
The first term represents the kinetic energy including the one-body level (chemical potential) term, while the second term is the Coulomb interaction in the present basis representation with electronic internal degrees of freedom such as $\mu$. Alternatively one can employ the basis of LDA eigenfunctions for Kohn-Sham equation, where $\mu$ specifies a LDA band and momentum.

The partition function of this whole electronic system is described by
$Z={\rm Tr} \exp [S]$ with $S$ being the action given by
\begin{eqnarray}
S[c^{\dagger},c]&=&\int{\cal L} d\tau, \\
{\cal L}&=&\int d{\bf r} c^{\dagger}\partial_{\tau}c + H[c^{\dagger},c],
\label{eqn:d2}
\end{eqnarray}
where $\cal L$ is the Lagrangian and $x$ denotes $({\bf r},\tau,\sigma)$ with the spatial coordinate $\bf r$, the spin $\sigma$ and the imaginary time $\tau$. Here we have suppressed electronic internal degrees of freedom such as the band and momentum index in the notation. 

In strongly correlated electron systems, the whole electronic Hamiltonians are in many cases rather well separated into two parts: One represents electrons in relatively well isolated bands near the Fermi level and the other represents the band degrees of freedom far from the Fermi level.  Then the Hamiltonian is rewritten in the form 
\begin{eqnarray}
H&=&H_L+H_H+H_{HL}.
\label{eqn:d3}
\end{eqnarray}
In the index, the part representing the bands far from the Fermi level (high-energy part) is denoted by the suffix $H$, while the ````thin-skin" part close to the Fermi level (low-energy part) is expressed by the suffix $L$. The coupling between the $H$ and $L$ parts are given by $H_{HL}$. 
Then the trace summation for the partition function is formally decomposed to the high- and low-energy parts as     
\begin{eqnarray}
Z&=&{\rm Tr}_L {\rm Tr}_H \exp [S].
\label{eqn:d4}
\end{eqnarray}
After the partial trace over the high-energy part, the high-energy degrees of freedom is eliminated leading to the action for the low-energy degrees of freedom only as 
\begin{eqnarray}
S_L[c_L^{\dagger},c_L]&=&\log {\rm Tr}_H \exp [S]. 
\label{eqn:d4}
\end{eqnarray}
Now the low-energy effective action $S_L$ represented only by the $L$ degrees of freedom and the resultant Lagrangian ${\cal L}_L\equiv \partial S_L/\partial \tau$ has been derived. 
When we formally expand  
\begin{eqnarray}
\tilde{H}_L[c_L^{\dagger},c_L]&\equiv&{\cal L}_L -\int c^{\dagger}\partial_{\tau}c d{\bf r}
\label{eqn:d5}
\end{eqnarray}
in terms of the creation and annihilation operators, it contains the $\tau$ dependence in general because of the retardation effect and the operator $\tilde{H}_L$ is not equal to $H_L$ because the partial trace summation over the $H$ degrees of freedom renormalizes $H_L$. The partial trace summation may, for instance, be performed by the perturbative treatment of the coupling $H_{HL}$ as we detail later.  

If the dependence on $\tau$ (namely, the retardation effect) can be ignored, $\tilde{H}_L$ may be regarded as an effective Hamiltonian. Then the effective Hamiltonian ${\cal H}\equiv \tilde{H}_L$ in general has the form
\begin{eqnarray}
{\cal H}&=&\sum_{\mu,\nu}h_{LK}(\mu,\nu)c_{\mu}^{\dagger}c_{\nu} \nonumber \\
&+&\sum_{\mu.\mu',\nu,\nu'}h_{LV}(\mu,\mu',\nu,\nu')c_{\mu}^{\dagger}c_{\mu'}^{\dagger}c_{\nu}c_{\nu'}+\cdots
\label{eqn:d6}
\end{eqnarray}
When terms of higher order than this expression (beyond the fourth order in the creation and annihilation operators) are small, this Hamiltonian is closed within the single-particle part and the two-body interactions.
If the $\tau$ dependence is appreciable, one has to treat it with the action $S_L$ or the Lagrangian. 

To understand the physical applicability of this downfolding, let us start from more physically transparent picture.
Even the electrons belonging to the low-energy part originally have the kinetic and interaction energies as in the Hamiltonian (\ref{eqn:d1}). These energies are, however, subject to the renormalization originating from the interaction with the electrons in the high-energy part. This effect appears as the "dressing" of the kinetic and interaction energies.  Actually, the interaction with the high-energy electrons (or holes) $H_{HL}$ reduces the bare interaction between low-energy electrons (holes), because polarizations of the high-energy electrons (holes) screen the interaction between low-energy electrons (holes). In addition, for instance, the effective mass is usually enhanced because of the dressing by the high-energy electrons (holes).  This effect is in more general expressed by the self-energy effect for the kinetic energy part. 
The real frequency dependence of the screened Coulomb interaction $h_{LV}(\omega)$ is obtained from the Fourier transform and analytic continuation of $h_{LV}(\tau)$. In the low frequency range that satisfies (1) $h_{LV}(\omega)\sim h_{LV}(\omega=0)$ where the frequency dependence can be ignored, and (2) the self-energy approximated as $\Sigma\sim {\rm Re}\Sigma(\omega=0)+\omega d{\rm Re}\Sigma/d\omega|_{\omega =0}$, the retardation effect can be ignored and the description by the Hamiltonian becomes adequate.  This corresponds to the case where the low-energy electrons can be adiabatically treated under the high-energy electrons moving fast.  The renormalization effect can be ascribed to the screening of the interaction and the mass enhancement  in the band dispersion given by the factor $1/(1-d{\rm Re}\Sigma/d\omega|_{\omega =0})$ multiplying the bare dispersion $\epsilon(\bk)$. 
In examples of the transition metal compounds, the $3d$ bands of the transition metal atom are relatively well isolated near the Fermi level from others as we already mentioned.  This makes the description by a Hamiltonian appropriate after ignoring the frequency dependence of the screening and the mass enhancement within the range of the $3d$ bandwidth.\cite{aryasetiawan04}

%%%%%%%%%%%%%%%%%%%%%%%%%%%%%%%%%
\subsection{Wannier functions} \label{Wannier functions}
\begin{figure}[htbp] 
\begin{center} 
\includegraphics[width=0.25\textwidth]{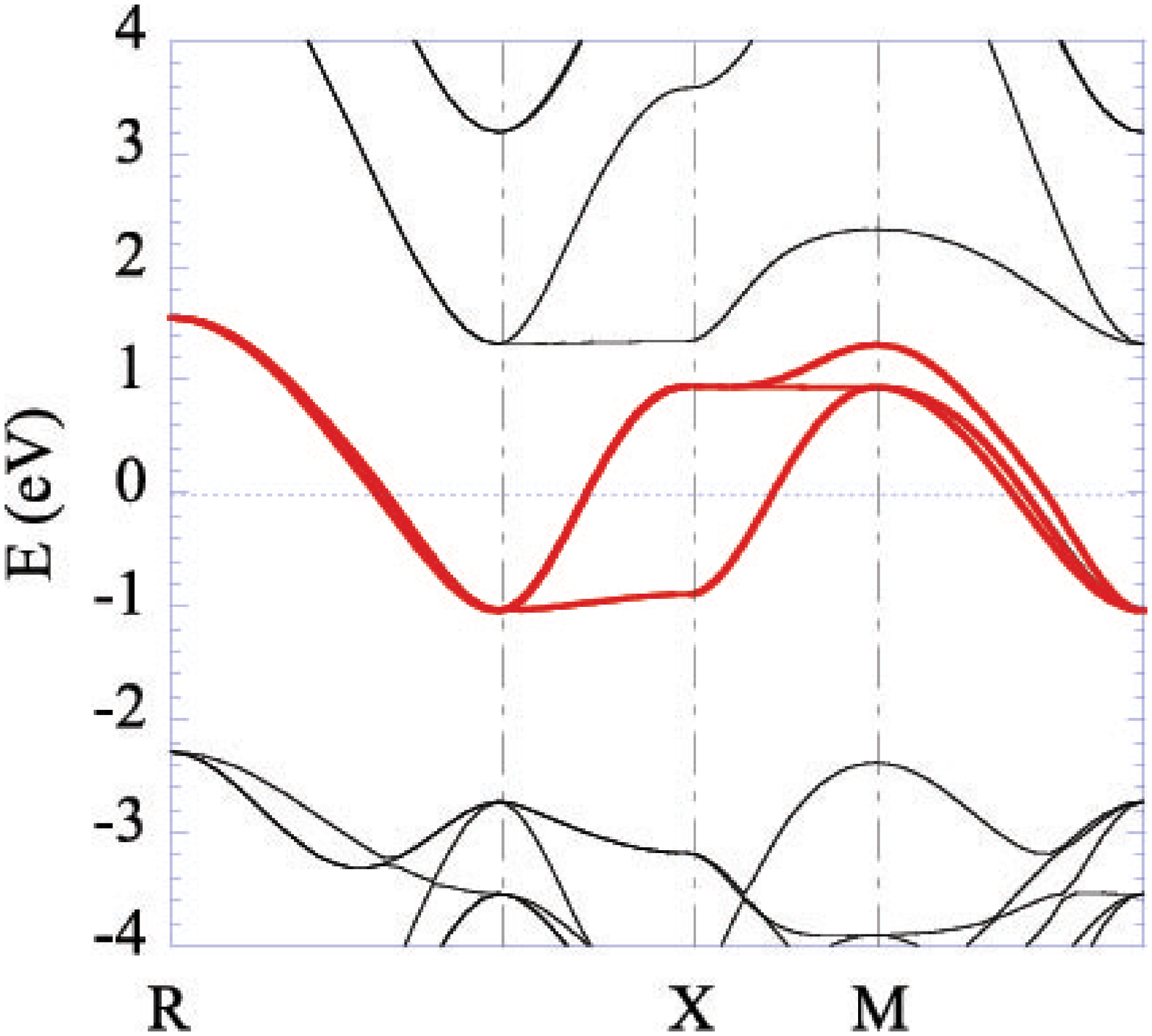}
\includegraphics[width=0.2\textwidth]{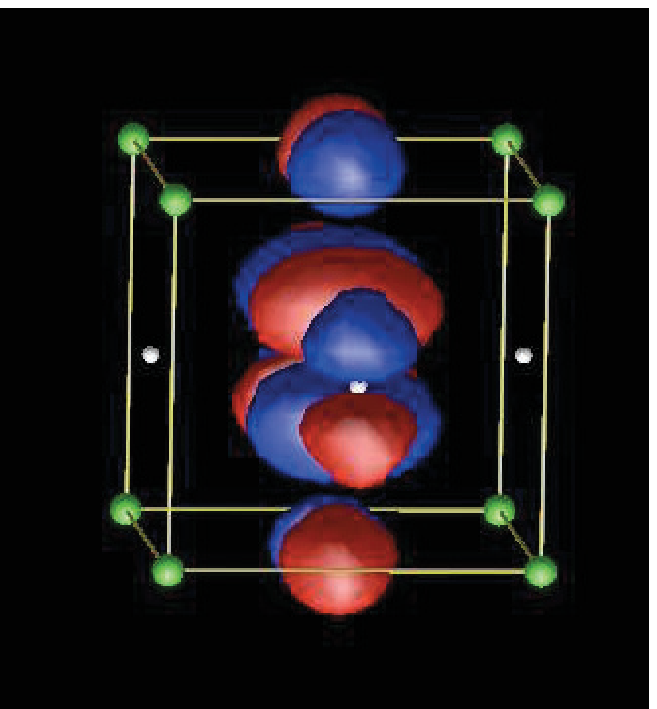}
\end{center} 
\caption{(Color online) 
%\textcolor{red}{
Electronic structure of SrVO$_3$ (left). 
The three states crossing the Fermi level (in red) can be treated as the low-energy part. 
The corresponding maximally localized Wannier function is shown 
in the right panel. The Wannier function having the $t_{2g}$ character 
is localized around the V atom, with a tail at the O sites.}
%}
\label{fig:mlwf} 
\end{figure}

To derive the low-energy effective model, 
one first needs to define and specify the low-energy Hilbert space. 
%Once one obtains the global electronic structure, 
%the next step is to 
In other words, the first step of the downfolding is to
construct a set of localized orbitals that 
span the Hilbert space of the low-energy electronic states.
In the case of SrVO$_3$, for example, three narrow states 
cross the Fermi level (Fig.\ref{fig:mlwf}). 
One may then wish to pick up three localized orbitals 
and construct three-orbital Hamiltonian 
that reproduces the (red) lines crossing the Fermi level in the figure. 
There are several ways for obtaining the localized orbitals. 
Here, we focus on maximally localized Wannier functions (MLWF) developed by
Marzari, Souza, and Vanderbilt \cite{marzari97,souza01} based on the
minimization of the quadratic extent of the orbitals. 
An alternative approach is to use the Wannier orbitals of Andersen 
\cite{andersen00}. The former is more general because it does not depend on 
any particular band-structure calculation method. Comparison between the two Wannier functions 
for some selected materials can be found in Ref.\citen{lechermann06}.
Although the Wannier basis can be chosen arbitrarily in principle and the final results of calculated physical quantities should not depend on the choice, it is better to find maximally localized orbitals to make the range of transfers and interactions in the effective lattice models as short as possible. 

Let $\{ \psi_{n{\bf k}} \}$ be the eigenfunctions of the low-energy 
states. Naively, the Wannier function is defined by 
\begin{equation}
\varphi_{n\mathbf{R}}({\bf r}) = \frac{V}{(2\pi)^{3}} \int e^{-i\mathbf{k}
\cdot\mathbf{R}} \psi_{n\mathbf{k}}({\bf r}) d^{3}\mathbf{k} \;. 
\label{eq:wannier}%
\end{equation}
This Wannier function is, however, ill-defined, because 
it depends on the choice of the phase factor at each {\bf k} point. 
Moreover, at the band-crossing points it is not clear which state should be taken. 
The MLWF utilizes this degrees of freedom. 
The MLWF with band index $n$ at cell $\mathbf{R}$ is defined by 
\begin{equation}
\varphi_{n\mathbf{R}}({\bf r}) = \frac{V}{(2\pi)^{3}} \int e^{-i\mathbf{k}%
\cdot\mathbf{R}} \psi_{n\mathbf{k}}^{\mathrm{(w)}} ({\bf r}) d^{3}\mathbf{k}\;. 
\label{eq:wannier2}%
\end{equation}
Here $ \psi_{n\mathbf{k}}^{\mathrm{(w)}} $ is {\it not} the eigenfunction of 
the Hamiltonian (e.g. Kohn-Sham wavefunction), but  it is a linear 
combination of the eigenfunctions as%
\begin{equation}
\psi_{n\mathbf{k}}^{\mathrm{(w)}} ({\bf r}) = \sum_{m} {\mathcal{U}}%
_{mn}(\mathbf{k}) \psi_{m\mathbf{k}} ({\bf r}) \;. 
\label{eq:umn}%
\end{equation}
The coefficients ${\mathcal{U}}_{mn}(\mathbf{k})$'s are numerically determined 
such that the spread 
\begin{equation}
\Omega = \sum_{n} [ \langle \varphi_{n{\bf 0}} | r^2 |\varphi_{n{\bf 0}}\rangle
- \langle \varphi_{n{\bf 0}} | {\bf r} |\varphi_{n{\bf 0}}\rangle^2 ] \:,
\end{equation}
is minimized. %
In contrast with $\psi_{n{\bf k}}({\bf r})$, 
the gauge of $\psi_{n{\bf k}}^{\rm (w)}({\bf r})$ is fixed, 
and it is a smooth function of ${\bf k}$. 
By representing the Hamiltonian in the MLWF basis, 
\begin{equation}
H_{mn}({\bf R}) = \langle \varphi_{m{\bf 0}} | H | \varphi_{n{\bf R}} \rangle \;,
\end{equation}
the on-site energy levels are obtained from $m=n, {\bf R}={\bf 0}$ component. 
Other matrix elements give the transfer integrals. 

%%%%%%%%%%%%%%%%%%%%%%%%%%%%%%%%%
\subsection{Screened Interaction} \label{Screened Interaction}
Now we discuss how to obtain the renormalization effects on the low-energy electrons near the Fermi level more concretely.  After the partial trace and the elimination of the high energy degrees of freedom given in eq.(\ref{eqn:d4}), the renormalization can be calculated perturbatively in terms of the interaction between the low- and high-energy electrons.

\begin{figure}
\begin{center}
\includegraphics[width=0.4\textwidth]{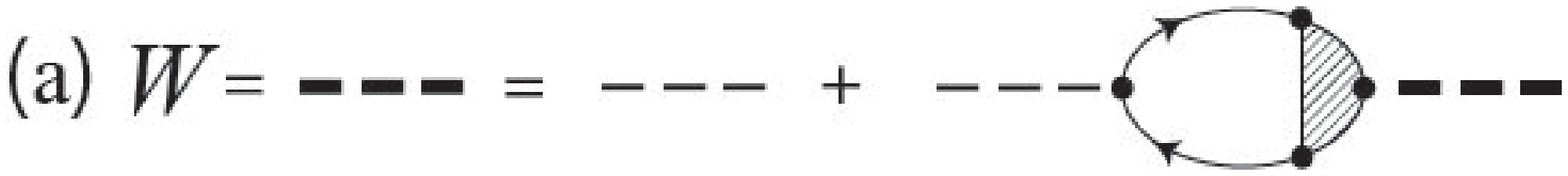}
\includegraphics[width=0.4\textwidth]{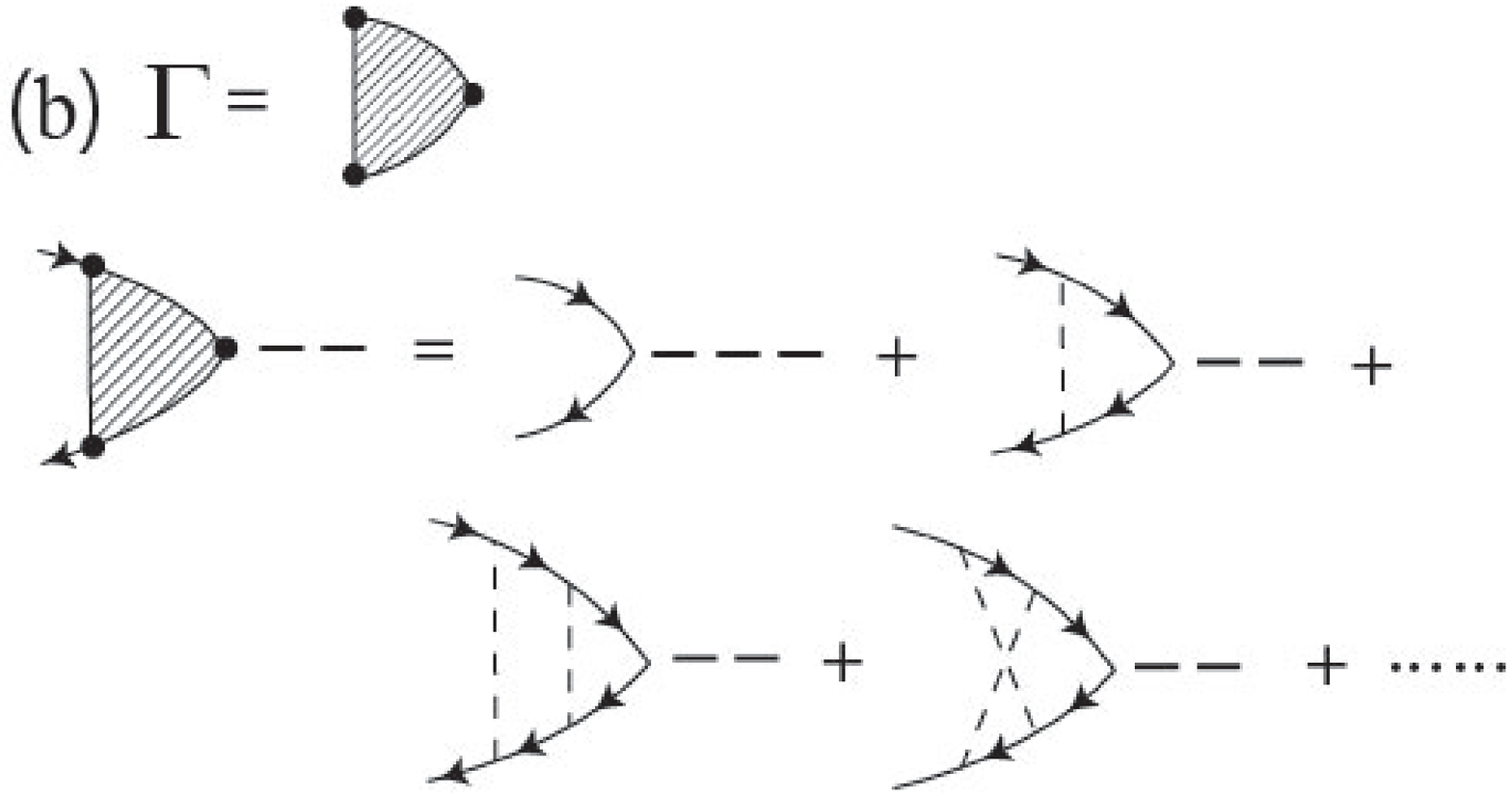}
\end{center}
\caption{
(a) Dyson equation for screened interaction.
Thin solid lines with arrows represent electron propagators (bare Green's functions) $G_0$ and thin dashed lines represent 
bare Coulomb interaction $V$.  Bold and bold dashed lines represent their corresponding renormalized Green's function (see Fig.\ref{fig:GDiagram}) and the 
screened interaction $W$, respectively.  Shaded triangle represents the three-point vertex illustrated in (b). 
(b) Three-point vertex.
}%
\label{fig:WDiagram}%
\end{figure}
\begin{figure}
\begin{center}
\includegraphics[width=0.35\textwidth]{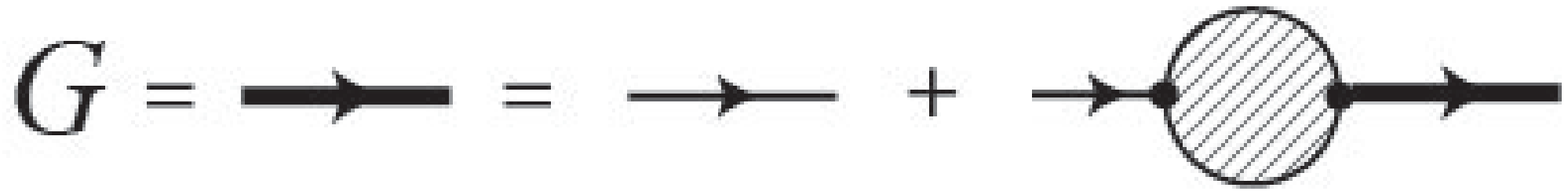}
\end{center}
\caption{
Dyson equation for renormalized Green's function. Shaded circle represents self-energy illustrated in Fig.\ref{fig:SigmaDiagram}. Notations are the same as Fig.\ref{fig:WDiagram}.
}%
\label{fig:GDiagram}%
\end{figure}
\begin{figure}
\begin{center}
\includegraphics[width=0.35\textwidth]{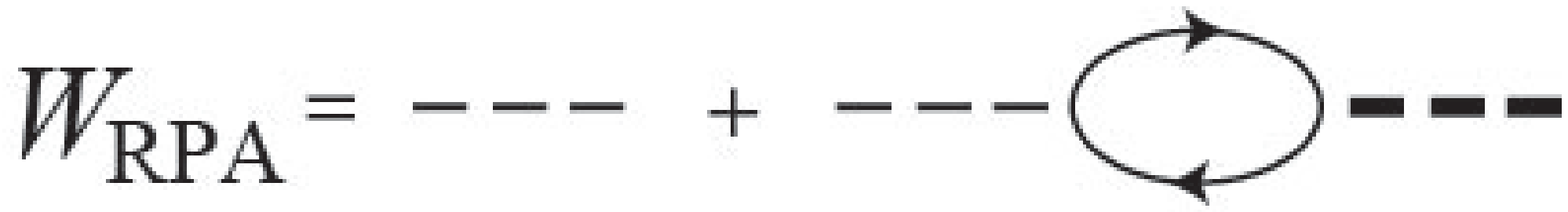}
\end{center}
\caption{
RPA diagram of screened interaction, which is obtained from $W$ in Fig.\ref{fig:WDiagram}(a) by ignoring the vertex correction $\Gamma$ in Fig.\ref{fig:WDiagram}. This is equivalent to the upper panel in Fig.~\ref{fig:gw}. Notations are the same as Fig.\ref{fig:WDiagram}.
}%
\label{fig:WRPADiagram}%
\end{figure}
In more general, the Dyson equation for the screened interaction is expressed in the diagram illustrated in Fig.\ref{fig:WDiagram}(a), where the three-point vertex is given in Fig.\ref{fig:WDiagram}(b) and the diagram representation of the dressed Green's function is illustrated in Fig.\ref{fig:GDiagram}.
If the vertex correction is small as we will discuss later, the screened interaction is reduced to the form of the standard RPA shown in Fig.\ref{fig:WRPADiagram}.
In the conventional RPA, we do not distinguish $L$ and $H$ degrees of freedom and the polarization (bubble in Fig.\ref{fig:WDiagram}(a) or Fig.\ref{fig:WRPADiagram}) contains contribution both from $H$ and $L$ electrons.

Now we extend this conventional RPA to the downfolding procedure.  In this extension, only the polarization containing the high-energy degrees of freedom may contribute in the screening and self-energy processes, because the polarization purely originating from the low-energy degrees has to be removed to keep it dynamical in the effective models. Such a RPA with restriction of the polarization channel is called the constrained RPA (cRPA).\cite{aryasetiawan04,aryasetiawan06} In this subsection we sketch how the interaction energy is renormalized by the screening in cRPA. In the next subsection we figure out how the kinetic energy (namely, dispersion) is renormalized as the self-energy effect.  

The bare Green's function is given by eq.(\ref{eq:g0})
%splitting into the summation over the occupied and unoccupied parts of the states:
%\begin{equation}
%G(\mathbf{r,r}^{\prime};\omega)=\sum^{occ}\frac{\psi%
%(\mathbf{r)}\psi^{\ast}(\mathbf{r}^{\prime})}{\omega-\varepsilon
%-i0^{+}}+\sum^{unocc}\frac{\psi(\mathbf{r)}\psi^{\ast}
%(\mathbf{r}^{\prime})}{\omega-\varepsilon+i0^{+}} \label{wholeG},%
%\end{equation}
while its low-energy part $G_L$ restricts the summation within the target low-energy bands as 
\begin{equation}
G_L(\mathbf{r,r}^{\prime};\omega)=\sum_{L}^{occ}\frac{\psi_{L}%
(\mathbf{r)}\psi_{L}^{\ast}(\mathbf{r}^{\prime})}{\omega-\varepsilon
_{L}-i0^{+}}+\sum_{L}^{unocc}\frac{\psi_{L}(\mathbf{r)}\psi_{L}^{\ast
}(\mathbf{r}^{\prime})}{\omega-\varepsilon_{L}+i0^{+}} \label{HubbardG} \;.%
\end{equation}
From eq.(\ref{HubbardG}) and below in this and the next subsections, the Green's functions are all the bare one but written by abbreviating the suffix 0 for a simple notation.  

The total polarization $P=P_0$ in eq.(\ref{eq:p0}) 
%Let $P$ be the total (bare) polarization, including the transitions between
%all bands:
%\begin{align}
%P(\mathbf{r,r}^{\prime};\omega)  &  =\sum_{i}^{occ}\sum_{j}^{unocc}\psi
%_{i}(\mathbf{r)}\psi_{i}^{\ast}(\mathbf{r}^{\prime})\psi_{j}^{\ast
%}(\mathbf{r)}\psi_{j}(\mathbf{r}^{\prime})\nonumber\\
%&  \times\left\{  \frac{1}{\omega-\varepsilon_{j}+\varepsilon_{i}+i0^{+}
%}-\frac{1}{\omega+\varepsilon_{j}-\varepsilon_{i}-i0^{+}}\right\}  \label{P}%
%\end{align}
%$P$ 
can be divided into: $P=P_{L}+P_{H}$, in which $P_{L}$ includes only $G_L$ (i.e limiting the summations in (\ref{eq:p0}) to $i,j\in\{\psi
_{L}\}$), and $P_{H}$ be the rest of the polarization.

Within cRPA, $h_{LV}$ is reduced to $W_H$ expressed by the bare Coulomb interaction $V$ as
\begin{eqnarray}
W_H(k)&=&\epsilon_H^{-1}(V,k)V(k) \label{eq:W} \;,\\
%\epsilon_{\alpha}(w,k) &=& 1- \int d3 P_{\alpha}(1,3) V(3-2),
%\epsilon_T &=& 1- \int dr_3 P(r_1,r_3) V(r_3-r_2) \nonumber \\
\epsilon_{\alpha}(w,k) &=& 1-  P_{\alpha}(k)w(k) 
\label{eqn:d7}
\end{eqnarray}
in the momentum space representation obtained from the Fourier transform of $\mathbf{r-r}^{\prime}$ in eqs.(\ref{HubbardG}), (\ref{eq:g0}) and (\ref{eq:p0}) into $\mathbf{k}$, where $\alpha$ represents either $L, H$ or $T$. $P_H$ is the polarization function that has contributions from the high-energy electrons and $k=(\bk,\omega)$ represents the both degrees of wavenumber and frequency. 
The RPA form of the screened interaction in general is illustrated in Fig.\ref{fig:WRPADiagram} obtained from the exact form of the screened interaction obtained from the Dyson equation in Fig.\ref{fig:WDiagram}(a) 
by ignoring the three-point vertex correction shown in Fig.\ref{fig:WDiagram}(b).
Here, in Fig.\ref{fig:WRPADiagram}, cRPA has to contain propagators including some of downfolded electrons, namely ``$H$"electrons. 
 In the lowest order, $P_H$ is given by using the Green's function for the low-energy electrons $G_L$ and the total Green's function $G$ (or low-energy polarization $P_L$ and the total polarization $P_T$ ) obtained from LDA as
\begin{eqnarray}
P_H(k)&=& P_T(k)-P_L(k) \label{eqn:d8} \;,\\
P_T(k)&=&-\int dk' G(k') G(k+k'), \\
P_L(k)&=&-\int dk' G_L(k') G_L(k+k'),
\end{eqnarray}
where we have divided the whole Green's function $G$ into the contribution from the low-energy band, $G_L$ and the rest $G_H$ as 
\begin{eqnarray}
G(k)&=& G_L(k) +G_H(k).
\end{eqnarray}

There exists a remarkable identity for the fully screened Coulomb interaction $W$ as\cite{aryasetiawan04}
\begin{eqnarray}
W(k)&=&\epsilon_T^{-1}(V,k)V(k) = \epsilon_L^{-1}(W_H,k) W_H(k)  \;. \label{eqn:d9} 
\end{eqnarray}
This identity proves that the fully screened interaction $W$ obtained from the full RPA by the whole polarization $P$ is the same as the Coulomb interaction obtained from RPA as if one takes $W_H$ were the bare Coulomb interaction and the low-energy polarization $P_L$ were the full polarization.
It assures that one can regard $W_H$ as the effective interaction in the effective low-energy model.
In this cRPA, frequency dependence of $W_H$ has been calculated for several transition metal and transition metal compounds, which revealed that the frequency dependence can be ignored within the order of the width of $d$ electron bands (typically several eV) (see an example for Ni in Fig.\ref{fig:u}).  This means that the Hamiltonian description with the effective interaction 
$U(r)=\lim_{\omega\rightarrow 0}W_H(r,\omega)$ becomes appropriate.\cite{aryasetiawan04}
%\textcolor{red}{小谷さんが計算したのは$W_H$ではなく、
%H-H間の遮蔽のみ考慮したものだったので、小谷さんの文献を削除}

%%%%%%%%%%%%%%%%%%%%%%%%%%%%%%%%%%%%%%%%%
\subsection{Self-energy correction} \label{Self-energy correction}
\begin{figure}
\begin{center}
\includegraphics[width=0.45\textwidth]{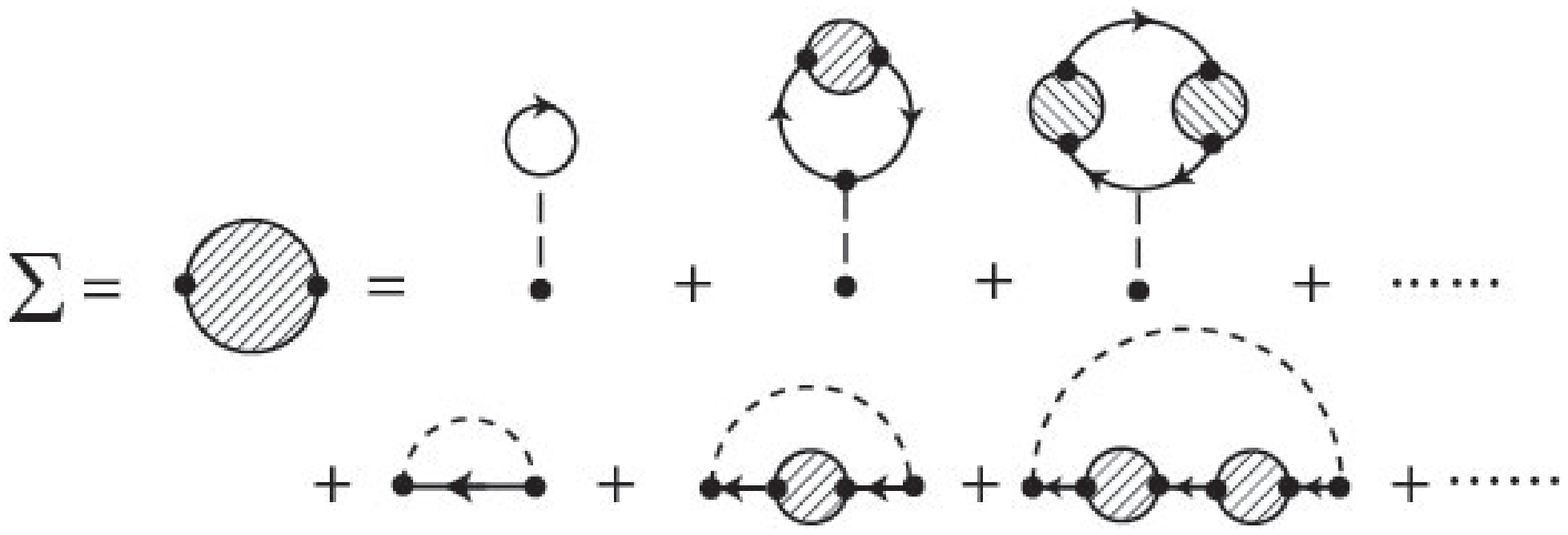}
\end{center}
\caption{
Self-energy diagrams expanded in terms of the interaction in general.
 Notations are the same as Fig.\ref{fig:WDiagram} and shaded circles represent self-energy. 
Diagrams in the first line in the right hand side represent the Hartree-type terms and the second line is for the Fock-type exchange contribution.   
}%
\label{fig:SigmaDiagram}%
\end{figure}
The renormalization of the kinetic energy is given by the self-energy within the level of cRPA.
Even when the frequency dependence of $W_H(k)$ is small in the energy range of the low-energy bands, the frequency dependence is still large beyond the energy scale of the screening channel in cRPA, because $W_H(k)$ eventually should come back to $V(k)$ in the large $\omega$ limit, where the screening does not occur.  
In the ordinary GW approximation the self-energy correction is given by
\begin{eqnarray}
\Sigma(k) &=&\int dk'G(k') W(k+k').
\label{eqn:Sigma-1}
\end{eqnarray}
This is the lowest order term of the self-energy expansion represented by the diagrams
in Fig.\ref{fig:SigmaDiagram}, where if $G$ and $W$ are replaced with their bare forms,
$G_0$ and $V$, respectively, it is reduced to the sum of two diagrams without shaded circles 
in the right hand side in Fig.\ref{fig:SigmaDiagram}.

The RPA-level self-energy can be formally rewritten as
\begin{eqnarray}
\Sigma(k) &=&\int dk'[G_{L}(k')W_{H}(k+k')+ G_{L}(k') \nonumber \\
\times (W(k+k')&-&W_{H}(k+k'))+G_{H}(k)W(k+k')]. 
\label{GW}%
\label{eqn:Sigma-2}
\end{eqnarray}
In the constrained GW approximation (or equivalently cRPA), the first term 
$G_{L}(k')W_{H}(k+k')$ is excluded because this part should be kept dynamical in the effective low-energy model.
Therefore, the renormalization of the LDA band dispersion in the downfolding is given from the self-energy correction as
\begin{eqnarray}
\Delta\Sigma(k) &=&\int dk'[G_{L}(k')(W(k+k')-W_{H}(k+k')) \nonumber \\
&+&G_{H}(k)W(k+k')]-V_{XC}. 
\label{eqn:Sigma-3}
\end{eqnarray}
The last (third) term $V_{XC}$ is introduced to exclude the double counting of the interaction already contained in the LDA level. 
In general the contribution from $G_HW$ in the second term is small as compared to the first term $G_L(W-W_H)$, because of $|G_L|>|G_H|$ symbolically and from the reason similar to the smallness of the vertex correction explained below.

The band dispersion $\omega = \epsilon_0(k) $ of $h_{LK}$ obtained from LDA is now renormalized by the self-energy $\Delta\Sigma(k) \sim \Sigma_0+\Sigma_1\omega$ as $\omega= \epsilon^* (k)=(\epsilon_0(k)+\Sigma_0)/(1-\Sigma_1) $ in the low-energy limit. This gives the momentum dependent flattening of the dispersions and mass enhancement.  The imaginary part of the self-energy is normally small which validates the hamiltonian description. 

The self-energy correction of the LDA band dispersion for the low-energy model has not been extensively studied so far and in many cases has been ignored, partly because this correction is in general not large.  Typically the correction reduces the bandwidth of the $3d$ bands of the transition metal oxides as large as 10-20\%.\cite{Solovyev1}  

Another issue of the self-energy effect is related to the double counting already discussed in the LDA+U method (\S \ref{LDA+U}). The interaction effect is already counted in the mean-field level in the LDA. It contains the Hartree contribution as well as  the exchange correlation.  Therefore, when we consider the self-energy effect, the corresponding part of the interaction effect in LDA has to be removed.  Usually the exchange contribution within LDA is expected to be small while the Hartree contribution becomes important for the multi-orbital systems, because it induces large shifts of the relative level of orbitals.  One simple way of subtracting this Hartree contribution is to adjust the level of orbitals after the Hartree aproximation of the effective low-energy model so as to have the same level with the LDA result.\cite{Misawa10}  

%%%%%%%%%%%%%%%%%%%%%%%%%%%%%%%%%%%
\subsection{Low-energy Hamiltonian} \label{Low-energy Hamiltonian} 
Low-energy effective Hamiltonian ${\cal H}$ in a restricted Hilbert space is thus derived by utilizing the hierarchical structure of the electronic structure in energy space especially applicable to strongly correlated electron systems.

After the Fourier transform and the analytic continuation from the imaginary time to the real frequency, if the frequency dependence of the screened interaction $W_H(\omega)$ is small and the self-energy can be well approximated by $\Sigma(\omega) \sim \Sigma(\omega=0) + \omega d{\rm Re}\Sigma/d\omega|_{\omega =0})$ in the energy range of the target bandwidth,    
the effective low-energy model is expressed by a Hamiltonian form of the extended Hubbard model,
\begin{eqnarray}
H&=&H_K+H_U \label{ham} \;, \\
H_K&=&\sum_{Rn,R^{\prime}n^{\prime}}c_{Rn}^{\dagger}t_{Rn,R^{\prime}n^{\prime}%
}c_{R^{\prime}n^{\prime}} \label{hamK} \;,\\
H_U&=&\frac{1}{2}\sum_{R,nn^{\prime},mm^{\prime}}%
c_{Rn}^{\dagger}c_{Rn^{\prime}}U_{nn^{\prime}R,mm^{\prime}R^{\prime}}c_{R^{\prime}m}^{\dagger
}c_{R^{\prime}m^{\prime}} \label{hamU}%
\end{eqnarray}
after the cRPA procedure. 
Here, $n, n^{\prime}, m, m^{\prime}$ denote both of spin and orbital degrees of freedom and $R$ is the spatial coordinate. 
Now we have arrived at the effective low-energy Hamiltonian by the downfolding to be solved by low-energy solvers discussed in \S \ref{Low-Energy Solver}. 

%%%%%%%%%%%%%%%%%%%%%%%%%%%%%%%%%%%%
\subsection{Vertex correction} \label{Vertex correction} 
In the standard downfolding scheme, the cRPA is an efficient way to derive the renormalized kinetic and 
interaction energies in the effective models.  
Usually the conventional RPA is justified for the case $|P_{T}(k)V(k)| <1$ in the denominator of eq.(\ref{eq:W}) with eq.(\ref{eqn:d7}) 
in the perturbative sense.
However, in the present case, the bare Coulomb interaction is typically as much as several tens of eV, while the polarization of the metallic target band is scaled by $P_T\sim \Delta E^{-1}$, where $\Delta E$ is the typical energy scale of 
the target bandwidth, because 
$P_T(k)\propto \int dk' G(k)G(k+k')$ is basically scaled by the energy denominator of the Green's function for the downfolded bands as eq.(\ref{eq:g0}), and
$G$ is given by eq.(\ref{HubbardG}).  
%\begin{equation}
%\end{equation}
Then $\Delta E$ is typically the inverse of eV. 
Therefore, $|P_{T}(k)V(k)| \gg1$ typically holds and clearly violates the above requirement meaning that the simple full RPA cannot be used.
  
Even in the case of cRPA, the polarization containing 
the high-energy bands ($H$ bands) is scaled by $P\sim \Delta E^{-1}$, where $\Delta E$ is now the typical energy scale of 
the downfolded band ($H$ bands) measured from the Fermi level, because 
$P_H$ is scaled by the energy denominator of $G_H$ for the downfolded bands.
%\begin{equation}
%\end{equation}
Then $\Delta E$ is typically the inverse of several eV. 
Therefore, $|P_{H}(k)V(k)| \gg1$ still holds.

Nevertheless, even in this region, cRPA can be a good and convergent approximation as we describe below:
We can show that cRPA becomes accurate and convergent when the vertex correction is small ($|\Gamma-1|\ll 1$), 
namely, the difference between the three-point vertex 
diagram $\Gamma$ shown in Fig.\ref{fig:WDiagram}(b) and its lowest order term $\Gamma_0\equiv 1$ (the first term in the right hand side) is small, 
because the dressed interaction with the full vertex correction $W$ shown in Fig.\ref{fig:WDiagram}(a) then becomes nearly the same as the RPA 
diagram in Fig.\ref{fig:WRPADiagram}.  Here in cRPA, at least some of the propagators have to contain the electrons 
in the downfolded band ($``H"$ electrons), because the diagram containing only the $"L"$ electrons should be excluded in cRPA. 

\begin{figure}
\begin{center}
\includegraphics[width=0.4\textwidth]{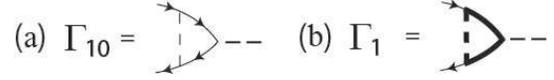}
\end{center}
\caption{
Three-point vertex diagrams. (a) Lowest order vertex (b) Dressed vertex represented with the renormalized interaction and Green's function. 
 Notations are the same as Fig.\ref{fig:WDiagram}.   
}%
\label{fig:GammaCor}%
\end{figure}
Let us consider the lowest order term of the vertex correction in Fig.\ref{fig:GammaCor}(a) 
Again, this lowest-order term is basically proportional to $P_HV\sim V/\Delta E$, which is much larger than  unity.
On the other hand, when we consider the dressed vertex correction shown in Fig.\ref{fig:GammaCor}(b), the renormalized interaction vertex and 
the renormalized propagators should be employed in a self-consistent fashion.
Then the renormalized interactions between two $H$ electrons and/or between a $L$ electron and a $H$ electron appear, together with propagators consisting of two $H$ electrons $G\sim \psi_H\psi_H$ and/or of $H$ and $L$ electrons $G\sim \psi_H\psi_L$.  Therefore, the vertex correction is expanded in terms of $W_H'/\Delta E$, where $W_H'$ is the interaction either between two $H$ electrons or between a $H$ electron and a $L$ electron, screened by $H$ electrons. Then $W_H'$ is in general even smaller than $W_H$ in 
eq.(\ref{eq:W}) because of more efficient screening by the electrons on the same band for the former and because of the interaction at distant $L$ and $H$ Wannier orbitals in the latter.    
%Furthermore, in the spirit of cRPA, in deriving $W_H'$, even the screening by $L$ electrons should contribute at low frequencies, in contrast to the derivation of $W_H$. 
Therefore, $W_H'/\Delta E$ can in general be a small parameter.  In addition, if $G\sim \psi_H\psi_L$ is contained, it multiples another small factor as a small matrix element in the numerator of the vertex correction, because the overlap of the wavefunction on different bands $\psi_L$ and $\psi_H$ is small.
These are in general justified if the coupling between 
the target and the downfolded bands are weak, in either sense of the hybridization ( or wavefunction overlap) and/or the interband interaction.

Even more important reason for the irrelevance of the vertex correction in cRPA is that in the counting of $W_H'$, the full screening channel by the $L$ electrons should be included, though this channel is excluded in the counting of $W_H$.  The gapless particle-hole excitations of the $L$ electrons very efficiently screen the Coulomb interaction as is known in the full RPA results shown in Fig.\ref{fig:u}~\cite{miyake09}.  In short, the vertex correction is irrelevant provided that $W_H'/\Delta E$ is small and this criterion is not perfectly but rather well satisfied in real strongly correlated electron materials. 
In typical transition metal oxides, $W_H'$ may be less than 1 eV while $\Delta E$ is several eV. 

Although it may be small, but the vertex correction arising from the $W_H'/\Delta E$ expansion quantitatively enhances the screening. On the other hand, the polarization from the transition between $L$ and $H$ electrons may be quantitatively reduced when the $L$ electrons are under strong correlation, because the energy for adding an $L$ electron and removing an $L$ electron may split as is typical in the Hubbard band splitting of the $L$ band to the upper and lower Hubbard bands. This splitting will reduce the screening and at least partially cancel the above vertex correction.  Therefore, the vertex correction may be even smaller.    

The self-energy coming from the $H$ electrons discussed in \S \ref{Self-energy correction} becomes also small from the same reason in the same situation. These small vertex correction and small self-energy are the reason why the cRPA offers a good approximation.  The irrelevance of the vertex and self-energy correction is understood more intuitively;  In the insulator or semiconductor where the density of states is zero around the Fermi level, the vertex and self-energy correction become small when the interaction is smaller than the gap.
%, because gapless particle-hole excitations are absent. 
The renormalization by cRPA is similar, where the gapless particle-hole excitations are excluded.
Though the correction is small, there must exist some finite correction from the vertex part, while its quantitative estimate has not been fully examined so far and is left for future studies.   

%In this paragraph, we discuss the condition for the justification of the cRPA in this procedure.
%When the energy of the downfolded band $\epsilon^r(k)$ is 
%far from the Fermi level, namely $E_{min}\equiv \min (|\epsilon^r(k)|)$
%is large as compared to 
%\epsilon_{\alpha}(k) &=& 1-  P_{H}(k)V(k). 
%\label{eqn:d7}

%%%%%%%%%%%%%%%%%%%%%%%%%%%%%%%%%%%%%%%
\subsection{Disentanglement} \label{Disentanglement} 
Although the cRPA method offers a general and accurate scheme for the downfolding, 
its applications to real systems still have
a serious technical problem. The problem arises when the narrow
band is entangled with other bands, i.e., if it is not completely isolated from
the rest of the bands, which is the case in many materials. 
Even in simple materials such as the 3$d$
transition metals, the 3$d$ bands mix with the 4$s$ and 4$p$ bands. Similarly, the 4$f$
bands of the 4$f$ metals hybridize with more extended $s$ and $p$ bands. For
such cases, it is not clear anymore which part of the polarization should be eliminated
when calculating the screened interaction using the cRPA method.

In this subsection, we take the notation of $``d"$ and $``r"$  symbolically, 
instead of $L$ and $H$ to specify the low-energy target bands and the high-energy downfolded bands, respectively 
employed in the previous sections, to deliver a more concrete image of the $d$ bands and the rest bands in transition metals and transition metal compounds. 
Since it does not mean the loss of generality, one may interchange the notations each other.  

\begin{figure}
\begin{center}
\includegraphics[height=50mm]{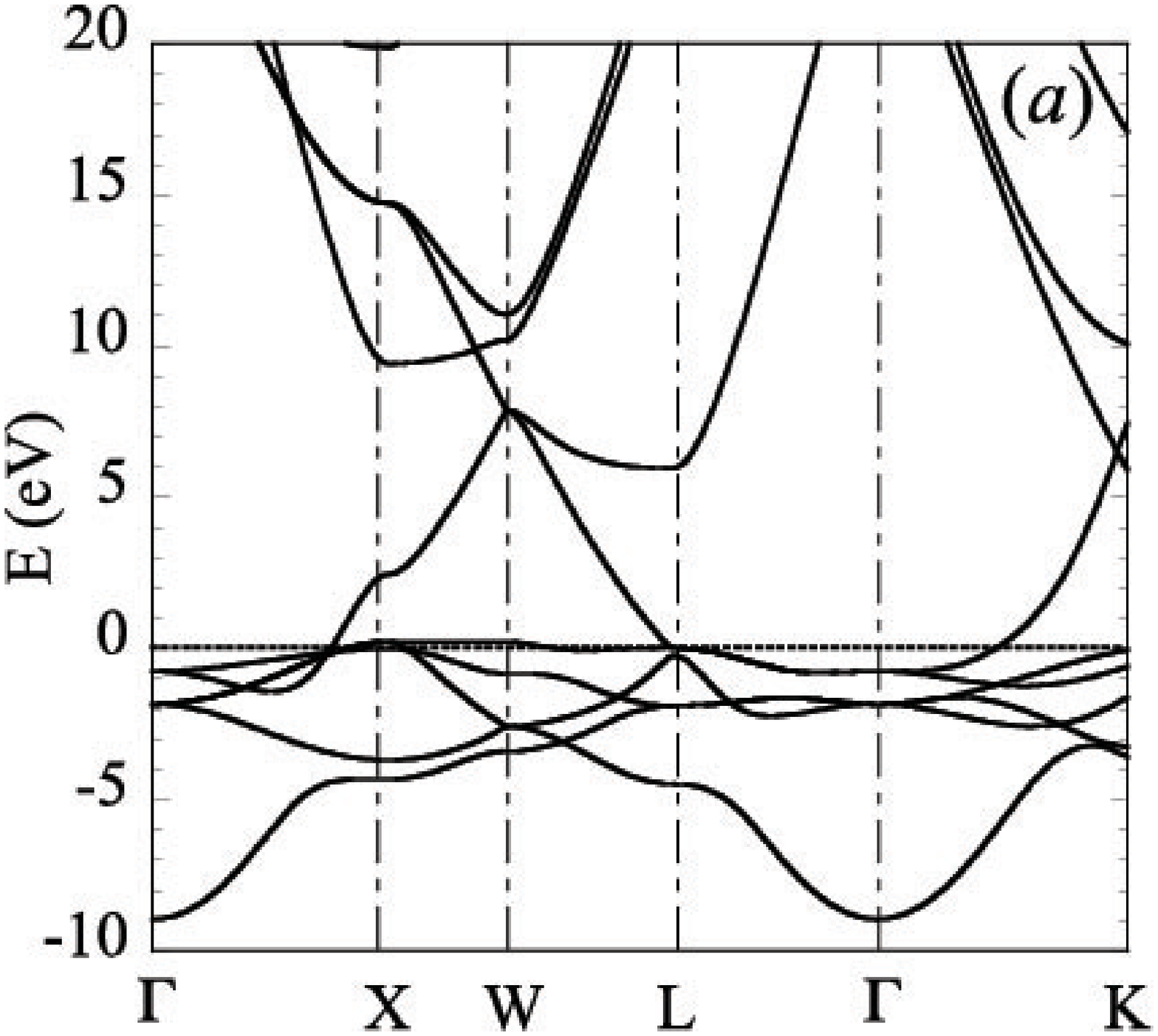}
\includegraphics[height=50mm]{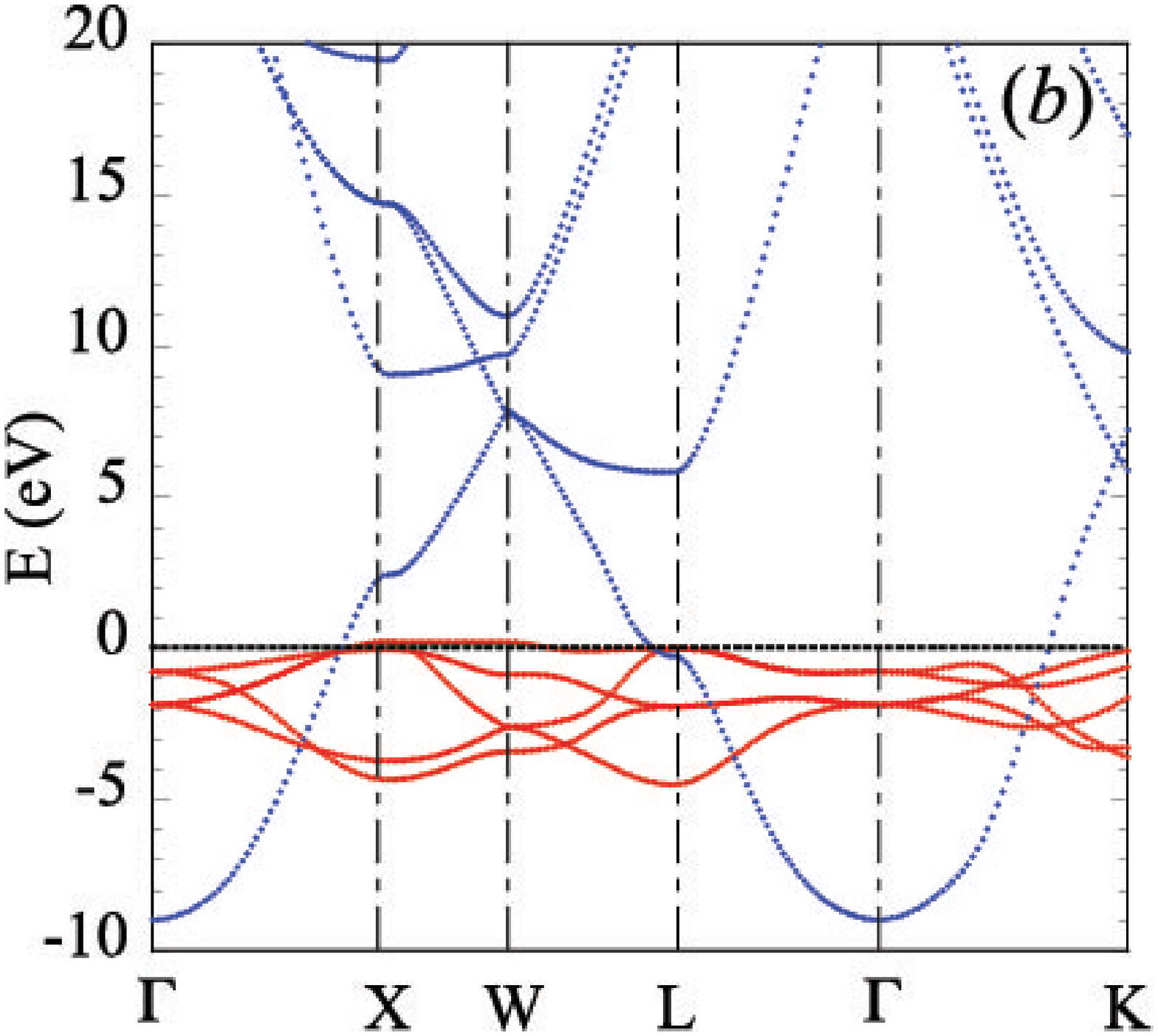}
\end{center}
\caption{(Color)
(a) Kohn-Sham band structure of nickel in LDA.
(b) Disentangled band structure with $d$-$r$ hybridization switched off. 
The red lines show the $d$ states obtained by 
the maximally localized Wannier scheme, 
while the blue lines are disentangled $r$ states. 
Energy is measured from the Fermi level.\cite{miyake09}
}%
\label{fig:disentangled_band}%
\end{figure}

If the $d$ subspace forms an isolated set of bands, as for example in the case
of the  $t_{2g}$ bands in SrVO$_{3}$ as we 
%\textcolor{red}{
saw in Fig.\ref{fig:mlwf}, 
%}
%see later in \S \ref{SrVO3}, 
the cRPA method can be straightforwardly
applied. However, in practical applications, the $d$ subspace may not always
be well identified. An example is 3$d$ transition
metal series such as Ni shown in Fig.\ref{fig:disentangled_band}(a), where the 3$d$ bands are entangled
with the 4$s$ and 4$p$ bands. 

To treat this complexity, a prescription for entangled bands was proposed recently.~\cite{miyake09} 
The essential point is that one has 
to strictly keep the orthogonality between the low-energy subspace contained in the model and 
the complementary high-energy subspace to each other. 
The orthogonalization by the projection technique enables a proper disentanglement 
of the bands. 
Although physical properties may not sensitively depend, 
we still have a freedom that the $d$ space 
somehow depends on the choice of the energy window when one constructs 
the Wannier functions. 
However, once the disentangled band structure is obtained, 
the constraint RPA method can be used to determine the partially screened 
Coulomb interaction uniquely. 
Numerical tests for 3$d$ metals show that the method is stable 
and yields reasonable results. 
The method is applicable to any system, and 
applications to more complicated systems.
%, such as interfaces of transition metal oxides are now under way.
We review this procedure in more details here.

We first construct a set of localized Wannier orbitals from a given set of
bands defined within a certain energy window. These Wannier orbitals may be
generated by the post-processing procedure of Souza, Marzari and Vanderbilt 
\cite{souza01,marzari97} 
or other methods, such as the preprocessing scheme proposed by Andersen {\it et al.} 
within the Nth-order muffin-tin orbital (NMTO) method \cite{andersen00}. 
We then fix this set of Wannier
orbitals as the generator of the $d$ subspace and use them as a basis for diagonalizing the
one-particle Hamiltonian, which is usually the Kohn-Sham Hamiltonian 
in LDA or in generalized gradient approximation (GGA). 
The obtained set of bands, defining the
$d$ subspace, may be slightly different from the original bands defined
within the chosen energy window, 
because hybridization effects between the $d$ and $r$ spaces are switched off. 
It is important to confirm that the dispersions near the 
Fermi level well reproduce the original Kohn-Sham bands. 

The wavefunctions are thus projected to the $d$ space by  
\begin{equation}
\vert \tilde{\psi}_{i} \rangle = {\hat{\cal P}}
\vert \psi_i \rangle
\;,
\end{equation}
where the projection operator ${\hat{\cal P}}$ is defined as 
\begin{equation}
{\hat{\cal P}} = \sum_{j=1}^{N_d} \vert \tilde{\psi}_j \rangle \langle \tilde{\psi}_j \vert
\;.
\end{equation}

We define the $r$ subspace by
\begin{equation}
\vert {\phi}_{i} \rangle =
(1 - {\hat{\cal P}}) \vert \psi_{i} \rangle
\end{equation}
which is orthogonal to the $d$ subspace constructed from the Wannier orbitals.
In practice it is convenient to orthonormalize $\{ \phi_i \}$ and prepare 
$N-N_d$ basis functions. 
By diagonalizing the Hamiltonian in this subspace 
a new set of wavefunctions $\{ \tilde{\phi}_i \}$ 
and eigenvalues $\{ \tilde{e}_i \}$ $(i=1,\cdots, N-N_d)$ are obtained. 
Namely, the Kohn-Sham Hamiltonian becomes block diagonal in the 
$d$ space and $r$ space separately, and the hybridization effects 
between them are neglected: 
\begin{equation}
H = 
\left(
\begin{tabular}{c | c}
$d$ space & 0 \\ \hline
0 & $r$ space \\
\end{tabular}
\right) \;.
\label{diagonaldr}
\end{equation}
As a consequence of orthogonalizing 
$\{  \tilde{\phi}_{i} \} $  and $\{ \tilde{\psi}_{j} \} $, 
the set of $r$ bands $\{ \tilde{e}_i \}$ are completely
disentangled from those of the $d$ space $\{  \tilde{\varepsilon}_{j} \} $, 
and they are slightly different from the original band structure $\{ \varepsilon_i \}$.  
As we will see later, however, the numerical tests show that 
the disentangled band structure is close to the original one. 

From the $``d"$ bands, we calculate the $``d"$ polarization $\tilde{P}_{d}$ as%
\begin{eqnarray}
\tilde{P}_{d}(\mathbf{r,r}^{\prime};\omega)&=&\sum_{i}^{\text{occ}}%
\sum_{j}^{\text{unocc}}\left[  \frac{\tilde{\psi}_{i}^{\ast}%
(\mathbf{r)}\tilde{\psi}_{j}(\mathbf{r)}\tilde{\psi}_{j}^{\ast}(\mathbf{r}%
^{\prime})\tilde{\psi}_{i}(\mathbf{r}^{\prime})}{\omega-\tilde{\varepsilon
}_{j}+\tilde{\varepsilon}_{i}+i\eta} \right. \nonumber \\
&-&\left. \frac{\tilde{\psi}_{i}(\mathbf{r)}%
\tilde{\psi}_{j}^{\ast}(\mathbf{r)}\tilde{\psi}_{j}(\mathbf{r}^{\prime}%
)\tilde{\psi}_{i}^{\ast}(\mathbf{r}^{\prime})}{\omega+\tilde{\varepsilon}%
_{j}-\tilde{\varepsilon}_{i}-i\eta}\right]
%\;,
\label{eq:pd}
\end{eqnarray}
where $\{  \tilde{\psi}_{i} \}$, $\{ \tilde{\varepsilon}_{i} \}  $  $(i = 1, \cdots N_d)$ are the
wavefunctions and eigenvalues obtained from diagonalizing the one-particle
Hamiltonian in the Wannier basis.

The effective screened interaction for the low-energy model is calculated 
according to eq.(\ref{eqn:d7}) with $P_{r}=\tilde{P}-\tilde{P}_{d}$, where 
$\tilde{P}$ is the full polarization calculated for the {\it disentangled} 
band structure. 
It is important to realize that the screening processes from the polarization $P_r$
include the Coulomb interaction between the $d$ space and the $r$ space, 
in calculating the screened interaction of $d$ bands (so called $U$ terms in eq.(\ref{hamU})), 
although the $d$-$r$ hybridization is cut off in the construction of 
the wavefunctions and eigenvalues. 

We also note that starting from the complete orthogonality between the $d$ and $r$ bands in Eq.(\ref{diagonaldr}) is crucially important to assure stable cRPA calculations. In fact, if the orthogonality is not perfect, the resultant frequency-dependent screened Coulomb interaction could have unphysically negative values in some frequency region.\cite{miyake09}
For example, if $P_{r}=\tilde{P}-\tilde{P}_{d}$ would be calculated from $P_d$
obtained from the above procedure while the total $P$ is calculated from the original LDA band, the resultant $P_r$ gives unstable behavior of the screened interaction $U$, because such a $P_r$ is not compatible with mutually orthogonal subspace of $r$ and $d$ and a small nonorthogonality between $d$ and $r$ subspaces implicitly assumed in this $P_r$ would yield a singular behavior at low energies. One has to use the total $P$ obtained after the disentanglement procedure to assure the orthogonality of the subspaces.   

This disentanglement procedure has been tested to work in 3$d$ transition metals. 
%The electronic structure calculations are done in the LDA \cite{kohn65}
%of density functional theory \cite{hohenberg64} 
%with the full-potential LMTO implementation \cite{methfessel00}. 
%The wavefunctions are expanded in the basis of $spdf+spd$ MTOs 
%and a $8\times8\times8$ {\bf k}-mesh is used for 
%the Brillouin zone summation. 
%As an illustrative purpose, spin polarization is neglected, 
%but this can be easily included. 
For technical details see the literature. 
\cite{miyake09,schilfgaarde06,miyake08a}
Figure \ref{fig:disentangled_band}(a) shows the Kohn-Sham band structure of nickel.\cite{miyake09} 
There are five orbitals having strong 3$d$ character at [-5 eV:1 eV], 
crossed by a dispersive state which is mainly of 4$s$ character. 
Using the prescription for the maximally localized Wannier function, and 
with the energy window of [-7 eV:3 eV], 
interpolated ``$d$'' bands are obtained. 
The subsequent orthogonalization procedure gives the orthocomplementary ``$r$'' bands. 
Comparing Fig.\ref{fig:disentangled_band}(b) with (a) we can see that 
there is no anti-crossing between the $d$ bands and the $r$ bands in (b)
contrary to (a).
Aside from this difference, the two band structures are very similar. 

\begin{figure}
\begin{center}
\includegraphics[width=70mm]{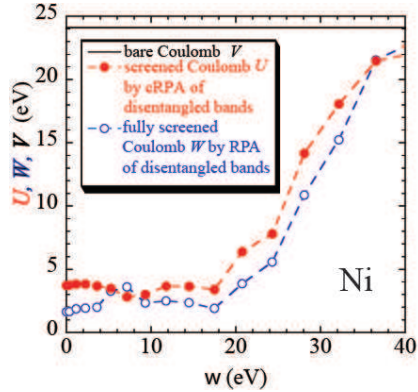}
\end{center}
\caption{(Color online)
Effective onsite Coulomb interaction $U$ calculated by cRPA (filled (red) circles)(eq.(\ref{eq:W}))
and fully screened onsite interaction $W$ by RPA (open (blue) circles) (eq.(\ref{eqn:d9}))for nickel 
as a function of frequency 
by disentanglement procedure\cite{miyake09}. 
The diagonal terms of screened interactions averaged over the $3d$ orbitals
in the Wannier basis are plotted.
}
\label{fig:u}%
\end{figure}

The effective screened interaction for the effective $3d$ model is calculated by the constrained RPA, namely, 
by eq.(\ref{eq:W}).
The results are shown in Fig.\ref{fig:u}. 
%There is no large fluctuation against frequency, 
%in contrast to the methods described in Sec.\ref{sec:method}, 
%and $U(\omega)$ shows a stable behavior. 
First, we observe that the frequency dependence of $U$ is small within the $3d$ bandwidth ($\sim$ 5 eV),
which justifies the treatment by an effective model Hamiltonian within this energy range.
Second, as is expected, the partially screened onsite interaction $U$ from eq.(\ref{eq:W}) ($\sim 4$eV)
is significantly larger than the fully screened RPA result $W$ from eq.(\ref{eqn:d9}) ($\sim 1-2$ eV).
At low frequencies (namely, for $<5$ eV). 
This implies proper elimination of $d$-$d$ screening processes has a large effect. 
This was also applied to a series of other $3d$ 
transition metals and was found to give stable and reasonable results.\cite{miyake09}

In the present formulation, 
small off-diagonal matrix elements of the Kohn-Sham Hamiltonian between 
the $d$ wavefunction $\vert \psi_i\rangle $ and 
the $r$ space $\vert \phi_j\rangle $ are ignored. 
However, if the energy of the $d$-$r$ hybridization point in the band dispersion is smaller than 
the energy scale of interest, one has to retain all of these hybridizing 
bands in the effective model, 
because the hybridization effect changes 
the band dispersion and the wavefunction significantly in the vicinity of 
the anti-crossing points. 
In many correlated materials with $d$ or $f$ electrons,  
the energy scale of interest determining material properties is typically of the order of 100 K or lower, 
which is smaller than the typical energy crossing points. 
Therefore, the low energy models constructed only from 
the $d$ or $f$ Wannier orbitals may give at least a good starting point of 
understanding the low energy physics.    

%%%%%%%%%%%%%%%%%%%%%%%%%%%%
%\textcolor{red}{
\subsection{Dimensional downfolding} \label{Dimensional downfolding}
Remarkable progress in understanding physics of materials with low-dimensional anisotropy such as cuprate \cite{Bednorz} and iron-based \cite{Hosono} superconductors have stimulated studies on electronic models in low dimensions.
In particular, 1D or 2D simplified models are frequently used and have greatly contributed 
in revealing 
characteristic low-dimensional physics with strong-correlation and fluctuation effects.\cite{ImadaRMP}

However, the {\it ab initio} downfolding method reviewed in this article
are formulated to derive {\it ab initio} models in 3D space.
Thus, we have to solve the derived model as a 3D model, but this requires significantly demanding
computation when we consider correlation effects with high accuracy.
To construct a low-dimensional model tractable by 
the widely employed theoretical approaches, 
it is crucial to bridge the 3D models to 
effective 1D or 2D models based on the {\it ab initio} derivation.    

Recently, a scheme of
downfolding a 3D model 
to lower-dimensional models from first principles has been formulated as 
a dimensional downfolding in real space.\cite{Nakamura10} 
This 
supplements the original band downfolding in energy space.
The formalism eliminates the degrees of freedom for layers (chains) other than the target layer (chain) 
after the interlayer/chain screening taken into account. This is useful when low-energy solvers for the effective models allow only the 2D models as tractable.  
The scheme is general and particularly works well 
for quasi-low-dimensional systems. 
The dimensional downfolding has another computational advantage that the range of the screened effective interaction becomes short-ranged when we take into account interlayer (interchain) screening for metals. Note that, the original band downfolding reviewed in previous sections leaves the effective screened interaction long-ranged even for metals because cRPA excludes metallic screenings in the target bands. 

\begin{figure}[h!]
	\begin{center}
	\includegraphics[width=0.3\textwidth]{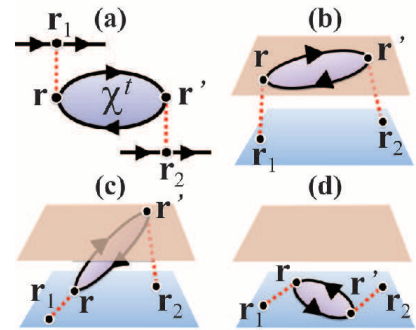}  
	\end{center}
\caption{(Color online)  
Schematic diagram for effective interactions between electrons at 
${\bf r}_1$ and ${\bf r}_2$ in the target layer/chain, 
screened by intra- and inter-layer/chain polarizations $\chi^t({\bf r}, {\bf r}')$.
(a) shows general second-order diagram for the screened interaction (b) shows the screening by a polarization in the other layers/chains, while (c) describes that by an interlayer polarization between the target and the other layers. 
(d) shows the screening 
by a polarization within the target layer/chain itself.
This process 
should be excluded in the present 2D-cRPA downfolding, while (b) and (c) are included.  
Notations are the same as Fig.\ref{fig:WDiagram}.\cite{Nakamura10}
}
\label{fig:interlayer}
\end{figure}

The polarization in the target band is decomposed into 
layer-by-layer contributions in the real space.  
The RPA polarizations except for those within the target layer/chain (namely the processes in 
Figs.\ref{fig:interlayer}(b) and (c) excluding the process in Fig.\ref{fig:interlayer}(d)) contribute to the interlayer/chain screening, which renormalizes the effective interaction between electrons within the target layer/chain. %} 
This screening deletes 
the long-range part of the interactions 
for the case of metallic systems and justifies 
the short-ranged models as 
effective {\it ab initio} models of real materials. 

As an example, it was applied to derive an effective 2D model
 for LaFeAsO.
It was found that the interlayer screenings reduce onsite 
Coulomb interactions by 10-20 \% 
and further remove the long-range part of the screened interaction.
This formalism justifies a multi-band 2D Hubbard model for LaFeAsO 
from first principles.

Here, we discuss vertex corrections and self-energy effects.
The band downfolding (3D-cRPA) becomes adequate basically when the downfolded band energy in the denominator of the propagator is far from the Fermi level as we discussed in \S \ref{Vertex correction}. 
In the case of the dimensional downfolding, the 2D-cRPA (or 1D-cRPA) becomes accurate not by a large energy denominator but by small numerators of the vertex expansion.  
If two neighboring layers or chains are far apart in space, the overlap of the wavefunction between two Wannier orbitals on different layers/chains are small. In this case, the Green's function (\ref{HubbardG}) with one $\psi$ being on one layer/chain and its partner being on another layer/chain (symbolically shown in Fig.\ref{fig:interlayer}(c) for the lowest order) hardly contributes to the polarization for the screening channel.  Even in this case, contribution of the polarization from the propagator with both of $\psi$ on the off-target layer/chain is large.  With this polarization, the 2D-cRPA diagram contains screened interlayer/chain interaction to connect the target and off-target layer/chain (see Fig.\ref{fig:interlayer}(b)).  This screened interlayer/chain interaction $W_{\rm inter}$ is again given as a consequence of efficient screening by the gapless particle-hole excitations on the target and off-target layers and it is normally smaller than $W_H'$ discussed in \S \ref{Vertex correction}. Since the vertex is roughly given by the expansion in terms of $P_LW_{\rm inter}\sim W_{\rm inter}/W_{\rm intra}$, where $W_{\rm intra}$ is a typical screened intralayer/chain interaction calculated from the full RPA.    
The ratio of these two interactions turns out to be the small parameter in the vertex correction for the dimensional downfolding.  The self-energy contribution in the dimensional downfolding has a similar feature where we have $G_LW_{\rm inter}$ as a small parameter in the expansion.  

Here, we note a more fundamental issue. Of course, the present dimensional downfolding does not mean that the 3D systems can be rigorously mapped to lower-dimensional models. For example it is obvious that we are unable to treat the 3D ordering process.  Nevertheless, within the properties and questions that allow the neglect of the energy scale of the interlayer/chain {\it effective} transfer and {\it effective} interaction, the present downfolded low-D models offer the best way of simplifying the problem to that for low-dimensional physics in an {\it ab initio} way.  An important point is that the interlayer/interchain interaction is more efficiently screened in the band downfolding procedure and can be much smaller than the original effective interaction for 3D systems.    

%%%%%%%%%%%%%%%%%%%%%%%%%%%%
\section{Low-Energy Solver} \label{Low-Energy Solver}
The next task of the three-stage scheme is to solve the effective models.\cite{ImadaRMP}
Historically, strongly correlated electron systems have been studied extensively by using an {\it ad hoc} theoretical models.  Such theoretical models contain only small number of bands and restricts the interaction range short as in the cases of the Hubbard model, Anderson model and Heisenberg model.  Such simplifications have made it possible to carry out high-accuracy calculations including ordering and quantum fluctuations beyond the mean-field level employed in DFT.  In fact, mechanisms of antiferromagnetic phase stabilized by the superexchange interaction, Kondo effect, and Mott transitions have been elucidated by using theoretical models.  In those studies in the long history, the interaction parameter $U$ in the Hubbard model has been chosen by hand to satisfy physical intuitions and/or comparisons with experimental results.  However, rapid progress in research for transition metal compounds including the copper oxide superconductors and rare earth compounds displaying heavy-fermion behavior as well as discoveries of many correlated and functional electron systems in materials research have strongly promoted studies of deriving models of real materials based on the first-principles calculations.  This trend is supported by the fact that one can not have a clue for understanding mechanisms of intriguing phenomena when an approximate solution of an {\it ad hoc} model does not reproduce experimental results; the disagreement could be ascribed to poor approximations in solvers, while it could equally be ascribed to a false of the model itself.  The present three-stage RMS scheme with the downfolding procedure has opened an avenue of studies on theoretical models on a firm basis of first principles to overcome such uncertainties.  

Several different low-energy solvers have been applied to solve effective lattice Fermion models derived by the three-stage RMS scheme. 
The present low-energy solvers are roughly classified into two streams.  One is the methods for solving lattice Fermion models as quantum many-body problems.  The other is the methods based on dynamical mean-field theory.

Dynamical mean field theory (DMFT) is formulated by ignoring the momentum dependence of the self-energy but considering the 
frequency dependence correctly.\cite{metzner89}  This becomes a good approximation when the spatial dimension increases and is proven to be exact in infinite dimensions.\cite{Muller-Hartman} In other words, the DMFT becomes exact when the coordination number becomes infinite.

%%%%%%%%%%%%%%%%%%%%%%%%%%%
%\subsection{Lattice model solver} \label{Lattice model solver}
For lattice Fermion solvers, various methods as exact diagonalization, auxiliary field Monte Carlo (AFMC),\cite{Hatsugai} path-integral renormalization group (PIRG),\cite{Kashima,Mizusaki} density matrix renormalization group (DMRG), many-variable variational Monte Carlo (mVMC)\cite{Tahara} and Gaussian-basis Monte Carlo\cite{GBMCAssaad,GBMCAimi} have been developed, which are applicable when the models have the Hamiltonian expression eq.(\ref{ham}).
In principle, these approaches allow taking account of spatial fluctuations equally with dynamical fluctuations in contrast to DMFT, while computational cost becomes demanding.
Functional renormalization group (fRG) has also been developed.\cite{RiceRG,Salmhofer,MetznerRG}
This method divides the whole Brillouin zone into patches and the renormalizations of the coupling constants in each patches are calculated with renormalization group transformation by approaching the Fermi energy window.  Since it gives how various coupling constants to orderings grow and which instability occurs first, it is suited in the weak-coupling region.
Within more biased framework, or within weak-coupling or mean-field framework, simple RPA or fluctuation exchange approximation (FLEX) have also been used as computationally tractable methods.\cite{Scalapino} 
Below we review the DMFT (either combined with GW or without it), mVMC and PIRG as typical and extensively examined solvers .

%%%%%%%%%%%%%%%%%%%%%%%%%%%
\subsection{Dynamical mean-field theory} \label{Dynamical mean-field theory}
Usual mean-field theories approximate effects of interacting electrons with a static effective field. 
As a result, the many-electron problem is reduced to a single-particle problem under the influence of 
the static mean field. 
In the dynamical mean field theory (DMFT),\cite{metzner89,georges} 
the many-electron problem is mapped onto an impurity problem in the following way. 
A chosen site in a periodic lattice is treated as an impurity and the surrounding sites are treated as a bath. 
The effective Coulomb interaction between electrons at the impurity is taken into account explicitly, 
whereas the self-energy arising from the surrounding bath is taken into account 
as a dynamic mean field. 
The impurity self-energy $\Sigma_{\rm imp}(\omega)$ is energy dependent but is assumed to be local, 
i.e., not {\bf k} dependent. 
Then, the self-energy of the original lattice model is replaced by the impurity self-energy on each site. 
Thus, the lattice Green's function is given by 
\begin{equation}
G(\mathbf{k,}i\omega)=\frac{1}{i\omega+\mu-
H_{0}(\mathbf{k)}-\Sigma_{\rm imp}(i\omega)} \;.
\label{eq:glat}
\end{equation}
The impurity Green's function $G_{\rm imp}$ 
\begin{equation}
G_{\rm imp}(\tau)=-\left\langle Tc_{0}(\tau)c_{0}^{+}(0)\right\rangle_{S_{\rm eff}}   
\label{eq:gimp}
\end{equation}
is calculated from an effective
action,
\begin{eqnarray}
S_{\rm eff} & =& -\int d\tau\int d\tau^{\prime}\sum_{\sigma}c_{0\sigma}^{+}(\tau)
\mathcal{G}_{0}^{-1}(\tau-\tau^{\prime})c_{0\sigma}(\tau^{\prime })  \nonumber \\
&+& U\int d\tau n_{0\uparrow}n_{0\downarrow}.
\label{eq:seff}
\end{eqnarray}
Here the subscript $0$ in eqs.(\ref{eq:seff}) and (\ref{eq:gimp}) denotes the impurity site 
and $c_{0}(\tau)$ is a Grassmann variable. 
The band index is omitted for simplicity. 
The dynamical mean field $\mathcal{G}_{0}^{-1}$ is given by
\begin{equation}
\mathcal{G}_{0}^{-1}(i\omega_{n})=G_{\rm loc}^{-1}(i\omega_{n})+\Sigma_{\rm imp}(i\omega_{n})  \,,
\label{Gmean-field}
\end{equation}
where the local Green's function is defined by 
\begin{equation}
G_{\rm loc}(i\omega_{n})=\sum_{\mathbf{k}}\frac{1}{i\omega_{n}+
\mu-H_{0}(\mathbf{k)}-\Sigma_{\rm imp}(i\omega_{n})}.   
\label{eq:gloc}
\end{equation}
To understand the meaning of $\mathcal{G}_{0}$,
consider the following expression:
\begin{equation}
\mathcal{\tilde{G}}(\mathbf{k})^{-1}=G(\mathbf{k})^{-1}+\Sigma_{imp}.
\end{equation}
Since $G$ is the full Green's function defined in eq.(\ref{eq:glat}), 
$\mathcal{\tilde{G}}$ is
the lattice Green's function excluding the effect of the
self-energy at the impurity site. 
$\mathcal{G}_{0}$ is the projection of $\mathcal{\tilde {G}}$ 
at the impurity site which contains the effects of the
self-energy from the rest of the sites. 
Consequently $\mathcal{G}_{0}$ is different from 
the non-interacting Green's function. 

The self-consistency condition requires that the local Green's function 
is equal to the impurity Green's function, 
\begin{equation}
G_{\rm loc}(i\omega_{n}) = G_{\rm imp}(i\omega_{n}) \;.
\label{eq:scg}
\end{equation}
Thus the self-consistent calculation is carried out in the following way.\cite{georges,ImadaRMP} 
\begin{enumerate}
\item For a given mean field $\mathcal{G}_{0}^{-1}$, 
the effective action $S_{\rm eff}$ is determined. 
\item The impurity problem eq.(\ref{eq:seff}) is solved, and 
the impurity Green's function $G_{\rm imp}$ 
and the impurity self-energy $\Sigma_{\rm imp}=\mathcal{G}_{0}^{-1}-G_{\rm imp}^{-1}$ 
are calculated. 
\item The impurity self-energy is used for the lattice self-energy. 
The lattice Green's function is computed from eq.(\ref{eq:glat}).
\item The local Green's function is calculated from eq.(\ref{eq:gloc}). 
\item If the self-consistency condition eq.(\ref{eq:scg}) is not satisfied, 
a new mean field $\mathcal{G}_{0}^{-1}$ is constructed from eq.(\ref{Gmean-field}), 
and the self-consistent cycle is continued.
\end{enumerate}
There are several techniques for calculating the impurity Green's function
eq.(\ref{eq:gimp}) from eq.(\ref{eq:seff}), such as iterative perturbation theory (ITP), 
exact diagonalization, numerical renormalization group and various quantum Monte Carlo methods.\cite{georges} 
Numerical renormalization group offers a high accuracy method at low energies.\cite{sakai-kuramoto94} 
A quantum Monte Carlo method with the Hirsch-Fye algorithm~\cite{Hirsch-Fye} is widely used for obtaining finite temperature properties.  In this algorithm, the imaginary time in the path integral must be discretized, which necessitates an additional extrapolation procedure for the continuum limit.
Recently by utilizing a continuous time algorithm~\cite{Wiese,Rombouts99}, an efficient solver has been developed in the weak coupling approach~\cite{Rubtsov05,Gull} as well as in the strong coupling approach~\cite{Werner}, where the exponential of the Hamiltonian is expanded~\cite{Handscomb,Sandvick} in terms of either the interaction or the kinetic energies and it is free of the discretization error.  A variation called diagrammatic Monte Carlo method has also been proposed.\cite{Assaad07}
The implementation of nonlocal correlation effects beyond DMFT has also been attempted from other approaches such as the dual Fermion method\cite{Rubtsov08,Rubtsov09} and dynamical vertex approximation\cite{ToschiHeld07,Sangiovanni10}

\begin{figure}[htb]
	\begin{center}
	\includegraphics[width=0.35\textwidth]{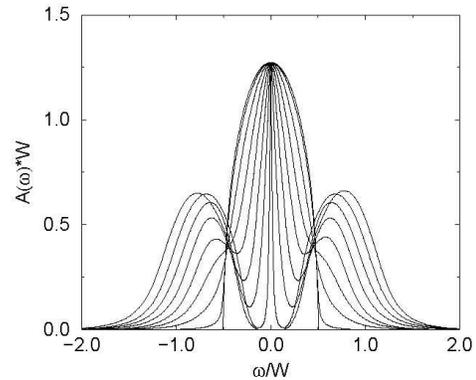}  
	\end{center}
\caption{ Density of states $A(\omega)=-{\rm Im}G$ by single-site DMFT at $T=0$ for half-filled Hubbard model at $U/W=0,0.2,0.4,.\cdots $ and 1.6 from top to bottom around $\omega=0$, where $W$ is the noninteracting bandwidth. The original noninteracting density of states is taken as semicircular. The calculation is done by numerical renormalization group. At $U/W=1.6$ is an insulator to which the coherent peak in metals becomes sharpened until $U/W=1.4$.\cite{vollhardt05} See also Fig.\ref{fig:ZhangClusterDMFT}(a)
}
\label{fig:GeorgesMottTransition}
\end{figure}

The DMFT successfully describes the correlation-driven metal-to-insulator transition 
in one consistent theoretical framework as we see the evolution of the density of states in Fig.~\ref{fig:GeorgesMottTransition}. 
As $U/t$ increases from a weakly correlated metallic regime, 
the quasiparticle peak gets narrower. 
At the same time, the upper and lower Hubbard bands evolve. 
As $U/t$ increases further, the quasiparticle peak disappears and 
the system becomes gapful. 
%\textcolor{red}{図があるとよい}

To improve the lack of ${\bf k}$ (momentum) dependence of the self-energy in the single-impurity DMFT, there are several attempts to include the ${\bf k}$ dependence in the self-energy. 
Most commonly used methods are the dynamical cluster approximation~\cite{maier05} and cellular DMFT~\cite{kotliar01,kotliar06}, in which the impurity problem is 
defined not for a single site but for a cluster including several sites (or alternatively Brillouin zone is divided into patches to allow the momentum dependence of the self-energy). 
Cluster perturbation theory was developed by S\'{e}nechal {\it et al.} to include intercluster coupling as an RPA type perturbation.\cite{Senechal}
Potthoff formulated a Baym-Kadanoff-type formalism as a functional of the self-energy 
instead of Green's function in case of DMFT,\cite{Potthoff1}
It was applied to a formalism for cluster degrees of freedom 
called variational cluster approach\cite{Potthoff2}, which bridges DMFT and the cluster perturbation theory.  
%\textcolor{red}{他の方法を含めて詳細を把握していません}

\begin{figure}[htb]
\begin{center}
\includegraphics[width=0.5\textwidth]{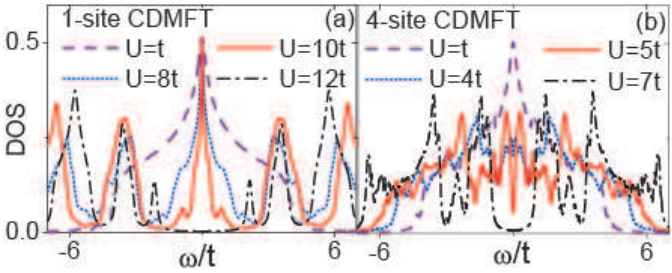}
\end{center}
\caption{(Color online) Density of states of Hubbard model on half-filled square lattice obtained from CDMFT.\cite{zhang07}
Results from the single-site DMFT (a) and 4-site cellular DMFT (b) are compared for four different choices of $U/t$. The broadening factor 0.1 is employed.
In contrast to (a), pseudogap is developed in (b) near the transition to the insulator. }
\label{fig:ZhangClusterDMFT}
\end{figure}

A typical effect of spatial correlations is seen in the cluster extension of DMFT for the Hubbard model.\cite{zhang07}   
It shows a suppression of the coherent peak
with a pseudogap formation in contrast to the sharpening of
the coherent peak at the Fermi level close to the metal-insulator transition in the single site DMFT 
(see Fig.~\ref{fig:ZhangClusterDMFT}).    
This indicates an appreciable role of intersite (spin) correlations ignored in the single site study depending on the
lattice structure. Differentiations of electrons in momentum space make the Mott transition strongly momentum dependent with distinction of more correlated and less correlated regions.  On the square lattice, the pseudogap opens first in the ``antinodal" region (around $(\pi,0)$ and $(0,\pi)$ regions) in metals separated from the opening of the real Mott gap eventually around the ``nodal" region at the Mott transition.\cite{Jarrell01,Maier02,Senechal04,Civelli05,Kyung06,Sakai09}  

%%%%%%%%%%%%%%%%%%%%%%%
A combination of DMFT and LDA, so-called 
LDA+DMFT, was developed to apply DMFT to real materials \cite{anisimov97,lichtenstein98,kotliar06,Held}. 
As is the same as the LDA+U method, LDA+DMFT is based on a many-body Hamiltonian, 
\begin{eqnarray}
H &=& \sum_{\{im\sigma\}}(H_{im,i^{\prime}m^{\prime}}^{\rm LDA}-
H_{\rm DC})a_{im\sigma}^{+}a_{i^{\prime}m^{\prime}\sigma}  \label{LDAUham} \nonumber \\
& +& \frac{1}{2}\sum_{imm^{\prime}\sigma}U_{mm^{\prime}}^{i}n_{im\sigma
}n_{im^{\prime}-\sigma}  \nonumber \\
& +& \frac{1}{2}\sum_{im\neq
m^{\prime}\sigma}(U_{mm^{\prime}}^{i}-J_{mm^{\prime}}^{i})n_{im%
\sigma}n_{im^{\prime}\sigma} \;,
\end{eqnarray}
where $H^{\rm LDA}$ is the Kohn-Sham Hamiltonian in the LDA. 
Hubbard terms for direct and exchange interactions for the correlated 
orbitals, e.g. $d$ or $f$ orbitals, are added on top of the LDA Hamiltonian.
In order to avoid double counting of the Coulomb interactions for these
orbitals, a correction term $H_{\rm DC}$ is subtracted. 
The resultant Hamiltonian (\ref{LDAUham}) is solved by the
DMFT by assuming that the many-body self-energy associated with 
the Hubbard interaction terms can be calculated from a multi-band impurity model.
As described above, the method can be applicable for a wide range of $U$, 
from metallic regime, such as Fe and Ni, to 
strongly correlated insulator, such as NiO, 
including the intermediate regime where both the coherent and the incoherent peaks exist 
(e.g. SrVO$_3$). 
On the other hand, the first-principles determination of $U$ and proper treatment of 
the double counting term are challenges as is discussed in this review in detail. 
%Recently developed cRPA technique is a good candidate for the former problem. 
\\

%%%%%%%%%%%%%%%%%%%%%%%%%%%
A combination of the GW approximation and DMFT is a new approach 
to go beyond the LDA+DMFT method. 
They were proposed both in a model context \cite{sun02} and 
within the framework of realistic electronic structure calculations 
\cite{biermann03}. 
The basic idea is simple. 
The DMFT is suitable to treat the onsite correlations, while 
the {\bf k} dependence in the self-energy and long-range interaction effects are 
not taken into account. The RPA, on which the GW approximation is based, is
generally good for handling long-range correlations. 
Hence, in the GW+DMFT approach, 
the on-site self-energy is taken to be the DMFT self-energy, 
while the off-site self-energy is calculated by the GW approximation. 
Viewed from the GW, the on-site GW self-energy is supplemented by that of DMFT, 
correcting the GW treatment of on-site correlations. 
Viewed from the DMFT, the off-site contributions to the self-energy 
approximated within the GW approximation give a momentum dependent self-energy.

The above idea can be formulated using the following free-energy functional \cite{coa99,chitra01}, 
\begin{eqnarray}
\Gamma(G,W) & =& Tr\ln G-Tr[(G_{H}^{-1}-G^{-1})G]  
 -\frac{1}{2}Tr\ln W \nonumber \\
&+&\frac{1}{2}Tr[(v^{-1}-W^{-1})W]+\Psi\lbrack G,W] \;,
\label{eq:LW}
\end{eqnarray}
where $G_{H}^{-1}=i\omega_{n}+\mu+\nabla^{2}/2-V_{H}$ is the Hartree
Green's function with $V_{H}$ being the Hartree potential. 
The functional is an extension of the functional by Luttinger and Ward (LW) \cite{luttinger60}, 
and has two variables, the Green's function $G$ and the screened Coulomb interaction $W$.
By taking functional derivatives of $\Psi$ with respect to $G$ and $W$, 
the stationary condition 
\begin{equation}
\frac{\delta\Gamma}{\delta G}=0, \ \ \frac{\delta\Gamma}{\delta W}=0
\label{Stationary}
\end{equation}
yields
\begin{eqnarray}
G^{-1}&=&G_H^{-1}-\Sigma, \ \ \Sigma=\delta\Psi/\delta G  \nonumber \;,\\
W^{-1}&=&V^{-1}-P, \ \ P=-2\delta\Psi/\delta W. 
\label{GinvWinv}
\end{eqnarray}

Now, the functional $\Psi$ is divided into two parts as 
\begin{equation}
\Psi=\Psi_{GW}^{\rm off-site}[G^{RR^{\prime}},W^{RR^{\prime}}]+%
\Psi_{\rm imp}^{\rm on-site}[G^{RR},W^{RR}],   \label{PsiGW+DMFT} % \;,
\end{equation}
where $R$ denotes a lattice site. 
The first term is approximated in the GW approximation as
\begin{equation}
\Psi_{GW}[G,W]=\frac{1}{2}GWG   \label{PsiGW} \;,
\end{equation}
while the second term is evaluated from the impurity problem defined by 
the following action, 
\begin{eqnarray}
S & =&\int d\tau d\tau^{\prime}\left[ -\sum c_{L}^{+}(\tau)\mathcal{G}%
_{LL^{\prime}}^{-1}(\tau-\tau^{\prime})c_{L^{\prime}}(\tau^{\prime}) \right. 
\nonumber \\
& + & \frac{1}{2}\sum:c_{L_{1}}^{+}(\tau)c_{L_{2}}(\tau):\mathcal{U}%
_{L_{1}L_{2}L_{3}L_{4}}(\tau-\tau^{\prime}) \nonumber \\
 & &\left. \times :c_{L_{3}}^{+}(\tau^{\prime})c_{L_{4}}(\tau^{\prime}):\right],
\label{eq:simp}
\end{eqnarray}
where the double dots denote normal ordering and $L$ is an orbital of
angular momentum on a given sphere where the impurity problem is
defined. 
Then the above stationary conditions
yield
\begin{eqnarray}
\Sigma&=&\Sigma_{GW}^{RR^{\prime}}(1-\delta_{RR^{\prime}})+\Sigma_{\rm imp}^{RR}
\delta_{RR^{\prime}}, \nonumber  \\
%\label{SigGW+DMFT}
%\end{equation}
%\begin{equation}
P&=&P_{GW}^{RR^{\prime}}(1-\delta_{RR^{\prime}})+P_{\rm imp}^{RR}\delta_{RR^{\prime }}.
\label{PGW+DMFT}
\end{eqnarray}
In momentum space, eqs.(\ref{PGW+DMFT}) are expressed as
\begin{equation}
\Sigma^{LL^{\prime}}(\mathbf{k},i\omega_{n})=\Sigma_{GW}^{LL^{\prime}}(%
\mathbf{k},i\omega_{n})-\sum_{\mathbf{k'}}\Sigma_{GW}^{LL^{\prime}}(\mathbf{k'}%
,i\omega_{n})+\Sigma_{\rm imp}^{LL^{\prime}}(i\omega_{n}),   \label{Sig_a}
\end{equation}
\begin{equation}
P^{\alpha\beta}(\mathbf{k},i\omega_{n})=P_{GW}^{\alpha\beta}(\mathbf{k}%
,i\omega_{n})-\sum_{\mathbf{k'}}P_{GW}^{\alpha\beta}(\mathbf{k'},i\omega
_{n})+P_{\rm imp}^{\alpha\beta}(i\omega_{n})   \;. \label{P_a}
\end{equation}

The outline of the self-consistent loop is the following. 
\begin{enumerate}
\item The impurity problem (\ref{eq:simp}) is solved for a given Weiss field 
$\mathcal{G}_{LL^{\prime}}$ and interaction $\mathcal{U}_{\alpha\beta}$, 
and the impurity Green's function $G_{\rm imp}$ and self-energy $\Sigma_{\rm imp}$ are 
obtained. 
The two-particle correlation function%
\begin{equation}
\chi_{L_{1}L_{2}L_{3}L_{4}}=\langle:c_{L_{1}}^{\dagger}(\tau)c_{L_{2}}(%
\tau)::c_{L_{3}}^{\dagger}(\tau^{\prime})c_{L_{4}}(\tau^{\prime}):\rangle_{S}
\label{eq:chi}
\end{equation}
is also evaluated.

\item 
The polarization function of the impurity problem is computed from 
the interaction $\mathcal{U}_{\alpha\beta}$ and the correlation function eq.(\ref{eq:chi}).

\item 
The full Green's function $G(\mathbf{k},i\omega_{n})$ and effective
interaction $W(\mathbf{q},i\nu_{n})$ are constructed
from eqs.~(\ref{Sig_a}) and (\ref{P_a}).  
Their local part is defined by 
\begin{eqnarray}
G_{\rm loc}(i\omega_{n})  &=& \sum_{\mathbf{k}} G(\mathbf{k},i\omega_{n})  \;, \label{Glocal} \\
W_{\rm loc}(i\nu_{n}) & =& \sum_{\mathbf{q}} W(\mathbf{q},i\nu_{n}) \;.  \label{Wlocal}
\end{eqnarray}

\item
The Weiss dynamical mean field $\mathcal{G}$ and
the interaction $\mathcal{U}$ are updated according to 
\begin{eqnarray}
\mathcal{G}^{-1} & =&G_{\rm loc}^{-1}+\Sigma_{\rm imp}  \label{update} \;,\\
\mathcal{U}^{-1} & =&W_{\rm loc}^{-1}+P_{\rm imp} \;.
\end{eqnarray}
\end{enumerate}
This cycle is iterated until self-consistency for $\mathcal{G}$ and $\mathcal{U}$ is 
achieved. 
When self-consistency is reached, $G_{\rm imp}=G_{\rm loc}$ and $W_{\rm imp}=W_{\rm loc}$ 
are satisfied. 
Therefore, the second term in eq.(\ref{Sig_a}) can be rewritten as 
\begin{equation}
\sum_{\mathbf{k}}\Sigma_{GW}^{LL^{\prime}}(\mathbf{k},\tau)=-%
\sum_{L_{1}L_{1}^{\prime}}W_{imp}^{LL_{1}L^{\prime}L_{1}^{\prime}}(%
\tau)G_{imp}^{L_{1}^{\prime}L_{1}}(\tau)   \label{Sig_correction}
\end{equation}
This shows that the on-site contribution of the GW self-energy is
precisely subtracted out, thus avoiding double counting. 

Eventually, self-consistency over the local electronic density can also be implemented. 
Using the new density from the Green's function at the end of the
convergence cycle above, the next iteration of the GW calculation can be 
carried out, until the self-consistency with respect to the charge density is achieved. 

So far, the full implementation of the above scheme is not yet done. 
Instead, a simplified scheme was applied to nickel \cite{biermann03}. 
In the application, the GW calculation was done only for one-shot. 
Also, the frequency dependence of $U$ was neglected, and its static value was used. 
Improvement over these simplifications and more applications are highly anticipated. 

%%%%%%%%%%%%%%%%%%%%%%%%%%%%%%%%---------------------------------
\subsection{Variational Monte Carlo method} \label{Variational Monte Carlo method} 
%\label{Ch:VWF}
%----------------------------------------------------------
Many-variable variational Monte Carlo method has been developed recently
as a choice of the low-energy solver.\cite{Tahara,SorellaSR}
Historically, the variational wavefunction has played important roles in understanding physics of 
interacting fermions. Superconductivity by the BCS wavefunction, 
the roton excitation of $^4$He by Feynman\cite{Feynman} and Laughlin's wavefunction for the fractional quantum Hall state\cite{Laughlin}
are typical examples.  

However, it is also known that, to capture the essence, one has to have a 
good physical intuition beforehand and the validity and accuracy of the wavefunctions strongly rely on 
this intuition and genius idea.  From the first principles point of view, it is desired to get around this bias inherent in the variational approach as much as possible
to have a versatile method in hand.   
We need to construct wave functions which enhance the capability of removing biases posed on the variational forms. This is achieved at least partially by introducing enormous number of parameters.
Recent development in the VMC method allows us to deal with a large number of parameters \cite{Tahara,SorellaSR,SorellaSRH,VMCOptLM02}. 
Numerical techniques for optimizing many parameters 
are described in \S \ref{Optimization method}. By many parameters in the part of refined single-particle wavefunctions as well as in the part 
of correlation factors such as Gutzwiller factor to punish two electrons on the same Wannier orbital,  
it opens a way of simulating
strongly correlated systems by substantially reducing the biases as we see below.
 
%%%%%%%%%%%%%%%%%%%%%%%%%%%%%%%-------------------------------
\subsubsection{Functional form of variational wave functions} \label{Functional form of variational wave functions} 
%----------------------------------------------------------
The general functional form of wave functions we employ is
\begin{equation}
  |\psi\ra = \sP \sL | \phi \ra,
\end{equation}
where $|\phi\ra$ is a Hartree-Fock-Bogoliubov type wave function called ``one-body part,''
$\sL$ is the quantum-number projector \cite{ManyBody, PIRGMizusaki} controlling symmetries of wave function, and 
$\sP$ is the Gutzwiller-Jastrow factor \cite{Jastrow,Gutzwiller} including many-body correlations.
In order to improve variational wave functions within the sector classified by quantum numbers, 
we employ $\sP$ that preserves symmetries of $\sL|\phi\ra$.
This means that $\sL$ and $\sP$ are commutable ($\sP\sL=\sL\sP$).

%%%%%%%%%%%%%%%%%%%%%%%%%%%%%%-----------------------------------
\subsubsection{One-body part} \label{One-body part}
%\label{Ch:VWF-Sec:One-body}
%----------------------------------------------------------
The one-body part usually corresponds to the mean-field Slater determinant with several variational parameters. Though the Gutzwiller-Jastrow factor introduces many-body correlations through $\cal{P}$, this variational hypothesis for the one-body part $|\phi\rangle$ strongly restricts flexibility of wave functions. 
The functional form of the one-body part has been reexamined and as many as 
possible variational parameters have been introduced in order to improve wave functions.
\par
Following this reexamination, we use a variational wave function in the form  
\begin{align} \notag
  |\phi_{\text{pair}}\ra &= 
  \Biggl[ \sum_{\vk\in\text{BZ}}
    \varphi^{(1)}(\vk)
    c_{\vk\uparrow}^\dag c_{-\vk\downarrow}^\dag
\\ \label{eq:VWF-one-bodyfull}
  &+\!\!\!
    \sum_{\vk\in\text{AFBZ}}
\!\!\!
    \varphi^{(2)}(\vk)
    \Bigl( 
        c_{\vk+\vQ\uparrow}^\dag c_{-\vk\downarrow}^\dag
      - c_{\vk\uparrow}^\dag c_{-\vk-\vQ\downarrow}^\dag
    \Bigr)
  \Biggr]^{N/2} \!\!| 0 \ra,
\end{align}
where $\varphi^{(1)}(\vk) $ and $\varphi^{(2)}(\vk)$ are $\vk$ dependent variational parameters
with the conditions
\begin{equation}
  \varphi^{(1)}(-\vk) = \varphi^{(1)}(\vk) \, , \ 
  \varphi^{(2)}(-\vk) = \varphi^{(2)}(\vk)
.
\end{equation}
This form allows explicitly representing antiferromagnetic mean-field state with the periodicity $\vQ$ 
by using the second term proportional to $\varphi^{(2)}(\vk)$.  
Not only the Fermi sea state, this form also allows representing the BCS-type superconducting wavefunction with the pairing amplitude
proportional to $\varphi^{(1)}(\vk)$. 
Therefore, dealing with $\varphi^{(1)}(\vk)$ and $\varphi^{(2)}(\vk)$ directly as $\vk$-dependent variational parameters allows us to express various states such as paramagnetic metals, antiferromagnetically ordered states, and superconducting states with any gap function within a single framework of $|\phi_{\text{pair}}\ra$.
Moreover, since the number of the variational parameters increases scaled by the system size, it allows taking account of fluctuation effects with short-ranged correlations.
We call $|\phi_{\text{pair}}\ra$ a ``generalized pairing function'' and $\varphi^{(1)}(\vk)$, and $\varphi^{(2)}(\vk)$ are called ``pair orbitals.''
Introducing all the possible ordered vectors $\vQ$ would further 
generalize $|\phi_{\text{pair}}\ra$. 
However, this extension 
substantially increases the number of variational parameters 
and computational costs ($\sim\sO(N)$). Therefore, 
one physically plausible $\vQ$ has been attempted so far.
\par
By using the $\vk$-dependent parameters $\varphi^{(1)}(\vk)$ and the Gutzwiller factor $\sP_{\text{G}}^{\infty}$ defined below, our variational wave function 
can also represent 
the resonating valence bond (RVB) basis \cite{LDA}, which 
is known to offer 
highly accurate variational wave functions in spin systems. 
We note that the RVB basis can represent the state with spin correlations decaying with arbitrary power laws for increasing distance.
%The relation between $\sP_{\text{G}}^{\infty} | \phi_{\text{pair}}\ra$ and the RVB basis is discussed in Appendix \ref{App:RVB}.
\par
%In quantum chemistry, Casula \etal{} have introduced a similar wave function called an antisymmetrized geminal power \cite{CasulaJCP}. Since the singlet pairs are only included in this wave function, it is not connected to the AF mean-field wave function. Our extension offers a clear representation to include the singlet pairing wave functions and the AF mean-field wave functions. 
\par
For actual numerical calculations, we rewrite $|\phi_{\text{pair}}\ra$ in a real space representation:
\begin{equation}
  |\phi_{\text{pair}}\ra = 
  \Biggl[ \sum_{i,j=1}^{\Ns}
    f_{ij} 
    c_{i\uparrow}^\dag c_{j\downarrow}^\dag 
  \Biggr]^{N/2} | 0 \ra
\end{equation}
with
\begin{align} \notag
  f_{ij} =& \frac{1}{\Ns} 
  \sum_{\vk\in\text{BZ}} \varphi^{(1)}(\vk) e^{i\vk\bdot(\vr_i-\vr_j)}
 \\ \label{eq:VWF-one-bodyfull-real-space}&+
  \frac{1}{\Ns} 
  \sum_{\vk\in\text{AFBZ}} \varphi^{(2)}(\vk) e^{i\vk\bdot(\vr_i-\vr_j)} \Bigl(e^{i\vQ\bdot\vr_i}-e^{-i\vQ\bdot\vr_j}\Bigr)
.
\end{align}
Here, BZ means the summation in the Brillouin zone and AFBZ represents the folded zone for translational symmetry broken states with the periodicity $vQ$.  One of the parameters $\{\varphi^{(1)}(\vk),\varphi^{(2)}(\vk)\}$ is 
not independent because of the normalization of the wave function.
We note that $f_{ij}$ depends only on $\vr_i-\vr_j$ because of the translational symmetry.
In practical calculations, the numbers of the variational parameters $f_{ij}$ can be decreased to reduce the computational load by restricting the range of $f_{ij}$ into only short-ranged combinations.   
%----------------------------------------------------------
%&\%%%%%%%%%%%%%%%%%%%%%%%%--------------------------------------
\subsubsection{Gutzwiller-Jastrow factors} \label{Gutzwiller-Jastrow factors}  
%\label{Ch:VWF-Sec:GutzJast}
%----------------------------------------------------------
In the variational study, the Gutzwiller-Jastrow type wave functions \cite{Jastrow,Gutzwiller} are often used to 
take account of 
many-body correlations. The Gutzwiller-Jastrow correlation factor $\sP$ allows us to go beyond a single Slater determinant and a linear combination of many Slater determinants are generated after the operation of $\sP$ to a one-body wave function $|\phi\ra$, which is crucial in representing strong correlation effects. 
Three types of many-body operators $\sP_{\text{G}}$, $\sP_{\text{d-h}}^{\text{ex.}}$, and $\sP_{\text{J}}$, which are called the Gutzwiller factor, the doublon-holon correlation factor, and the Jastrow factor, respectively, have been employed for the low-energy solvers so far.
%----------------------------------------------------------
\par
%----------------------------------------------------------
Gutzwiller has introduced a basic and efficient correlation factor $\sP_{\text{G}}$ \cite{Gutzwiller}, which punishes double occupancy of up and down electrons at the same Wannier orbital as
\begin{equation}
  \sP_{\text{G}} = \exp \biggl[
    -g\sum_{i} n_{i\uparrow} n_{i\downarrow}
  \biggr]
=
  \prod_{i} \Bigl[ 1 - (1-e^{-g}) n_{i\uparrow} n_{i\downarrow} \Bigr]
,
\end{equation}
where $g$ is a variational parameter and $i$ represents the site and orbital indices. In the limit $g\to\infty$, $\sP_{\text{G}}$ fully projects out the configurations with finite double occupancy as 
\begin{equation} 
  \sP_{\text{G}}^{\infty} = \prod_{i} \Bigl[ 1 -  n_{i\uparrow} n_{i\downarrow} \Bigr]
.
\label{eq:VWF-GutzInf}
\end{equation}
$\sP_{\text{G}}^{\infty}$ is used for the Heisenberg model and the $t$-$J$ model. 
In Hubbard-type models with finite $U/t$ with the onsite interaction $U$ and the typical transfer $t$, 
the double occupancy at the same orbital and site is nonzero 
even in the insulating state. Thus we deal with $\sP_{\text{G}}$ at a finite $g$.
%----------------------------------------------------------
\par
%----------------------------------------------------------
In order to 
take account of 
many-body effects beyond the Gutzwiller factor, the doublon-holon correlation factor \cite{Kaplan, YokoyamaShiba3} is implemented in the wave function. 
This factor comes from the idea that a doublon (doubly occupied site) 
and a holon (empty site) are bound in the insulator for large $U/t$.
An example of the short-ranged correlation factor has a form
\begin{equation}
  \sP_{\text{d-h}} = \exp \biggl[
    -\ga_1 \sum_{i} \xi_{i(0)}^{(1)} 
  \biggr]
,
 \label{eq:d-h01}
\end{equation}
where $\ga_1$ is a variational parameter. Here, $\xi_{i(0)}^{(1)}$ is written by
\begin{equation}
  \xi_{i(0)}^{(1)} = d_i \prod_{\tau}^{\text{n.n.}} (1-h_{i+\tau}) + h_i \prod_{\tau}^{\text{n.n.}} (1-d_{i+\tau})
,
\end{equation}
where the product $\prod_{\tau}^{\text{n.n.}}$ runs over nearest-neighbor sites, and $d_i=n_{i\uparrow}n_{i\downarrow}$ and $h_i = (1-n_{i\uparrow})(1-n_{i\downarrow})$ are doublon and holon 
operators, respectively. 
This factor takes into account the attraction of a doublon and a holon at the nearest neighbor site.
The doublon-holon correlation factor $\sP_{\text{d-h}}$ given by eq. (\ref{eq:d-h01})
or by slightly different forms has been employed in several VMC studies \cite{Kaplan,YokoyamaShiba3,Liu,WataVMC01,YokoVMCdiag}. 
%Recently, there is a proposal to extend eq.(\ref{eq:d-h01}) by introducing 
%many $\xi_{i(m)}^{(\ell)}$ \cite{KobayashiYokoyama}. We take $\sP_{\text{d-h}}^{\text{ex.}}$ as
%\[
%  \sP_{\text{d-h}}^{\text{ex.}} = \exp\biggl[
%   - \sum_{m=0}^{2}\sum_{\ell=1,2} \ga_{(m)}^{(\ell)} \sum_{i} \xi_{i(m)}^{(\ell)}
%  \biggr]
%,
%\]
%where $\ga_{(m)}^{(\ell)}$ are variational parameters. 
%It is in principle possible to include operators with $m=3,4, \ldots$, 
%but contributions of higher $m$ parts have turned out to be negligible while have induced instabilities in our optimization procedure. 
%Therefore we confine ourselves to $m$ up to $2$.
The correlation between doublons and holons at further distance may also be considered in more sophisticated forms.\cite{Tahara} 
%----------------------------------------------------------
\par
%----------------------------------------------------------
Jastrow has introduced a long-ranged correlation factor for continuum systems \cite{Jastrow}. This factor takes into account correlation effects through two-body operators. In the Hubbard model at quarter filling, Yokoyama and Shiba have discussed the effects of the Jastrow-type correlation factor \cite{YokoyamaShiba3}. Recently, Capello \etal {} have claimed a necessity of this factor to describe the Mott transition.\cite{CapelloPRL01} The Jastrow factor $\sP_{\text{J}}$ in lattice models has the following form:
\begin{equation}
  \sP_{\text{J}} = \exp \biggl[
    - \frac{1}{2} \sum_{i\neq j} v_{ij} n_{i} n_{j}
  \biggr] 
\end{equation}
with two-body terms, 
where $n_{i}=n_{i\uparrow}+n_{i\downarrow}$ is a density operator and $v_{ij}=v(\vr_i-\vr_j)$ are variational parameters depending on the displacement $\vr_i-\vr_j$.

In addition to the correlation factor, one can also operate Hamiltonian matrix $\sH$ to further approach the true ground state.  This is the idea to implement Lanczos-type diagonalization partially.  The first order correction is realized by operating $1+\alpha \sH$ with a variational parameter $\alpha$, which corresponds to the first order Lanczos step.\cite{Tahara,HeebRice}.  

%%%%%%%%%%%%%%%%%%%%%%%%%%-----------------------------------------
\subsubsection{Quantum-number projection} \label{Quantum-number projection} 
%----------------------------------------------------------
In general, quantum many-body systems have several symmetries related to the Hamiltonian such as translational symmetry, point group symmetry of lattice, $U(1)$ gauge symmetry, and $SU(2)$ spin-rotational symmetry. 
While symmetry breaking occurs in the thermodynamic limit, these symmetries must be preserved in finite many-body systems.
\par
Variational wave functions constructed from one-body parts and the Gutzwiller-Jastrow factors do not often satisfy inherent symmetry properties, because the Hartree-Fock-Bogoliubov type one-body part comes from symmetry broken mean-field treatment. 
Even in the generalized pairing wave function $|\phi_{\text{pair}}\ra$, the spin-rotational symmetry is broken by the orbital $\varphi^{(2)}(\vk)$ which enables to include the mean-field AF state.
\par
The quantum-number projection technique \cite{ManyBody} enables to control symmetries of wave function. This technique has been used successfully in the PIRG method \cite{PIRGMizusaki} and the Gaussian-basis Monte Carlo method \cite{GBMCAssaad, GBMCAimi}. 
By using the quantum-number projection together with the Gutzwiller-Jastrow factor, one can construct variational wave functions with controlled symmetries and many-body correlations.
The quantum-number projection operator $\sL$ is constructed by superposing transformation operators $T^{(n)}$ with weights $w_{n}$:
\begin{equation} \label{eq:QPgeneral}
  \sL |\phi\ra = \sum_{n} w_{n} T^{(n)} |\phi\ra = \sum_{n} w_{n} |\phi^{(n)}\ra
,
\end{equation}
where $|\phi\ra$ and $|\phi^{(n)}\ra$ are the original one-body part and the transformed one-body parts, 
respectively. When $\sL$ restores some continuous symmetry, the summation $\sum_{n}$ is replaced by the integration over some continuous variable.
%----------------------------------------------------------
\par
%----------------------------------------------------------
For example, the $SU(2)$ spin-rotational symmetry preserved in many effective models is restored by superposing wave functions rotated in the spin space. 
The spin projection operator $\sL^{S}$ which filters out $S^z=0$ component of $|\phi\ra$ and generates a state with total spin $S$ and $S^z=0$ has a form
\begin{equation}
  \sL^S = \frac{2S+1}{8\pi^2} \int d\varOmega \, P_{S}(\cos \beta) R(\varOmega)
,
\label{eq:projection}
\end{equation}
where $\varOmega=(\ga,\gb,\gc)$ is the Euler angle and the integration is performed over whole range of $\varOmega$. The weight $P_{S}(\cos\beta)$ is the $S$-th Legendre polynomial. The rotational operator $R(\varOmega)$ is defined as
\begin{equation}
  R(\varOmega) = R^z(\ga) R^y(\gb) R^z(\gc) = e^{i\ga S^z} e^{i\gb S^y} e^{i\gc S^z}
,
\end{equation}
where $S^y$ and $S^z$ are total spin operators of $y$ and $z$ directions, respectively.
\par
The one-body part introduced in this article contains 
only $S^z=0$ component $|\phi\ra = [\sum_{ij} f_{ij} c_{i\uparrow}^\dag c_{j\downarrow}^\dag]^{N/2}|0\ra$. Then, the integration over $\gc$ and $\ga$ can be omitted and $\sL^S|\phi\ra$ is written as 
\begin{align} \notag
  \sL^S|\phi\ra &= \sum_{x}^{S^z=0} | x\ra\la x|\sL^S|\phi\ra
\\&=
  \sum_{x}^{S^z=0} | x\ra 
  \frac{2S+1}{2}
  \int_{0}^{\pi} \!\!\!\!d\gb
  \, \sin\beta\, P_{S}(\cos\beta) 
  \la x|R^y(\gb)|\phi\ra
.
\end{align}
with the complete basis set of the real space representation $| x\ra$. 
The integration over $\beta$ is evaluated efficiently by the Gauss-Legendre quadrature in actual numerical calculations \cite{NumRec}. 
Typically, for $S=0$ of the half-filled electron system for the single-band Hubbard model on square lattices with the sizes $4\times 4$ and $14\times 14$, $10$ and $20$ mesh points are sufficient, respectively.
\par
%
%----------------------------------------------------------
\par
%----------------------------------------------------------
There are also other quantum-number projections to restore symmetries \cite{PIRGMizusaki}. 
The total momentum projection and the lattice symmetry projection restore the translational symmetry and the point group symmetry of lattice, respectively. 
The momentum projection is given by taking 
$T^{(n)}$ as the operator to shift the state with the translation vector $\vR_n$.
The state with the total momentum $\vk$ is obtained by employing $w_n=\exp(i\vk\cdot\vR_n)$ with the summation over $n$ in eq.(\ref{eq:projection}) for all the possible spatial translations in the finite size system.
The momentum projections can be redundant with the spin projection, if the translational symmetry is automatically restored 
by the spin projection, where a superposition of the spin-rotated wave functions includes a superposition of spatially translated wave functions. In such cases, the momentum projection does not make a better wavefunction any more and can be omitted.

With the quantum number projections, not only the ground state but also excited states with specified quantum  numbers are obtained.  It is useful, for instance, in obtaining the dispersion of quasiparticles by taking the momentum projection.  

%%%%%%%%%%%%%%%%%%%%%%%%--------------------------------
\subsubsection{Optimization method} \label{Optimization method} 
Here, we summarize the basic idea of wave function optimizations. The stochastic reconfiguration (SR) method\cite{SorellaSR} introduced by Sorella has been employed in many-parameter optimization.
%----------------------------------------------------------
%\subsection{Basic idea of wave function optimization ---Steepest Descent method and Newton method}
%----------------------------------------------------------

\noindent
{\bf [1] Wave function optimization by energy minimization} \\
We first recollect the conventional way of minimizing the variational energy 
$E_{\vga} = \la \psi_{\vga} | \Ha | \psi_{\vga}\ra / \la \psi_{\vga} | \psi_{\vga}\ra$ 
estimated from the wave function $| \psi_{\vga}\ra$ with variational parameters $\{\ga_k | k=1,\cdots,p\}$. 
Here $\vga$ denotes the initial 
vector in the $p$-dimensional parameter space.
\par
The energy $E_{\vga+\vgc}$ is expanded up to the second order around $\vga$:
\begin{equation}
  E_{\vga+\vgc} = E_{\vga} + \sum_{k=1}^{p} g_k \gc_k 
                 + \frac{1}{2} \sum_{k,\ell=1}^{p} h_{k\ell} \gc_k \gc_\ell
                 + \mathcal{O}(\vgc^3),
\end{equation}
where $\vgc$ is the vector for parameter variations,
\begin{equation}
  g_k = \frac{\partial}{\partial \ga_k} E_{\vga}
\quad
  (k=1,\cdots,p)
\end{equation}
are the energy gradient vector $\bm{g}$, and
\begin{equation}
  h_{k\ell} = \frac{\partial^2}{\partial \ga_k \partial \ga_\ell} E_{\vga}
\quad
  (k,\ell=1,\cdots,p)
\end{equation}
are the elements of the energy Hessian matrix $\mh$.
\par
The steepest decent (SD) method 
gives the updated variational parameter by 
\begin{equation}
  \ga'_{k} = \ga_{k} + \bar{\gc}_{k},
\end{equation}
where the change from the initial value $\ga_{k}$ should be 
%\begin{equation} \label{eq:Opt.004}
%  \bar{\gc}_k = -\varDelta t \, g_k
%\quad
%  (\bar{\vgc} = -\varDelta t \, \bm{g})
%,
%\end{equation}
%with a small constant $\varDelta t$.
%Combination with the Hessian leads to the Newton method. From the stationary condition $\partial E_{\vga} / \partial \ga_k = 0$ ($k=1,\cdots,p$), 
%the best parameter change $\bar{\vgc}$ is 
%\begin{equation} \label{eq:Opt.005}
%  \bar{\gc}_k = -\sum_{\ell=1}^{p} h_{k\ell}^{-1} g_\ell
%\quad
%  (\bar{\vgc} = - \mh^{-1} \bm{g}).
%\end{equation}
%\par
%Let us generalize eqs. (\ref{eq:Opt.004}) and (\ref{eq:Opt.005}) with suitably chosen nonsingular matrix $\mX$:
\begin{equation} \label{eq:Opt.006}
  \bar{\gc}_k = -\sum_{\ell=1}^{p} X_{k\ell}^{-1} g_\ell
\quad
  (\bar{\vgc} = - \mX^{-1} \bm{g}).
\end{equation}
with a suitably chosen nonsingular matrix $\mX$. 
%Equations (\ref{eq:Opt.004}) and (\ref{eq:Opt.005}) are reduced from eq. (\ref{eq:Opt.006}) 
This general form reduces to the steepest descent (SD) method in the choice $X_{k\ell}=\delta_{k\ell} / \varDelta t$ and 
to the Newton method by taking the hessian for $X$ as $X_{k\ell} = h_{k\ell}$, respectively.
In general, $\mX$ should be properly chosen to accelerate the optimization.

%----------------------------------------------------------
%\subsection{Stochastic Reconfiguration method}
%----------------------------------------------------------
\noindent
{\bf [2] Stochastic reconfiguration method }\\
Sorella has introduced the SR method \cite{SorellaSR} for a better choice of $\mX$ 
in optimizing many variational parameters.
For the normalized wave function
\begin{equation}
  |\bar{\psi}_{\vga}\ra = \frac{1}{\sqrt{ \la \psi_{\vga}|\psi_{\vga}\ra }} |\psi_{\vga}\ra,
\end{equation}
up to the first order around $\vga$, $|\bar{\psi}_{\vga+\vgc}\ra$ is expanded as
\begin{equation}
  |\bar{\psi}_{\vga+\vgc}\ra = |\bar{\psi}_{\vga}\ra 
  + \sum_{k=1}^{p} \gc_k |\bar{\psi}_{k\vga}\ra
  + \mathcal{O}(\vgc^2),
\end{equation}
where $|\bar{\psi}_{k\vga}\ra$ ($k=1,\cdots,p$) are the derivatives of $|\bar{\psi}_{\vga}\ra$:
\begin{align}\notag
  &|\bar{\psi}_{k\vga}\ra = \frac{\partial}{\partial \ga_k} |\bar{\psi}_{\vga}\ra
\\ \label{eq:Opt-SR-049}
&=
  \frac{1}{\sqrt{\la \psi_{\vga}|\psi_{\vga}\ra }}
  \biggl(
    \frac{\partial}{\partial \ga_k} |\psi_{\vga}\ra
  - \frac{\la \psi_{\vga}|(\partial / \partial \ga_k)|\psi_{\vga}\ra}{\la \psi_{\vga}|\psi_{\vga}\ra} 
    |\psi_{\vga}\ra
  \biggr).
\end{align}
The wave function set $\{ |\bar{\psi}_{k\vga}\ra | k=1,\cdots,p \}$ forms nonorthogonal basis in the $p$-dimensional parameter space. 
The norm change from $|\bar{\psi}_{\vga}\ra$ to $|\bar{\psi}_{\vga+\vgc}\ra$ is
\begin{align}\notag
  \varDelta_{\text{norm}}^{2} 
&=
  \Bigl\Vert |\bar{\psi}_{\vga+\vgc}\ra - |\bar{\psi}_{\vga}\ra \Bigr\Vert ^2
\\ \label{eq:Opt.010}
&=
  \sum_{k,\ell=1}^{p} \gc_k\gc_\ell \la\bar{\psi}_{k\vga}|\bar{\psi}_{\ell\vga}\ra
=
  \sum_{k,\ell=1}^{p} \gc_k\gc_\ell S_{k\ell}.
\end{align}
%Since $S_{k\ell} = \la\bar{\psi}_{k\vga}|\bar{\psi}_{\ell\vga}\ra$ is the overlap matrix in the parameter space, 
%$\mS$ becomes positive definite even with a finite number of samples. 
Equation (\ref{eq:Opt.010}) shows that $\mS$ whose $(k,\ell)$ element is $S_{k\ell}$ is the metric matrix in the parameter space.
\par
The SR method chooses $\mS$ as the matrix $\mX$ in eq. (\ref{eq:Opt.006}), namely
\begin{equation} \label{eq:Opt.011}
  \bar{\gc}_k = -\varDelta t \sum_{\ell=1}^{p} S_{k\ell}^{-1} g_\ell
\quad
  (\bar{\vgc} = -\varDelta t \, \mS^{-1}\bm{g})
,
\end{equation}
where $\varDelta t$ is a small constant.
%As the overlap matrix $\mS$ does not have any information about the energy Hessian, the SR method is close to the SD method. 
%The main difference is that 
The SR method takes into account the variation of the wave function in addition to the SD method. 
We can derive eq. (\ref{eq:Opt.011}) by minimizing the functional 
$
  \mathcal{F}_{\text{SR}}
= \varDelta E_{\text{lin.}} + \lambda \varDelta_{\text{norm}}^{2}
$
with a Lagrange multiplier $\lambda$. Here $\varDelta E_{\text{lin.}} = \sum_{k} g_k\gc_k$ is the linear change of the energy. The stationary condition $\partial \mathcal{F}_{\text{SR}} / \partial \gc_k=0$ ($k=1,\cdots,p$) leads to the SR formula (\ref{eq:Opt.011}) with $\varDelta t = (2\lambda)^{-1}$. 
%%%%%%%%%%
%The SD method can be obtained in a similar way. We can derive eq. (\ref{eq:Opt.004}) with $\varDelta t = (2\lambda)^{-1}$ by minimizing the functional 
%$
%  \mathcal{F}_{\text{SD}}
%= \varDelta E_{\text{lin.}} + \lambda \varDelta_{\text{SD}}^{2}
%$
%with 
%$
%  \varDelta_{\text{SD}}^{2} = 
%  \sum_{k} \gc_{k}^2
%,
%$
%where $\varDelta_{\text{SD}}$ is the Cartesian distance in the parameter space.
%The advantage of the SR method compared with the SD method is the following: Sometimes small change 
%of the variational parameters corresponds to a large change of the wave function, 
%and conversely a large change of the variational parameters corresponds to a small change of the wave function. 
%This leads to uncontrolled changes of the wave function if one takes $S_{k\ell}=\delta_{k\ell}$ (SD method). 
%When the change of the wave function exceeds a threshold, the iteration  for the optimization becomes unstable. 
%To suppress this instability, one needs to keep $\varDelta t$ small enough for the event of the largest change of the wave function. 
%If one can control change of the wave function, $\varDelta t$ can be taken large. 
%The SR method takes into account this effect through a better definition of the distance $\varDelta_{\text{norm}}$. 
%Thus, the SR method is more stable than the SD method. 
%Finally, we can choose larger $\varDelta t$ to accelerate the convergence.
The SR method is more stable than the conventional method, because the SD and hessian methods sometimes cause a large change in the wavefunction even though the changes in the variational parameters are small.  This large change in the wavefunction causes an instability in the iteration, which requires to keep $\Delta t$ very small and the iteration becomes inefficient. The SR method solves     
this difficulty.  To avoid the numerical instability possibly caused by an extremely large $\mS^{-1}$ in eq.(\ref{eq:Opt.011}), it is also useful to take $(1+\varepsilon) S_{kk}$ instead of $ S_{kk}$ for its expression in eq.(\ref{eq:Opt.011}) with a small constant $\varepsilon$.\cite{CasulaJCP, SorellaJCP} 

The positive definite matrix of $\mS$ may have an eigenvector with very small eigenvalues after the diagonalization. The variation in the direction in such an eigenvector is redundant and may be truncated for a better efficiency.  
Typically, the parameters are taken as $\varDelta t = 0.1$, $\varepsilon = 0.2$, and $\varepsilon_{\text{wf}} = 0.001$. 
For more details including the practical implementation of the optimization, readers are referred to Ref.\citen{Tahara}.  

\subsubsection{Benchmark} \label{Benchmark} 
\begin{figure}[t]
\centering
\includegraphics[width=0.45\textwidth]{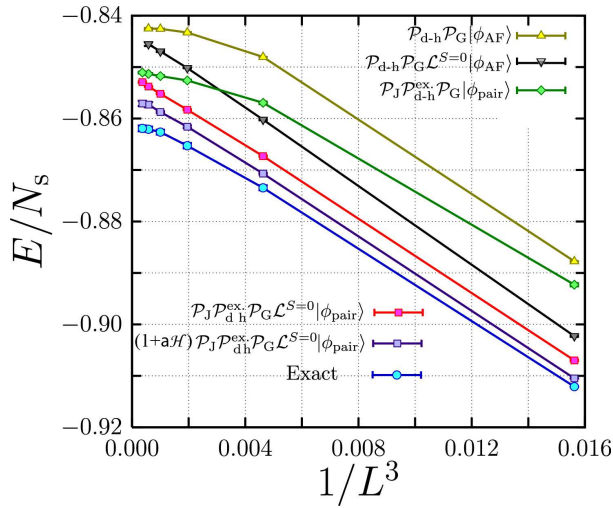}
\caption{%
(Color online) Total energy $E/\Ns$ as a function of $1/L^3$ for the Hubbard model on a square lattice with $\Ns=L\times L$ lattice with the transfer $t=1$ to the nearest neighbor only, and the onsite interaction $U=4$ at half filling $n=1$.\cite{Tahara} The exact values are calculated by the exact diagonalization ($L=4$) and AFMC ($L=6,8,10,12,14$).
Error bars are comparable to the symbol size. The accuracy is enhanced with improving variational forms.  The best results by (purple) square are obtained by operating the spin quantum number projection to $S=0$, Gutzwiller projection, doublon-holon correlation factor, Jastrow factor and the 1st order Lanczos step to the paired singlet function $|\phi_{pair}\rangle$. In this case, the relative error is smaller than 0.5 \% irrespective of the system size.
}
\label{fig:Res-SizeEn}
\end{figure}

\begin{figure}[t]
\centering
\includegraphics[width=0.45\textwidth]{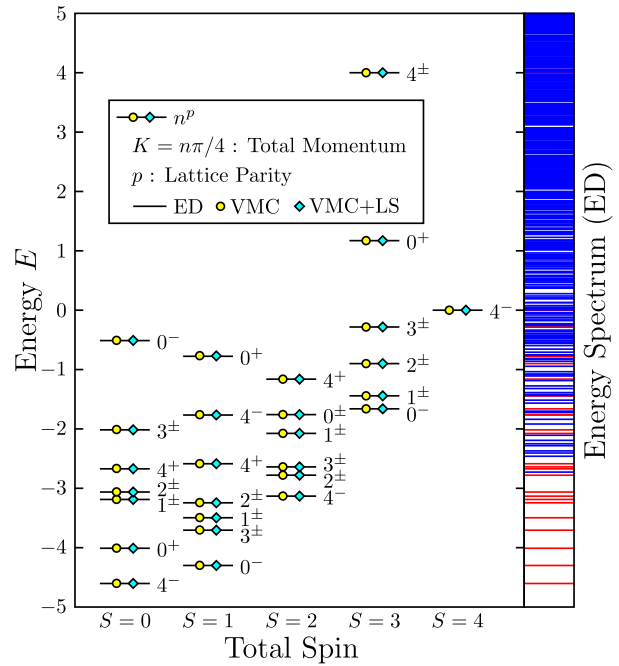}
\caption{%
(Color online) Excitation spectra of 1D Hubbard model compared with the exact diagonalization result (ED).
VMC can reproduce the lowest energy states of each quantum number. VMC+LS represents the results obtained by the form corresponding to the squares in Fig.\ref{fig:Res-SizeEn}. Energies are sufficiently accurate without the Lanczos step as we see for circles.
}
\label{fig:spect01o}
\end{figure}

The accuracy of the multi-variable VMC method has been critically tested in various cases and in many cases has proven to be a good low-energy solver as comparably accurate as PIRG reviewed below.
In Fig.\ref{fig:Res-SizeEn}, we show the energy accuracy for the case of the Hubbard model on the square lattice with various system sizes at the onsite interaction $U=4$ and the nearest-neighbor transfer $t=1$.\cite{Tahara} 
%In Fig.~\ref{fig:rerror31o} we show the relative accuracy of energies.
Excitations are also tested in Fig.\ref{fig:spect01o}, where different total spin $S$, momentum and parity states are obtained by the present VMC for the Hubbard model on a 1D $8\times 1$ lattice at $U=4$ and $t=1$. 
%\begin{figure}[t]
%\centering
%\includegraphics[width=0.4\textwidth]{comp01o.eps}
%\caption{%
%(Color online) Peak value of spin structure factor $S(\bm{q}_{\text{peak}})$ as a function of the doping concentration $\delta$ for for the Hubbard model on a square lattice with $t'/t=0$ and $U/t=4$. Diamonds are AFQMC results reported in ref. \citen{FurukawaImada}.  Triangles and circles are VMC results obtained by using the variational wave function $\mathcal{P}_{\text{J}}\mathcal{P}_{\text{d-h}}^{\text{ex.}} \mathcal{P}_{\text{G}} \mathcal{L}^{S=0} | \phi_{\text{pair}} \rangle$. Error bars are comparable to the symbol size. The solid curve is the fitting given in Fig. 13(a) of ref. \citen{FurukawaImada}.
%}
%\label{fig:Res-DopeSq}
%\end{figure}
%\begin{figure}[t]
%\centering
%\includegraphics[width=0.45\textwidth]{rerror31o.eps}
%\caption{%
%(Color online) Relative error of energy
%}
%\label{fig:rerror31o}
%\end{figure}

%%%%%%%%%%%%%%%%%%%%
\subsection{Path-integral renormalization group} \label{Path-integral renormalization group} 
Now we introduce PIRG method applied for low-energy effective models.\cite{Kashima} 
Since reviews are also found in the literature,\cite{ImadaPIRGReview} we review only the essential part here.
This method allows to approach the ground state of the model with a high accuracy.
It starts, as in other Monte Carlo and projection methods, with a relation 
\begin{equation}
| {\psi _g} \rangle=\lim_{\tau \rightarrow \infty} e^{-\tau {H} }| {\phi _{\rm initial}} \rangle. 
\end{equation}
for an arbitrary chosen state  $| {\phi _{\rm initial}} \rangle $ that is not orthogonal to the ground state $| {\psi _g} \rangle$.
By following the Feynman path integral, the operation of $\exp[-\tau H]$ is decomposed with $\Delta\tau$ as $\exp[-\tau H] \sim [\exp[-\Delta\tau H_K]\prod_i \exp[-\Delta\tau H_{U_i}]]^{\cal N}$.  Here we take a sufficiently large $N$ so as to satisfy $\tau={\cal N}\Delta\tau$. For simplicity, we have taken an example of the Hubbard type model with the onsite interaction $H_{U_i}$ at the $i$th site.
When we take a Slater determinant $| {\phi _{\rm initial}} \rangle $, the operation of 
$\exp[-\Delta\tau H_K]$ to the Slater determinant simply generates another single Slater determinant.  However, if we operate $\exp[-\Delta\tau H_{U_i}]$ to a single Slater determinant, the result is a linear combination of two Slater determinants when this interaction is transformed by the discrete Stratonovich-Hubbard transformation.\cite{Hirsch} To approach the ground state we need to operate $\exp[-\Delta\tau H_{U_i}]$ many times, which requires a linear combination of an exponentially large number of Slater determinants. It easily exceeds the accessibility by computers.  Then by restricting within the computationally tractable range, a linear combination of $L$ Slater determinants in a partial Hilbert space truncated from the original Hilbert space as  
\begin{equation}
| {\psi ^{(L)}} \rangle =\sum_{\alpha =1}^L {c_\alpha }| 
{\phi^{(L)}_\alpha } \rangle, 
\end{equation}
is employed in PIRG.  The coefficients $c_{\alpha}$ and the choice of nonorthogonal basis $\phi^{(L)}_{\alpha}$ are numerically optimized.  Then by increasing $L$ systematically, the ground state is speculated from the extrapolation to large $L$ (essentially the limit $l\rightarrow \infty$) 
When the Hilbert space is truncated, it can lose the original symmetries of the Hamiltonian that are guaranteed by the conservation law, while the ground state should be an eigenstate of good quantum numbers that are conserved in the original Hamiltonian.  For instance the total spin and total momentum are normally good quantum numbers in the Hubbard-type Hamiltonian and in the ground state, one of the quantum numbers has to be chosen for each conservation law.  To restore the original symmetry, quantum number projection may be imposed to keep the quantum number of the truncated state.   This quantum number projection was combined with PIRG that has proven much better accuracy.\cite{Mizusaki} Since PIRG does not suffer from the negative sign problem known in AFMC, a high-accuracy calculations have been made possible.  We will refer to applications to ab initio calculations in \S \ref{Applications}.

%%%%%%%%%%%%%%%%%%%%%%%%%%%%
\section{Applications} \label{Applications} 
%%%%%%%%%%%%%%%%%%%%%%%%%%%%%%
Applications of the three-stage RMS formalisms combining {\it ab initio} electronic structure calculations, downfolding and low-energy solvers are diverse and it is not possible to cover all of them in this review. Here, we just pick up several examples to demonstrate the efficiency, accuracy and versatility of the method in practical applications. 
 \subsection{Dynamics of semiconductors} \label{Dynamics of semiconductors}
\begin{figure}[htbp] 
\begin{center} 
\includegraphics[width=0.4\textwidth]{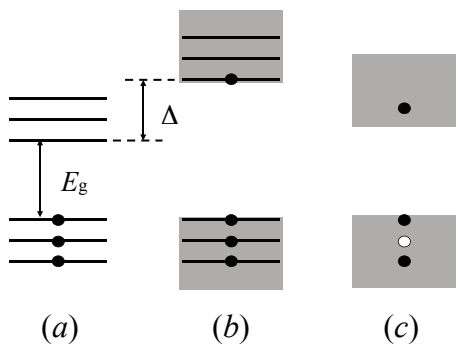}
\end{center} 
\caption{
%\textcolor{red}{
(a) Kohn-Sham energy is
the energy level of a non-interacting electron system, where 
the lowest $N$ states are occupied. 
(b) Quasiparticle energy is the electron removal / addition energy to the ground state of the $N$ electron 
system, which is different from (a) when electron interaction is considered. 
Optically excited state is shown in (c). Both an electron and a hole exist, and 
their interaction brings another many-body effect.}
%} 
\label{fig:gap} 
\end{figure}

Despite its success in $sp$ semiconductors and insulators, 
the LDA is not satisfactory when it comes to electron excited states. 
The LDA band gap is 1/2$\sim$2/3 of measured values in a wide range of materials. 
The cause of the deviation is not ascribed to approximations in the exchange correlation functional. 
In fact, the GGA yields similar results as the LDA and does not cure the band gap problem. 
The problem originates from the fact that Kohn-Sham eigenvalues do not 
correspond to observable quantities. 
Schematically it is expressed as Fig.\ref{fig:gap}.
The Kohn-Sham eigenvalues represent energy levels of a $N$ electron system, 
where the lowest $N$ levels are occupied. 
On the other hand, what is observed in (inverse) photoemission measurements is 
electron addition/removal energy, which is the total energy difference 
between the $N$ electron system and the $N\pm1$ electron system.
Theoretically, they are obtained by the spectral function of the one particle Green's function, 
which can be computed from first-principles, e.g. in the GW approximation.

The band gap of silicon is, for example, 1.17 eV experimentally. 
The LDA gap is 0.5 eV, much smaller than the experiment. 
In the Hartree-Fock approximation, the gap increases substantially to $>$ 6 eV 
because of the nonlocal exchange term. 
In the GW approximation, the exchange term is replaced with 
the screened one, thereby the increase of the gap is suppressed.
As a result, the GW yields the band gap close to the experimental one (Table \ref{tab:gap}). 
%\textcolor{red}{
The same trend is observed also in diamond.
%}
\begin{table}[tbp]
\begin{center}
\begin{tabular}{|ccc|}
\hline
& Silicon &  Diamond\\
\hline
LDA & 0.5 & 4.0  \\
HF & 6.4 \cite{gygi87}, 6.3 \cite{massidda93} & 12.9 \cite{gygi87}, 12.4 \cite{massidda93} \\
Hartree & -1.19 \cite{sodeyama08} & 1.90 
\cite{sodeyama08} \\
GW & 1.21 \cite{hybertsen85}, 1.24 \cite{godby} & 5.43 \cite{hybertsen85}, 5.33 \cite{godby} \\
Expt. & 1.17 & 5.48 \\
\hline
\end{tabular}
\end{center}
\caption{Band gap of silicon and diamond in 
LDA, Hartree-Fock, Hartree and GW approximations. 
The negative values mean the band overlap in metals. 
The energies are given in units of eV.}
\label{tab:gap}
\end{table}

The GW self-energy is nonlocal and energy dependent. 
The nonlocality increases the band gap substantially and solves the underestimation of the gap in LDA. 
The energy dependence, on the other hand,  reduces the gap and 
partially cancels the nonlocal effect.
Spatial and energy dependence of the dielectric function
$\epsilon({\bf r},{\bf r}';\omega)$ is also important. 
If the dielectric function is approximated as a function of  $|{\bf r}-{\bf r}'|$, 
the value of the gap changes significantly (local field effect). 
The energy dependence cannot be neglected either, but 
a simplification using the plasmon pole model is valid in weakly correlated materials. 
In the plasmon model, the imaginary part of the dielectric function is 
approximated as a single delta function. 
This enables us to extrapolate the energy dependence of the dielectric function 
from its static value. 
Many GW approximations adopt this approximation in order to improve computational efficiency. 

\begin{figure}[htbp] 
\begin{center} 
\includegraphics[width=0.4\textwidth]{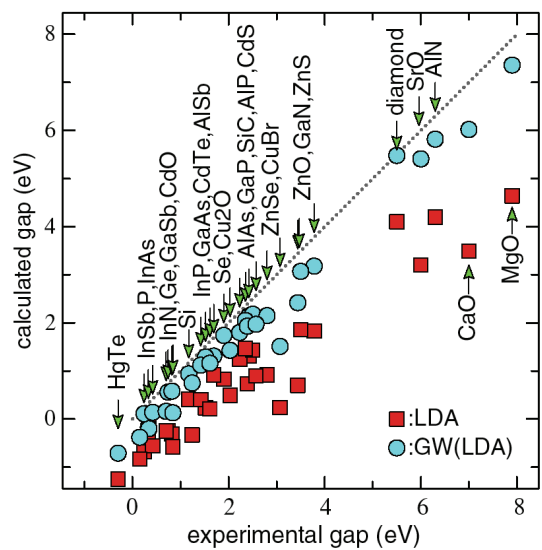}
\end{center} 
\caption{(Color online) Band gap of semiconductors and insulators.\cite{mvs06} 
%\textcolor{red}{
The dotted line is the ideal line that the theoretical gap agrees with the experiment. 
The LDA gap is systematically smaller than the experimental one, which is 
greatly improved by adding the GW self-energy correction.}
%}
\label{fig:gwgap} 
\end{figure} 

After a seminal work by Hybertsen and Louie \cite{hybertsen85}, 
many groups applied the GW method to a number of materials, 
including solids, surfaces, molecules, clusters and nanostructures. 
By today the GW method is established as a reliable method 
to predict the band gap of weakly and moderately correlated materials 
within 10-15 \% error to experimental values (Fig.\ref{fig:gwgap} \cite{mvs06}).

On the other hand, the conventional GW method is not satisfactory for strongly correlated materials. 
This is partly because the DFT gives a poor starting Hamiltonian. 
The band gaps are too small, and sometimes insulators are wrongly described to be metallic. 
Moreover, localized electron levels are too shallow in many cases. 
These drawbacks result in over-screened Coulomb interaction $W$, which in turn 
yields inaccurate self-energy. In principle, the dressed Green's function 
(including the GW self-energy correction) can be used to recalculate the self-energy. 
The calculation is continued until the self-consistency is achieved.
The fully self-consistent GW calculations have not been performed 
extensively up to now, partly because they are both computationally and technically demanding. 
The fully self-consistent GW calculations for the electron gas, however, have been
performed in detail with a rather discouraging result with regard to the excitation spectrum 
\cite{holm98}. On the other hand, the total energy is found to be in
almost perfect agreement with the quantum Monte Carlo \cite{holm99}.
Applications to silicon have been performed with four different methods,  
but the results are inconsistent %\textcolor{red}{
\cite{schoene98,ku02,zein06,kutepov09}.
Further systematic calculations are anticipated for deeper insight into self-consistency. 

Another approach to reach self-consistency is to use the GW self-energy 
to update the one-particle wavefunctions and eigenvalues. 
Due to the energy dependence of the self-energy it is not clear,
however, how this should be done. Consider the self-energy in the Kohn-Sham basis, 
\begin{equation}
\langle \psi_{{\bf k}n} | \Sigma(\omega) | \psi_{{\bf k}m} \rangle \;.
\end{equation}
To use this self-energy in place of the LDA exchange-correlation potential, 
it is necessary to determine the energy in some way. 
We note that the self-energy matrix is required to be Hermitian 
in order to produce a set of orthonormal wavefunctions.
There are several choices. 
The simplest one is to fix the energy at some chosen energy, say, 
the Fermi energy or the center of the band of interest. 
Another choice is to take the average \cite{faleev04}
\begin{eqnarray}
\langle \psi_{{\bf k}n} | V_{\rm xc}  | \psi_{{\bf k}m} \rangle = 
&\frac{1}{2}& \left[ \langle \psi_{{\bf k}n} | \Sigma(\epsilon_{{\bf k}n}) | \psi_{{\bf k}m} \rangle \right. \nonumber \\
&+& \left.
\langle \psi_{{\bf k}n} | \Sigma(\epsilon_{{\bf k}m}) | \psi_{{\bf k}m} \rangle \right].
\end{eqnarray}
This scheme is called quasiparticle self-consistent GW (QSGW) method. 
The scheme was applied to many materials including transition metal 
mono-oxides and $f$ electron systems, and improvement over LDA was confirmed 
in strongly correlated materials \cite{faleev04,mvs06,kotani07}. 
A self-consistent scheme based on the static COHSEX approximation (eq.(\ref{eq:cohsex}))
is also reported \cite{bruneval}.
Recently, Sakuma {\it et al.} proposed an alternative scheme \cite{sakuma09}. 
In their scheme, the quasi-particle equation is solved with neglecting the imaginary part of the self-energy,
\begin{equation}
\det | \omega - H_{0}({\bf k}) - \Re \Sigma({\bf k},\omega) | = 0 \;.
\end{equation}
The obtained quasiparticle wavefunctions $\{ \psi^{\rm QP}_{{\bf k}n} \}$ are not orthonormal 
because of energy dependence of 
the self-energy. The wavefunctions are then orthonormalized using L\"{o}wding's scheme \cite{lowdin}, 
that generates the closest set of orthonormal orbitals $\{ \psi_{{\bf k}n} \}$ to $\{ \psi^{\rm QP}_{{\bf k}n} \}$, where
\begin{eqnarray}
\psi &=& \psi^{\rm QP} C \;, \\
C C^{\dagger} &=& S \;. \\
S_{mn} &=& \langle \psi_{{\bf k}m} | \psi_{{\bf k}n} \rangle \;.
\end{eqnarray}
 \begin{figure}[htbp] 
\begin{center} 
\includegraphics[width=0.5\textwidth]{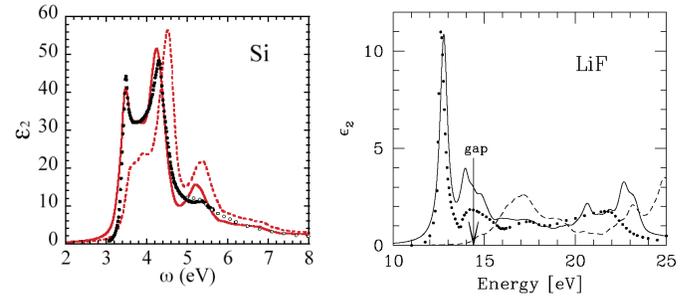}
\end{center} 
\caption{(Color online) Optical absorption spectra of (a) silicon 
and (b) LiF \cite{rohlfing-louie98}. The dashed line is the calculated spectrum 
in the independent particle approximation using Kohn-Sham wavefunctions 
and GW quasiparticle energies, the solid line includes the electron-hole interaction 
in the GW-BSE scheme, and open and closed circles are experimental data. 
%\textcolor{red}{
The BSE spectrum is red shifted compared to the independent 
particle approximation due to electron-hole interaction, and 
agrees well with the measurement.}
%}
\label{fig:si_lif} 
\end{figure} 
 
Optical absorption spectrum is another quantity where electron interaction 
is crucial both quantitatively and qualitatively. 
Figure \ref{fig:si_lif}(a) shows the optical absorption spectrum of crystalline 
silicon. The dashed line is the spectrum in the independent particle approximation, 
\begin{eqnarray}
\epsilon_2^{(0)} (\omega) &=& \left(\frac{2 \pi e}{m \omega} \right)^2
\sum_{v}^{\rm occ}\sum_{c}^{\rm unocc}
| \langle \psi_{v} | {\bf e}_{\lambda} \cdot{\bf v} | \psi_{c} \rangle |^2 \nonumber \\
&&\times \delta ( \omega - (E_{c} - E_{v} ) ),
\label{eq:ipa}
\end{eqnarray}
where $E_v$ and $E_c$ are the GW quasiparticle energies. 
As the GW gap is accurate, the threshold of the spectrum is close to 
the experimental one. 
However, the peak position is too high compared to the experiment. 
Moreover, the first peak at 3.5 eV is not 
reproduced in the calculation. These discrepancies clearly show 
that many-body effects are crucial for the optical absorption spectrum. 

The spectrum 
%\textcolor{red}{
including many-body effects
%} 
is obtained  by the imaginary part of the macroscopic 
dielectric function,
\begin{equation}
\epsilon_{\rm M}(\omega) = \lim_{{\bf q}\rightarrow 0}
\frac{1}{\epsilon^{-1}_{{\bf G}=0,{\bf G}'=0}({\bf q},\omega)} \;.
\label{eq:epsm}
\end{equation}
The independent particle approximation eq.(\ref{eq:ipa}) 
adopts two approximations on top of eq.({\ref{eq:epsm}). 
One is that the ${\bf G} \neq {\bf G'}$ components are neglected when $\epsilon$ is inverted. 
Namely, the local field effect is neglected in the calculation. 
Analysis revealed that this effect is minor in silicon \cite{louie75}. 
The other approximation is the RPA for the polarization eq.({\ref{eq:p0}).
To go beyond the RPA and include electron-hole interaction, we start with Hedin's equation. 
Firstly, the vertex function is evaluated in the following way. 
The GW approximation is adopted in the second term of eq.(\ref{eq:gamma}). 
Then it follows that $\delta\Sigma / \delta G = i W + i G (\delta W / \delta G)$. 
Assuming that the screening effect is not affected by electron-hole excitations, 
the $\delta W / \delta G$ contribution is safely neglected.
Putting thus obtained vertex function into eq.(\ref{eq:p}), 
we can compute an improved polarizability. 
The final form called Bethe-Salpeter equation is written as\cite{hanke78,onida02}
\begin{equation}
\epsilon_{M}(\omega) = 1 - \lim_{{\bf q}\rightarrow 0}
\left[
 v_{\bf G=0}({\bf q}) \bar{P}_{\bf G=G'=0}({\bf q},\omega)
\right]
,
\label{eq:hanke}
\end{equation}
\begin{equation}
\bar{P}(1,2) = {^{4}\bar{P}}(1,1,2,2),
\end{equation}
\begin{eqnarray}
&^{4}\bar{P}(1,2,3,4)& = {^{4}P}_0(1,2,3,4) + 
\int {^{4}P}_0(1,2,5,6) \nonumber \\
&\times&K(5,6,7,8) ^{4}\bar{P}(7,8,3,4) {\rm d}(5678), \label{eq:p4bse}
\end{eqnarray}
which is diagrammatically illustrated in Fig.~\ref{fig:bse}. 
Here 
${^4}\bar{P}$ is a correlation function between the electron and the hole. 
The first term in eq.(\ref{eq:p4bse}) is the correlation function in the 
independent particle approximation, whereas the second term 
represents the electron-hole interaction. It is characterized by 
the electron-hole interaction kernel $K$, given by 
\begin{eqnarray}
K(1,2,3,4) &=& \delta(1,2)\delta(3,4^-) \bar{v}(1,3) \nonumber \\
   &-& \delta(1,3)\delta(2,4) W(1^+,2) \;,
\label{eq:kernel}
\end{eqnarray}
%\textcolor{red}{場所を移動
where $\bar{v}$ is a modified bare Coulomb interaction, in which 
${\bf G}=0$ component in the Fourier representation is replaced with 0.
% }
The second term, called direct term, represents electron-hole attraction 
(excitonic effect), while the first term (exchange term) is the local field effect.
\begin{figure}[htbp] 
\begin{center} 
\includegraphics[width=0.5\textwidth]{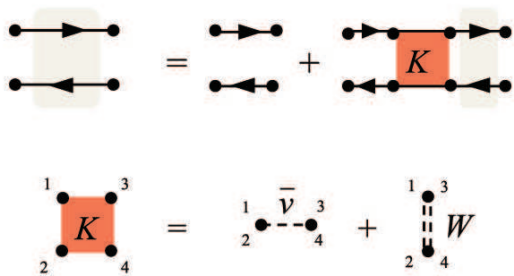}
\end{center} 
\caption{(Color online) Feynman diagrams representing Bethe-Salpeter equation. 
%\textcolor{red}{
The second term in the right hand side is the electron-hole interaction (top panel). 
The interaction is characterized by the kernel (bottom panel), where the second term 
represents the excitonic effect.}
%}
\label{fig:bse} 
\end{figure} 

Further approximations are introduced to make first-principles calculations to real materials feasible. 
The excited state is approximated to be 
a linear combination of single electron-hole excitations (Tamm-Dancoff approximation), 
\begin{equation}
| S \rangle = \sum_{c}^{\rm elec}\sum_{v}^{\rm hole} 
A_{cv}^{S} \hat{a}^{\dagger}_c \hat{b}^{\dagger}_v | 0 \rangle
\; ,
\end{equation}
where $\hat{a}^{\dagger}_c$ and $\hat{b}^{\dagger}_v$ are 
creation operators for an electron and a hole, respectively. 
In addition, the frequency dependence of the screened Coulomb interaction is neglected 
and its static value is used. With these approximations, the Bethe-Salpeter equation (\ref{eq:p4bse}) 
is reduced to the following eigenvalue problem, 
\begin{equation}
(E_c^{\rm QP} - E_v^{\rm QP})A_{cv}^{S} + \sum_{c'v'}K_{cv,c'v'}A_{c'v'}^{S} 
= \Omega_{S} A_{cv}^S
\; ,
\label{eq:bse} 
\end{equation}
\begin{eqnarray}
K_{cv,c'v'} &=& \int \psi_c^*({\bf r}_1) \psi_v({\bf r}_2) 
K(1,2,3,4)
\nonumber \\
&\times& \psi_{c'}({\bf r}_3) \psi_{v'}^*({\bf r}_4) 
{\rm d}(1234). 
\end{eqnarray}
Using the eigenstates and eigenvalues of eq.(\ref{eq:bse}), 
the macroscopic dielectric function is obtained by 
\begin{equation}
\epsilon_2(\omega) = \left(\frac{2 \pi e}{m \omega} \right)^2
\sum_{S} | \langle 0 | {\bf e}_{\lambda} \cdot {\bf v} | S \rangle |^2
\delta(\omega - \Omega_S)
\;.
\end{equation}

Many-body theory for optical absorption is seen already in 1960's \cite{sham66}. 
In the 70's, semi-quantitative calculation was reported by Hanke \cite{hanke}. 
First {\it ab-initio} calculation was carried out in 1995 for sodium cluster \cite{onida95}. 
The scheme was applied to solids  2-3 years after \cite{albrecht,benedict98,rohlfing-louie98}.

The Bethe-Salpeter-Equation (BSE) spectrum of Si is shown by the solid line in Fig.\ref{fig:si_lif}(a). 
The first peak appears and the second peak is red shifted by the electron-hole interaction. 
The spectrum compares quite well with the experiment. 
Good agreement with experiment is also seen in an insulator with a large band gap. 
Figure \ref{fig:si_lif}(b) shows the spectra for LiF. 
In sharp contrast to silicon, a bound exciton peak is formed in the gap by the 
inclusion of electron-hole interaction. Consequently, the threshold of the spectrum 
is red shifted substantially.\cite{rohlfing-louie98}  

A few different approaches have also been developed for the optical absorption spectra. 
Time Dependent Density Functional Theory (TDDFT) \cite{gross} is studied intensively 
in the last decade. 
It was pointed out by Runge and Gross that 
by taking the {\it time-dependent} one electron density as a basic variable, 
DFT can be rigorously extended to treat dynamic response of many electron systems. 
The many-body effects are included in the exchange-correlation kernel, which is the density derivative of 
the exchange correlation potential. 
Most applications to real materials adopt the adiabatic local density approximation (ALDA) 
that neglects the nonlocal effects in both time and space. 
The TDDFT using the ALDA works well for finite systems. 
As the system size becomes larger, discrepancy from the experiment becomes significant. 
Improvement of the exchange-correlation kernel is needed for application to solids. 

Nakamura {\it et al.} followed the three-stage scheme for the optical absorption spectrum of GaAs \cite{nakamura08}. 
Starting with the GGA band structure, they derived a low-energy tight-binding Hamiltonian using the maximally localized 
Wannier function procedure. The onsite Coulomb interaction is evaluated by the constrained DFT method. 
Solving the derived model using the Hartree-Fock approximation supplemented by the single-excitation 
configuration-interaction method considering electron-hole interactions, they obtained the spectrum 
in good agreement with experiments. 

%%%%%%%%%%%%%%%%%%%%%%%%%%%%%%%%%%%%%%%%
\subsection{3$d$ transition metal and its oxides}
\subsubsection{Transition metal}
Simple substances of the 3$d$ transition metals are reasonably described by the LDA. 
They are classified as moderately correlated materials. 
The value of $U$ is estimated to be 3-5 eV in the cRPA \cite{aryasetiawan06,miyake08a}, 
which is consistent with the above picture. 

One of the problems in the LDA band structure is too wide $d$ band width. 
A GW calculation of Ni by Aryasetiawan \cite{aryasetiawan92} showed that 
the GW self-energy correction raise the bottom of the $d$ band by about 1 eV, 
resulting in a band narrowing, in agreement with experiments. 
However, experimentally observed satellite at -6 eV below the Fermi level is not reproduced even in the GW. 
In addition, the exchange splitting does not change significantly compared to LDA, and larger  
than measured values. 
In fact, the satellite originates from short-range correlations, which is not properly described 
in the GW approximation. The problem was solved by a T-matrix calculation \cite{aryasetiawan98}. 
The problem of the exchange splitting, as well as the 6 eV satellite and band narrowing, was 
settled down later by the LDA+DMFT~\cite{lichtenstein01}, and GW+DMFT calculation \cite{biermann03}. 

Another well-known problem in the LDA is magnetism. 
In iron, for example, the nonmagnetic hcp structure becomes more stable than the ferromagnetic bcc in the LDA. 
The GGA correctly describes the ground state in this particular case \cite{singh91,leung91,asada92}, but 
in general careful analysis is needed for magnetic properties. 
Looking at finite-temperature properties, 
one needs a formalism that takes into account the existence of 
local magnetic moments above the Curie temperature $T_c$. 
Many-body effects which incorporate the local atomic character of the electrons are essential.
This can be achieved by the LDA+DMFT method. 
Applications to Fe and Ni reproduced semi-quantitatively 
the ferromagnetic susceptibility above $T_c$ and temperature dependence of the ordered moment
below $T_c$  \cite{lichtenstein01}. 

%%%%%%%%%%%%%%%%%%%%%%%%%%%%%%%%%%%%%%%%%%
\subsubsection{SrVO$_3$} \label{SrVO3}
Transition metal perovskite compounds exhibit various intriguing electronic and magnetic properties. 
Among them, SrVO$_3$ can be regarded as a prototype. 
The material is cubic. There is no GdFeO$_3$-type lattice distortion, 
and only one formula unit is contained in the unit cell. 
%If we take the V position at the origin, oxygen atoms are located at 
%($\pm\frac{1}{2}$,0,0), (0,$\pm\frac{1}{2}$,0), (0,0,$\pm\frac{1}{2}$), forming an octahedron 
%with the V atom sitting at the center. The Sr atom is located at ($\frac{1}{2}$,$\frac{1}{2}$,$\frac{1}{2}$).

Experimentally, the compound is a paramagnetic metal \cite{onoda91}. 
Fujimori {\it et al.} found by photoemission spectroscopy (PES) 
that the occupied $d$ band has a double-peak structure, one within about 1 eV of the Fermi level 
with a sharp Fermi cutoff, whereas the other centered at $\sim$1.5 eV below the Fermi level\cite{fujimori92}. 
This suggests that SrVO$_3$ is a correlated metal, where the peak around the Fermi level 
corresponds to the quasiparticle peak, and the other peak is the lower Hubbard band. 
%\textcolor{red}{場所を移動
The inverse photoemission spectrum shows a peak at 2.5-3 eV, which can be 
interpreted as an upper Hubbard band.\cite{morikawa95}
%}

Fujimori {\it et al.} also studied other $d^1$ electron systems including 
ReO$_3$, VO$_2$, SrVO$_3$, LaTiO$_3$, YTiO$_3$, and 
found that as the bandwidth decreases, deviation from the band structure calculation becomes 
substantial, and the weight near the Fermi level is transferred to the higher binding energy.

Comparison between SrVO$_3$ and CaVO$_3$ has also attracted much attention.
Both compounds are metallic perovskites having a single $d$ electron. 
A noticeable difference is that the CaVO$_3$ has a distorted structure. 
This leads to reduction in hopping between the $t_{2g}$ orbitals of neighboring V atoms 
mediated by the O-$p$ orbital, which would enhance the correlation effects. 
Early PES experiment for Ca$_{1-x}$Sr$_{x}$VO$_3$  reported
that the spectral weight is indeed transferred from a coherent to incoherent peak as 
$x$ decreases \cite{inoue95}. 
However, later high-energy photoemission experiment casted doubt on this conclusion, 
finding that the bulk spectra is insensitive to $x$ \cite{sekiyama04}. 
A more recent low-energy PES measurement supported this result \cite{eguchi06}. 
%\textcolor{red}{
In the measurement, it was also found that the spectral intensity is suppressed near the Fermi level. 
%}
This is consistent with the cluster extension of DMFT for the Hubbard model showing the suppression of the coherent peak
with a pseudogap formation as we discussed and illustrated in Fig.~\ref{fig:ZhangClusterDMFT}).\cite{zhang07}    

\begin{figure}[ptb]
\begin{center}
\includegraphics[width=0.4\textwidth]{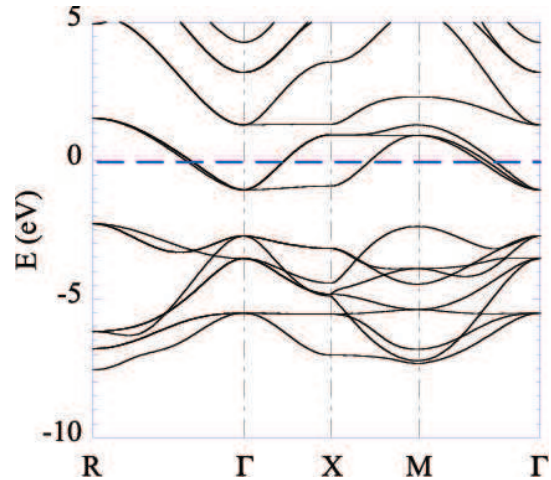}
\end{center}
\caption{LDA band structure of SrVO$_3$. 
%\textcolor{red}{
The three states crossing the Fermi level have strong V-$t_{2g}$ component. 
There are nine states at [-8eV:-2eV], which is mainly of oxygen 2$p$ character.} 
%}
\label{fig:svo}
\end{figure}
In the LDA band structure (Fig.\ref{fig:svo}), three states having strong V-$t_{2g}$ character 
cross the Fermi level. They form an isolated band, below which is an oxygen band at [-8eV:-2eV], 
while the V-$e_{g}$ band is just above the $t_{2g}$ band. 
Two valence electrons are transferred from V to oxygen, thus a single electron occupies 
the $t_{2g}$ band. 
The LDA correctly reproduces paramagnetic metal nature.
However,  the $t_{2g}$ band is much broader than the experimental 
quasiparticle peak: The latter is about 60 \% of the former in width. 
In addition, the satellite structure at -1.5 eV does not exist in LDA. 
LDA+U cannot reproduce these features neither. 
A standard 1-shot GW reduces the band width by about 30 \%, 
in reasonable agreement with the experiment. %\cite{miyake_svo}.

Coexistence of coherent and incoherent peaks can be reproduced only 
by going beyond static mean-field treatment of electron correlation effects. 
DMFT is a possible solution for this. 
Liebsch performed a DMFT calculation of SrVO$_3$ and CaVO$_3$ 
using the tight-binding Hamiltonian fitted to the LDA $t_{2g}$ bands \cite{liebsch03}. 
For a reasonable choice of $U \sim 4 eV$, a narrowing of quasiparticle peak and 
evolution of Hubbard bands are observed, in agreement with experiments. 
It was also found that the orthorhombic distortion causes a weak transfer 
of spectral weight from the coherent to the incoherent peak. 
Later on, Pavarini {\it et al.} studied SrVO$_3$ and other three $d^1$ perovskites, 
CaVO$_3$, LaTiO$_3$ and YTiO$_3$, ranging from correlated metal to magnetic insulator.
They first extracted $t_{2g}$ bands using NMTO-Wannier procedure.\cite{pavarini04} 
Then the derived low energy Hamiltonian, with several values of $U$ between 3-6 eV, 
was solved by DMFT. They found that the main features of the photoemission spectra 
for all four materials, as well as the correct values of the Mott-Hubbard gap for 
the insulators were reproduced by taking $U$ to be 5 eV.
Both SrVO$_3$ and CaVO$_3$ are correlated metals, while 
the quasiparticle peak disappears and the system becomes insulating in LaTiO$_3$ and YTiO$_3$, 
as can be seen in Fig.\ref{fig:svoPavarini}.
It was also revealed that the lattice distortion leads to reduction of not only 
band width but also of effective orbital degeneracy, which plays an important 
role in the metal-insulator transition. 
\begin{figure}[ptb]
\begin{center}
\includegraphics[width=0.4\textwidth]{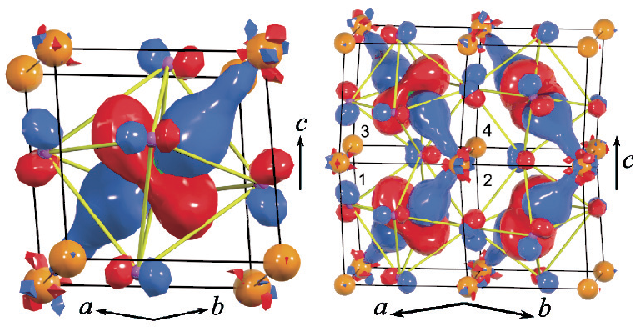}
\includegraphics[width=0.4\textwidth]{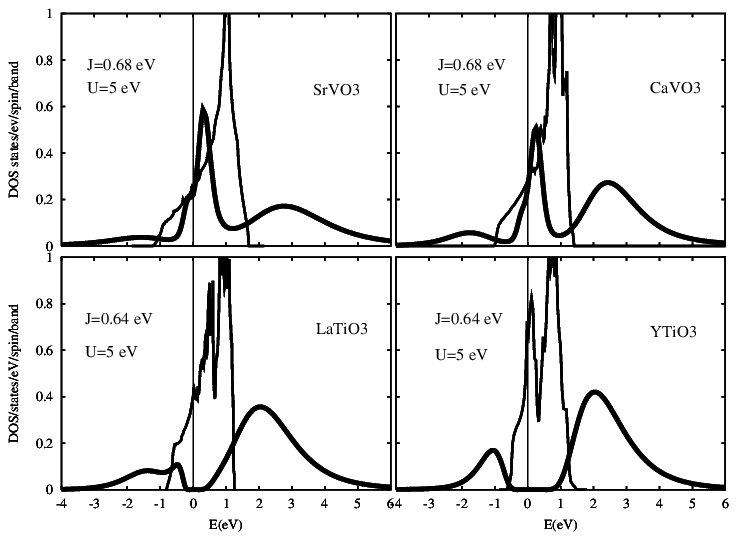}
%\vspace{7cm}
\end{center}
\caption{(Color online) (a) Occupied Wannier orbital of LaTiO$_3$ in primitive cells (right) and a subcell (left) obtained by LDA+DMFT calculations.  La atoms are on the corners of the cubes (orange). (b) DMFT density of states (DOS) at T=770K (thick lines) at onsite interaction of Ti 3d orbitals, $U$ and the exchange interaction $J$ compared with LDA DOS (thin lines)\cite{pavarini04} }
\label{fig:svoPavarini}
\end{figure}

For a full quantitative treatment, ab-initio determination of $U$ is important. 
The value is estimated to be 3.0-3.5 eV in cRPA \cite{aryasetiawan06,miyake08a}, 
which is smaller than that used in previous LDA+DMFT calculations. 
This discrepancy implies that spatial correlations significantly enhance the electron correlation effects and
tendency for the Mott insulator.
Careful {\it ab initio} analysis on {\it e.g.,} long-range interaction and non-local 
self-energy effects, is an open issue.

%%%%%%%%%%%%%%%%%%%%%%%%%%%%%%%%%%%%%%%%%%%%%%%%%
\subsubsection{VO$_2$}
Vanadium dioxide is a material under debate for many years. As the
temperature decreases, the material shows metal-insulator transition at 340
K from the high temperature metallic phase in the rutile structure to the
low temperature insulating phase with the monoclinic (M1) structure \cite%
{Morin}. There has been long discussion about the transition, with
particular interest on the role of electron correlations in forming a gap.

\begin{figure}[ptb]
\begin{center}
\includegraphics[width=0.25\textwidth]{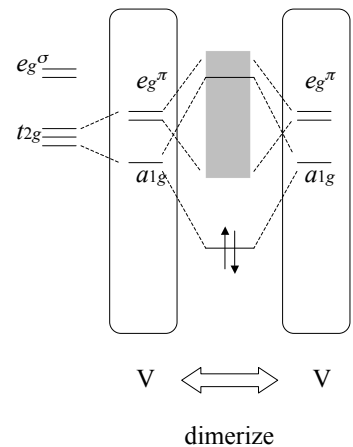}
\end{center}
\caption{ Orbital energy diagram VO$_{2}$ in monoclinic phase.
%\textcolor{red}{
The degeneracy of the V-${t_{2g}}$ states are lifted by the non-cubic crystal symmetry. 
The Peierls distortion leads to a coupling between  two $a_{1g}$ states each on  
the neighboring  V sites, forming the bonding and the anti-bonding states. 
The bonding state accommodates two electrons, and the system becomes 
gapful.}
%}
\label{vo2level}
\end{figure}
Seen from the band picture, the electronic states are understood as Fig.\ref{vo2level}
\cite{Goodenough}. The low-energy states near the Fermi level are of strong
vanadium 3$d$ character. The crystal field makes the $d$ states split into $%
t_{2g}$ and $e_{g}$. Since the structure is not cubic, the $t_{2g}$ states
are lifted further into $e_{g}^{\pi}$ and $a_{1g}$ state. The isolated vanadium atom has
three $d$ electrons. Two of them are transferred to oxygen 2$p$ orbitals in VO$%
_{2}$, thereby VO$_{2}$ is a $d^{1}$ system. The remaining $d$ electron
partially occupies the $a_{1g}$ band so that the system is metallic in the
rutile phase. In the M1 phase, two vanadium atoms form a dimer. This Peierls
distortion causes strong hybridization between the $a_{1g}$ orbitals of the
two vanadium atoms. Then the bonding state is fully filled, which opens a
gap between the bonding $a_{1g}$ and unoccupied $e_{g}^{\pi}$ band. The
overall feature was confirmed by first-principles calculations \cite%
{Wentzcovitch,Eyert} in the LDA.
However, it is also found that the bonding $a_{1g}$ band overlaps with 
the $e_{g}^{\pi}$ band, yielding metallic behavior in contrast with the experiment. 
This may be ascribed to the band gap problem of LDA. If we include many-body effects,
the $e_{g}^{\pi}$ may shift up and the gap would open. 
On the other hand, there is another phase in which 
one half of vanadium atoms dimerize, while 
the other half form chains with equal space. 
This M2 phase is also insulating, which 
suggests that VO$_2$ may be a Mott insulator.
Some authors claimed that the electron correlation plays a major role \cite%
{Zylbersztejn,Rice}. 
The controversy is not yet settled down and correlation
effects beyond LDA are discussed with various techniques such as LDA+DMFT 
\cite{Liebsch,Laad,Biermann05}, simplified GW scheme \cite%
{Continenza} or full \textit{ab-initio} GW \cite{gatti07,sakuma08}.

Here we show how the GW works. Figure \ref{vo2band}(a)
shows electronic structure of M1 phase obtained by LDA. As described above,
the bonding $a_{1g}$ orbitals are near the Fermi level, which are located in
[-0.5 eV: 0 eV]. These states are almost fully filled, but there is an
overlap with the $e_{g}^{\pi}$ band that is located just above the $a_{1g}$.
As a result, there is a small hole (electron) pocket in $a_{1g}$ ($%
e_{g}^{\pi}$) band, and the system becomes metallic in LDA.
\begin{figure}[ptb]
\begin{center}
\includegraphics[width=0.46\textwidth]{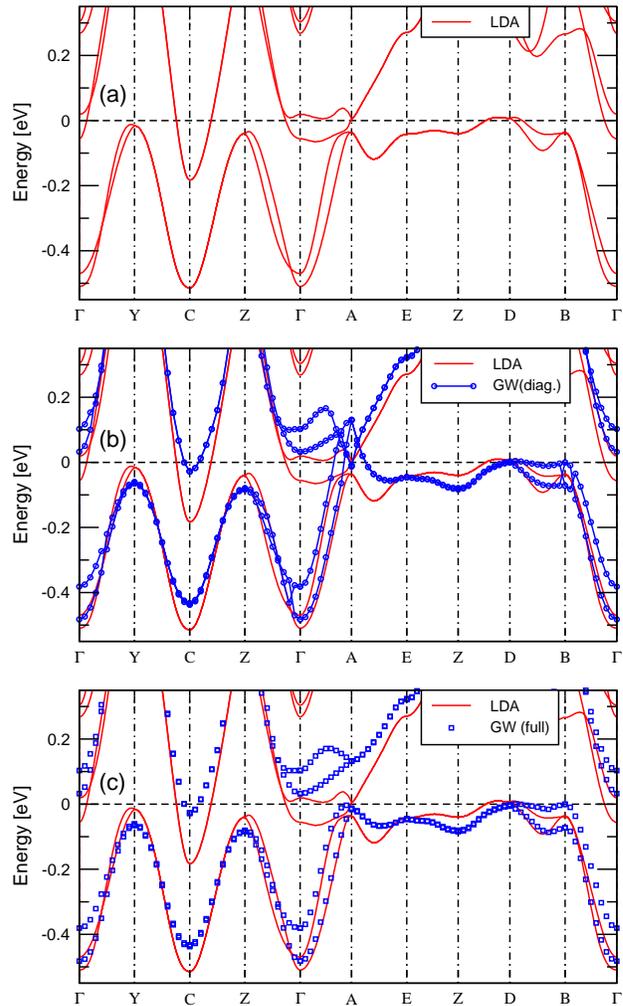}
\end{center}
\caption{(Color online) 
%\textcolor{red}{
Electronic structure of VO$_{2}$ in the monoclinic phase. 
(a) The LDA band shows metallic character. 
(b) The GW self-energy correction with diagonal elements (in the Kohn-Sham basis)  
pulls up the conduction band, while the bands are still entangled at around the A point. 
(c) The bands are disentangled by including the off-diagonal elements, 
consequently the bonding $a_{1g}$ band is isolated \protect\cite{sakuma08}. }
%}
\label{vo2band}
\end{figure}
\begin{figure}[ptb]
\begin{center}
\includegraphics[width=0.35\textwidth]{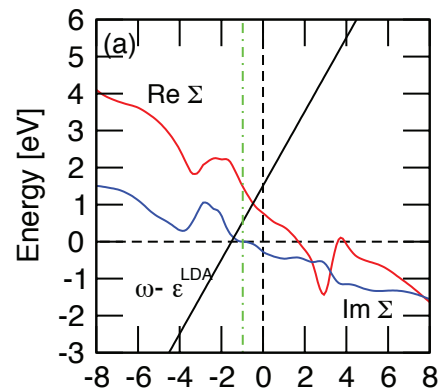}
\end{center}
\caption{(Color online) GW self-energy of VO$_{2}$ in the monoclinic phase at $\Gamma$ point in the $a_{1g}$ band 
\protect\cite{sakuma08}. The self-energy is not a smooth function of frequency, which yields a satellite structure 
in the spectral function. The straight line with a positive slope is $\omega - \epsilon^{LDA}_{\vk}$. 
The intersection between the line and the real part of the self-energy gives the quasiparticle energy.}
\label{vo2sig}
\end{figure}
\begin{figure}[ptb]
\begin{center}
\includegraphics[width=0.35\textwidth]{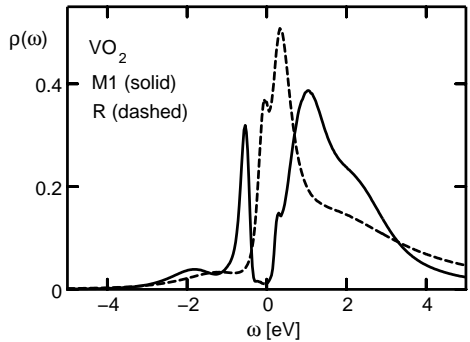}
\end{center}
\caption{ 
%\textcolor{red}{
Spectral function of $t_{2g}$ orbitals for rutile and M1 phases obtained by 
the LDA+DMFT calculation \cite{Biermann05} 
The lower Hubbard band is located at -1.2 eV in the R phase, 
while the prominent peak at -0.8 eV in the M1 phase is the quasiparticle peak.
 }
%}
\label{BiermannVO2}
\end{figure}

Now we add the self-energy correction to the LDA Kohn-Sham energies. Figure %
\ref{vo2sig} shows the self-energy as a function of energy for selected
states. The self-energy does not decrease monotonically but has dips and
peaks. This behavior is quite different from weakly correlated
semiconductors, such as silicon. Because of the peculiar energy dependence
we need to treat full energy dependence of the self-energy. In fact if we
compute the GW quasiparticle band by linearizing the energy dependence of
the self-energy, as most \textit{ab-initio} GW calculations assume, the $%
a_{1g}$ band gets too narrow. 
Also the non-linearity yields a weak satellite structure in the one electron spectral function 
%\textcolor{red}{
above the Fermi level, but not below, in contrast with the LDA+DMFT result \cite{Biermann05}. 
%}

Quasiparticle band structure is plotted in Fig.\ref{vo2band}(c) by circles.
We can see that band overlap between the $a_{1g}$ and $e_{g}^{\pi}$ is
removed by the self-energy correction. It should be noted that the Kohn-Sham
wavefunctions have too much hybridization between $a_{1g}$ and $e_{g}^{\pi}$
near the A point, therefore off-diagonal elements are essential to
disentangle the bands. If we neglect the off-diagonal self-energy elements,
the $a_{1g}$ and $e_{g}^{\pi}$ overlap each other even in the GW level, as
shown in Fig.\ref{vo2band}(b).
The quasiparticle $a_{1g}$ band is now isolated and direct gap opens.
%\textcolor{red}{
For the opening of the fundamental gap, 
the effect of self-consistency plays a crucial role. 
%以下数行を削除}
%However, the fundamental gap is almost zero and it is hard to tell whether
%the system is metallic or insulating. This implies that the one-shot scheme,
%which is used in almost all \textit{ab-initio} GW calculations, is not
%enough and self-consistency has big impact. The starting Kohn-Sham solution
%is metallic in the present case. Thus, screening effect is overestimated and
%so the self-energy correction may be underestimated. If we perform the GW
%calculation starting with a gapful state, the self-energy correction gets
%larger and the quasiparticle gap would increase \cite{sakuma08}.

Biermann {\it et al.} studied the compound by LDA combined with a cluster extension of DMFT \cite{Biermann05}. 
Starting from the LDA band structure, they extracted three $t_{2g}$ states per V atom, and 
constructed a multi-band Hubbard model. 
The model was solved using cluster DMFT including all off-diagonal terms in orbital space. 
More precisely, instead of calculating the self-energy from 
a local impurity model embedding one single atom in a self-consistent bath, 
a pair of V atoms in a bath is considered. 
This is important because the formation of singlet pairs resulting
from the strong dimerization can be captured only in a cluster extension.

They carried out calculations for both rutile phase and M1 phase. 
The calculated spectral function is shown in Fig.\ref{BiermannVO2}.
For the rutile phase, the results of single-site and
cluster-DMFT calculations are very similar. 
A clear quasiparticle peak is found near the Fermi level, 
and many-body effects reduce the bandwidth. 
Hence, the rutile phase can be characterized as a metal with intermediate correlation.
Hubbard satellites are observed at high energies at -1.5 eV below 
and 2.5-3 eV above the Fermi level.
In the M1 phase, nonlocal self-energy opens up a gap of about 0.6 eV
(for $U$=4 eV and $J$=0.68 eV), 
in reasonable agreement with experiments. 
There is a sharp coherent peak at -0.8 eV. 
Below this peak is a weak lower Hubbard satellite at -1.8 eV, 
whereas the broad peak centered at 2.2 eV is the upper Hubbard band.
Charge distribution is modified significantly, and 
the single electron occupies almost entirely the $a_{1g}$ orbital. 
The low-energy nature of the insulator is quite different 
from that of a standard Mott insulator in which local moments are formed. 
In fact, at low frequency, the onsite component of the self-energy 
for the $a_{1g}$ orbital behaves linearly as a function of frequency
in contrast to the $1/\omega$ behavior for the local moment Mott insulator. 
Based on these results, they concluded that  at low energy, the compound is 
a Peierls insulator assisted by strong Coulomb correlation.

%\textcolor{red}{
The gap in the M2 phase appears to be a correlation-origin Mott gap, but it is not a settled issue. 
%}
%Self-energy was also calculated by a self-consistent GW calculation starting from a proper 
%initial state.  Again metallic states for the rutile and insulating states with a gap ($\sim 0.6 $ eV)
%for the monoclinic M1 structure were correctly reproduced. Fig.\ref{vo2sig} shows the frequency dependence of 
%the self-energy, where the real part has $\omega$ linear dependence.\cite{sakuma08}

%%%%%%%%%%%%%%%%%%%%%%%%%%%%%%%%%%%%%%%%%%%%%%
\subsubsection{Sr$_2$VO$_4$}
Sr$_2$VO$_4$ has a layered perovskite structure and is isomorphic to the mother compounds of a cuprate superconductor La$_2$CuO$_4$.\cite{Cyrot,ISTEC}  This compound has one 3$d$ electron per V site ($d^1$ system) with strong two-dimensional anisotropy and has a dual relation to the one 3$d$ hole per Cu sites ($d^9$ system) of the cuprates.  The duality is, however, not perfect because, in Sr$_2$VO$_4$, the orbital degeneracy of $d^1$ electron remains between $d_{yz}$ and $d_{zx}$ orbitals.  The crystal field splitting of $d_{xy}$ orbital is also rather small ($\sim 0.08$ eV in the LDA calculation), which evokes us importance of orbital physics.  

Since the $3d$ $t_{2g}$ bands are located near the Fermi level and are rather isolated from other bands as we see in Fig.\ref{Fig.SVOLDADispersion}, a low-energy effective model of the form (\ref{ham}) for the $t_{2g}$ Wannier orbitals has been derived.\cite{Imai1,Imai2} After the downfolding, the onsite interactions among the Wannier orbitals of intraorbital $xy, yz(zx)$ and interorbital $xy$-$yz(xy$-$zx)$ and $yz$-$zx$ combinations are $U=2.77, 2.58, 1.35$ and 1.28 eV, respectively. The onsite exchange interactions between $xy$-$yz(xy$-$zx)$ and $yz$-$zx$ orbitals are 0.65 and 0.64 eV, respectively.  The nearest neighbor transfers between $xy$-$xy, yz$-$yz$ and $zx$-$zx$ orbitals in $x$ direction are -0.22, -0.05 and -0.19 eV, respectively. 
In order to monitor the Coulomb interaction effects, the scale-factor dependence has been studied by multiplying all the matrix elements $U_{nn'mm'}$ in eq.(\ref{hamU}) with a factor $\lambda$.  Namely, the realistic value corresponds to $\lambda=1$. 
The effective model (\ref{ham}) with the above parameters has been solved by PIRG. Technical details are found in Refs. \citen{Imai1,Imai2}.

It has turned out that this compound shows very severe competitions as we see in Fig.\ref{energy.eps}.  First, it lies on the verge of the Mott transition.  Second, the ferromagnetic state is rather close in energy to the true ground state with the antiferromagnetic order. Third, spins and orbitals order in a complicated pattern in the ground state as we see in Fig.\ref{orb_order2.eps} and candidates of the spin-orbital order are in severe competitions each other in the order of 100K in energy. They have revealed rich orbital-spin physics arising from the competitions. 

Experimentally, transport and optical properties of this compound indicate either a very small Mott insulating gap or semiconducting property with rapidly increasing resistivity with decreasing temperature~\cite{Cyrot,ISTEC,Matsuno} in agreement with the above calculated results.   
The gap amplitude is nearly zero and 
it can easily be metallized by La doping~\cite{Matsuno}.
Recent experiments by dc susceptibility and X-ray diffraction\cite{Zhou} have suggested a transition around 100K into a phase with antiferromagnetic and orbital coupled order  below this temperature, again in essential agreement with the above theoretical prediction.  Recently, it has been proposed~\cite{Khaliullin} that the orbital-spin coupled order essentially described by the octupole order frequently discussed for $f$-electron systems~\cite{Kuramoto} might be stabilized when the spin-orbit coupling ignored in the available first-principles study are considered.

In sharp contrast to this nearly insulating transport properties, the LDA calculation predicts a good metallic behavior (Fig.\ref{Fig.SVOLDADispersion}).  
On the other hand, the Hartree Fock approximation (HFA) predicts a clear ferromagnetic insulating phase at the realistic parameter values (see Fig.~\ref{energy.eps}). LDA+U approach predicts results similar to HFA. 
The failure of single-Slater-determinant approximations as HFA and LDA+U is naturally understood because they relatively well describe a simple ferromagnetic state, while not the antiferromagnetic state with a nontrivial periodicity. Such a phase with large quantum fluctuations can be
described only by a more accurate solver such as PIRG beyond a single Slater determinant. 

All of the above agreement between the experiments and the present theory indicate that the approach of PIRG combined with the downfolding by using the LDA-GW scheme works well as a method for strongly correlated materials. 
From the viewpoint of the computational methods, Sr$_2$VO$_4$ appears to offer a very severe and good benchmark for testing the accuracy in taking account of the correlation effects because of the severe competing orders. 
\begin{figure}
\begin{center}
%$$ \psboxscaled{600}{SVOLDA3Dispnew.eps} $$
\includegraphics[width=8.5cm]{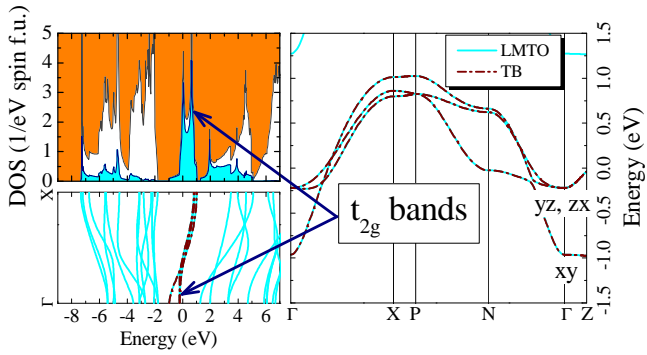}
\end{center}
\caption{(Color online) (Left panel) Electronic structure of Sr$_2$VO$_4$ in LDA.\cite{Imai1,Imai2}  (Right panel) Enlarged behavior of $t_{2g}$ bands computed from LMTO basis functions (solid light blue curves) and downfolded tight-binding bands (dot-dashed brown curves).  The corresponding bands in the left panel are shown by arrows. The symbols denote the character of $t_{2g}$ bands in the $\Gamma$-point.  The Fermi level is at zero energy.}
\label{Fig.SVOLDADispersion}
\end{figure}

%\begin{figure}
%\includegraphics[width=8.5cm]{SVOsigmanew.eps}
%\caption{(Color online) Diagonal matrix elements of the screened Coulomb interaction $W_r(\omega)$ (left panel) and the self-energy, $\Sigma(\omega)$ (right panel). The inset shows the high-frequency part of $W_r(\omega)$ (the lines corresponding to the $xy$ and $yz$ orbitals are indistiguishably close). The real and imaginary parts are shown by solid and dashed curves, respectively.}
%\label{Fig.SVOsigma}
%\end{figure}

\begin{figure}
\begin{center}
\includegraphics[width=7cm]{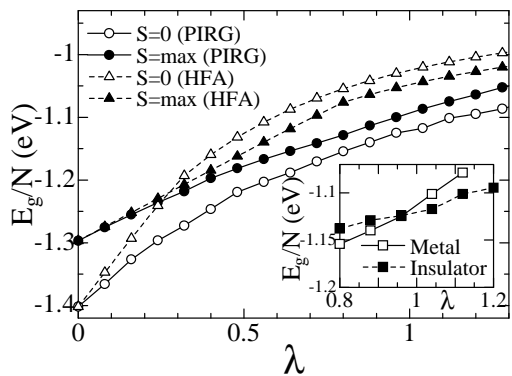}
\end{center}
%$$ \psboxscaled{350}{lambda_Eg3.eps} $$
\caption{Lowest energies per unit cell of total S=0 (open symbols) and ferromagnetic states (filled symbols) by quantum-number projected Hartree-Fock (triangles) and PIRG (circles) calculations for the downfolded model of Sr$_2$VO$_4$.\cite{Imai1,Imai2}  (Inset):Lowest energies per unit cell of metallic (open squares) and insulating states (filled squares) by quantum-number projected PIRG for the downfolded model of Sr$_2$VO$_4$.}
\label{energy.eps}
\end{figure}

\begin{figure}
\begin{center}
\includegraphics[width=6cm]{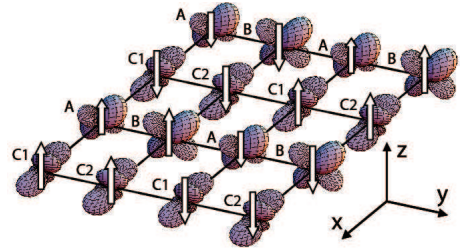}
\end{center}
%$$ \psboxscaled{240}{spin-orbital_new.eps} $$
\caption{(Color online) Ordered spin-orbital patterns in plane for Sr$_2$VO$_4$ clarified by PIRG.\cite{Imai1,Imai2}  Ordered spin moment is proportional to the length of arrows. At each site, occupied orbitals can be specified by a 3-dimensional unit vector in the basis of $t_{2g}$ Wannier orbitals. Its $xy$, $yz$, and $zx$ components are given by (0.70,0.60,0.39), (0.51,0.80,0.31),(0.40,0.04,0.92), and (0.33,0.06,0.94), for the sites A, B, C1 and C2, respectively.}
\label{orb_order2.eps}
\end{figure}

%%%%%%%%%%%%%%%%%%%%%%%%%%%%%%%%%%%%%%%%%%%%%%%%
\subsubsection{YVO$_3$}
YVO$_3$ belongs to the family of transition-metal oxides
with two valence electrons in the 3$d$ orbitals ($t_{2g}$ manifold).\cite{ImadaRMP}
The lattice structure is an orthorhombically distorted
perovskite with the space group $Pbnm$ 
(four vanadium sites in a unit cell)
at room temperatures.
The GdFeO$_3$-type distortion, rotation and tilting
of the VO$_6$ octahedra are present,
where the reduced V-O-V angle makes the narrow $t_{2g}$ bands.
With lowering the temperature,
it undergoes two successive phase transitions
in both spin and orbital sectors.
First, the $G$-type orbital ordering (OO)
appears at 200K with a structural change to the $P2_{1}/a$ symmetry,
where a site with the $d_{xy}$ and $d_{yz}$ orbitals occupied
and one with the the $d_{xy}$ and $d_{zx}$ are alternately
arranged in three dimensions.
The magnetic structure also shows the $C$-type 
spin ordering (SO) below 116K,
where spins are aligned antiferromagnetically in the $a$-$b$ plane
and ferromagnetically along the $c$-axis.
With further lowering the temperature,
the SO and OO simultaneously change at 77K, and
the ground-state is the $C$-type OO with the $G$-type
SO.~\cite{Kawano,Miyasaka2003}
The crystal structure recovers the $Pbnm$ symmetry. 
%as illustrated in Fig.~\ref{fig:crystal}.
In the charge sector, 
YVO$_3$ is a typical Mott insulator with a large charge gap ($\sim$ 1eV).
This is partly attributed to a large GdFeO$_{3}$-type distortion, 
which reduces the bandwidth effectively.
In addition, coupling to Jahn-Teller distortions is
important in determining the orbital states.

Electronic structure of YVO$_3$ has been studied by the three-stage RMS scheme.\cite{Otsuka}
The DFT-LDA calculations by using the local muffin-tin orbital basis
has been applied to derive the global band structure. The band structure shown in Fig.\ref{fig:YVObands} shows an isolated group of bands near the Fermi level mainly consisting of V $3d$ $t_{2g}$ atomic orbitals.  The electron degrees of freedom
far from the Fermi level are eliminated by a downfolding procedure
leaving only the V $3d$ $t_{2g}$ Wannier bands as the low-energy degrees of freedom, 
for which a low-energy effective model is constructed.
This low-energy effective Hamiltonian is solved exactly by
the PIRG method.\cite{Otsuka}
It is shown that the ground state has
the $G$-type spin and the $C$-type orbital ordering as we see in Fig.~\ref{fig:HF-02} in agreement with
experimental indications.
%The indirect charge gap is estimated to be around 0.7 eV, which prominently 
% improves the
%previous estimates by other conventional methods.
\begin{figure}[htbp]
 \begin{center}
  \includegraphics[width=8.5cm,clip]{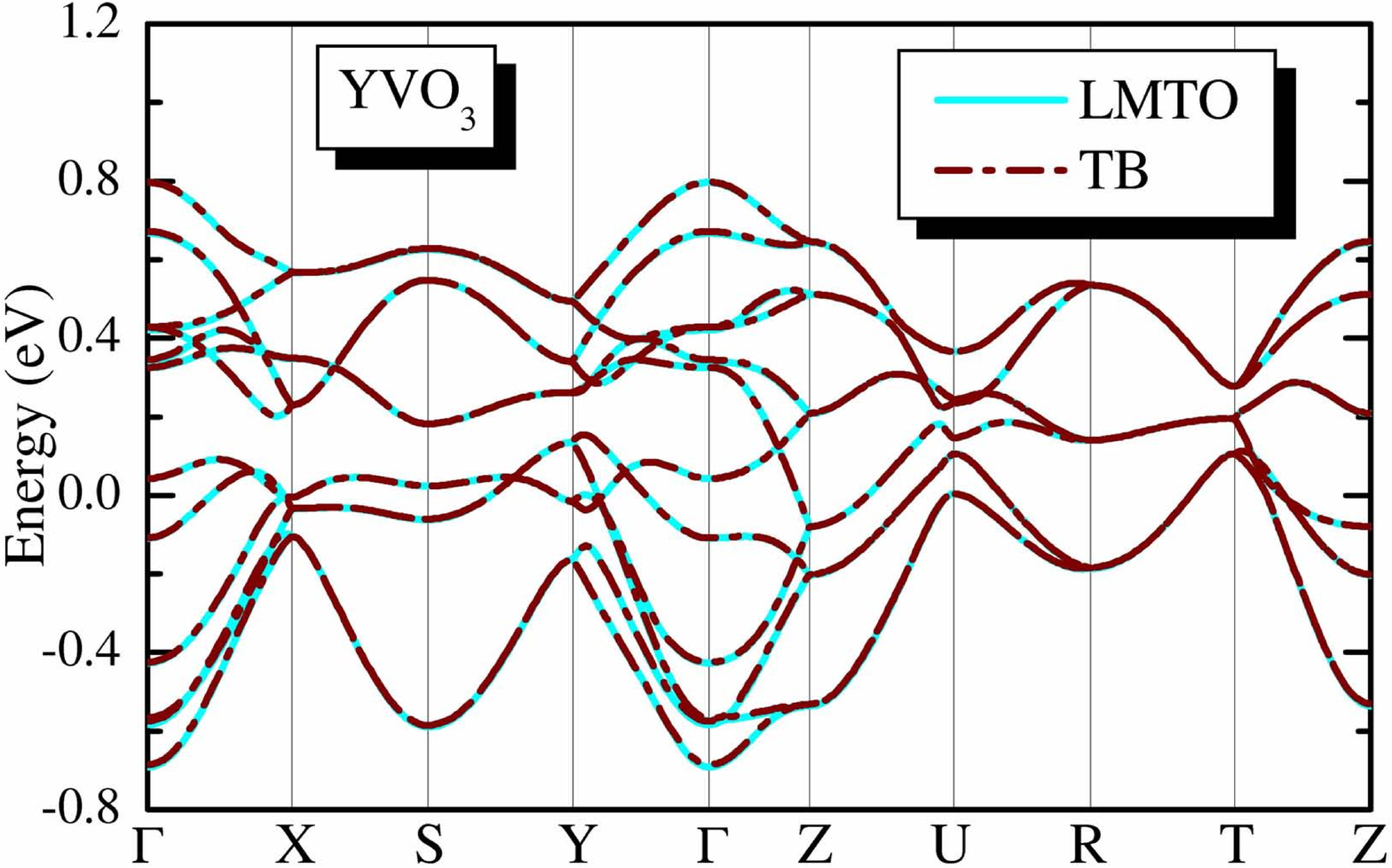}
  \caption{(Color online)  Comparison of the band structure of $3d$ $t_{2g}$ orbitals
  computed from LMTO calculations (solid light blue lines)
  with the 
  downfolded tight-binding model (dashed-dotted brown lines)
for YVO$_3$.\cite{Otsuka}
  \label{fig:YVObands} 
  }
 \end{center}
\end{figure}
\begin{figure}[htbp]
 \begin{center}
 \includegraphics[width=5.5cm,clip]{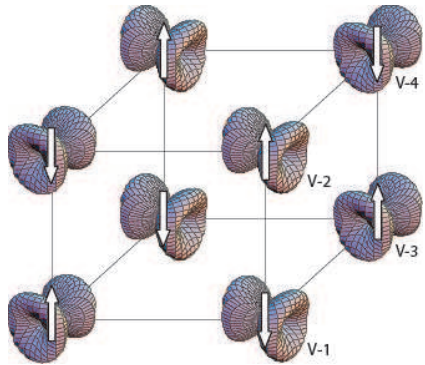}
  \caption{(Color online)
  Ordered spin- and orbital- patterns 
  in the ground state of the PIRG solution for YVO$_3$.
  The arrows represent the magnetic local moment at each vanadium
  atom. 
  The spins order antiferromagnetically in the $G$ type, while the orbitals
  order in the $C$ type in agreement with the experiments.\cite{Otsuka} 
  The orbital states are shown in the form of 
  the spatial electron distribution.
 \label{fig:HF-02} 
}
\end{center}
\end{figure}

%The obtained spin and orbital state is consistent with the experiment.
% for the HF and PIRG calculations.
The indirect-charge gap is estimated to be 0.70 $\pm$ 0.07 eV, 
which is smaller than the inferred experimental optical (direct) gap,
but is consistent each other because the experiment has measured the direct gap while the gap in the calculation is the indirect gap.  
It has prominently improved the estimation compared to 
the LDA or GGA method.
So far the indirect gap is not available experimentally. 

The LDA+PIRG results are thus all consistent with the available experimental results.
In fact, this is the first result that reproduces an experimental charge gap 
as well as the spin and orbital ordering of YVO$_3$ from the first-principles
calculations.

YVO$_3$ and LaVO$_3$ have also been studied by the combination of the downfolding scheme and DMFT.\cite{PavariniAndersen07} It has been shown that the Jahn Teller and GdFeO$_3$ type distortions are both crucial in reproducing the experimental orbital and magnetic orders at low temperatures. 

%%%%%%%%%%%%%%%%%%%%%%%%%%%%%%%%%%%%%%%%%%%%%%
\subsubsection{Iron-based superconductors}
Recent discovery of iron based superconductors has renewed interest 
on high temperature superconductivity.\cite{Hosono,Hosono-Nakai-Ishida}  
Several families of compounds are identified as superconducting materials, where 
Fe-3$d$ conduction bands  
are commonly located near the Fermi level according to LDA band-structure calculations~\cite{Mazin,Miyake10} and their electrons are likely to form Cooper pairs. 
So far,  
the mechanism of superconductivity is not well understood.  
In the family with ZrCuSiAs-type structure (called 1111 hereafter),  
SmFeAs(O,F) has the record of the highest superconducting critical temperature $T_{\rm c} \sim 56$ K (ref.~\citen{Ren})  
when fluorine is substituted with $\sim 20\%$ of oxygen as electron doping.
There exist other families. BaFe$_2$As$_2$ with ThCr$_2$Si$_2$-type structure (called 122)  
shows the highest $T_{\rm c}\sim 38$ K, when potassium is substituted for $\sim 40 \%$ of Ba as hole doping.\cite{Rotter}  LiFeAs and NaFeAs (called 111) with the PbFCl-type tetragonal structure indicate $T_c \sim 18$ K.\cite{LiFeAs_Wang,LiFeAs_Pitcher,LiFeAs_Tapp} 
FeSe$_x$Te$_{1-x}$ (called 11) also shows 
superconductivity at $T_{\rm c}\sim 10K$~\cite{11,Li}) at ambient pressure and at $T_{\rm c}\sim $37 K under pressure (7 GPa).\cite{Margadonna}  
 
A common aspect of iron-based superconductors  
is the existence of antiferromagnetic order close to the superconducting region except for the 111 family.  
However the ordered moment and pattern of the antiferromagnetism are strongly material dependent: LaFeAsO shows antiferromagnetic  
long-range order of the stripe type below $T_N\sim 130$ K with the Bragg point at $(\pi,0)$  
in the extended Brillouin zone with a strongly reduced ordered moment $\sim$0.36-0.63 $\mu_B$ as compared to the nominal saturation moment 4 $\mu_B$ for the high-spin 3$d^{\rm 6}$ state.\cite{Mook,Qureshi} 
Furthermore, LaFePO does not show an antiferromagnetic order and instead it undergoes a transition to the superconducting state at $\sim 4$ K.\cite{HosonoP} 
On the other hand, the 122-type (BaFe$_2$As$_2$) shows a relatively large ordered moment 
 $\sim 0.9$ $\mu_B$ (refs.~\citen{Huang}~and~\citen{Matan}) and the 11-type (FeTe) indicates an even larger ordered moment $\sim 2.0$-$2.25$ $\mu_B$ at a different Bragg point, $(\pi/2,\pi/2)$.\cite{Bao,Li}  
%Even for the 1111 family, NdFeAsO shows larger ordered moment ($\sim 0.9$ $\mu_B$),  
%although the moment is apparently reinforced by 
%Nd moment ($\sim 1.55$ $\mu_B$).\cite{Nd}  
 
Conventional LDA calculations of the 1111-type,~\cite{Lebegue,Singh,Hirschfeld,Terakura,Ma,Kuroki,Nekrasov1111} 122-type,~\cite{Singh_122,Nekrasov122} 
111-type,~\cite{Nekrasov111} and 11-type compounds~\cite{Subedi_SinghFeSe,MaFeSe} 
show a very similar band structure of the Fe $3d$ bands for all the compounds as we see in Fig.\ref{fig:Fe_Based_band},~\cite{Miyake10} 
where small electron pockets around M point and hole pockets around $\Gamma$ point  
constitute semimetallic Fermi surfaces and the total widths of ten-fold Fe-3$d$ bands are mostly around 4.5 eV.  
The local spin density approximation (LSDA) commonly predicted the  
antiferromagnetic order for mother materials.\cite{Hirschfeld,Terakura,Ma}  
The stripe-type antiferromagnetic order is  
correctly reproduced for the 1111-type.\cite{Terakura,Ma} However, the calculated ordered  
moment is unexpectedly too large (from 1.2 to 2.6 $\mu_{\rm B}$).\cite{Hirschfeld,Terakura,Ma,Mazin2}   
in contrast to much smaller ordered moment discussed above. 
The bicolinear order for FeTe is reproduced  
in the LSDA with the ordered moment $\sim 2.25$ $\mu_{\rm B}$) in agreement with the experimental  
results.\cite{MaFeSe} 
Diversity of the ordered moment ranging from zero to 2 $\mu_{\rm B}$ is surprising and not easily explained from the  
very similar band structure with semimetallic small pockets of the Fermi surface. 
Broad peak structures of magnetic Lindhard function calculated  
by using the LDA/GGA Fermi surface suggest severe competitions of   
different orderings.\cite{Mazin,Dong,Kuroki,KurokiPRB,Yildirim}   
\begin{figure}[htbp] 
\begin{center} 
\includegraphics[width=0.5\textwidth]{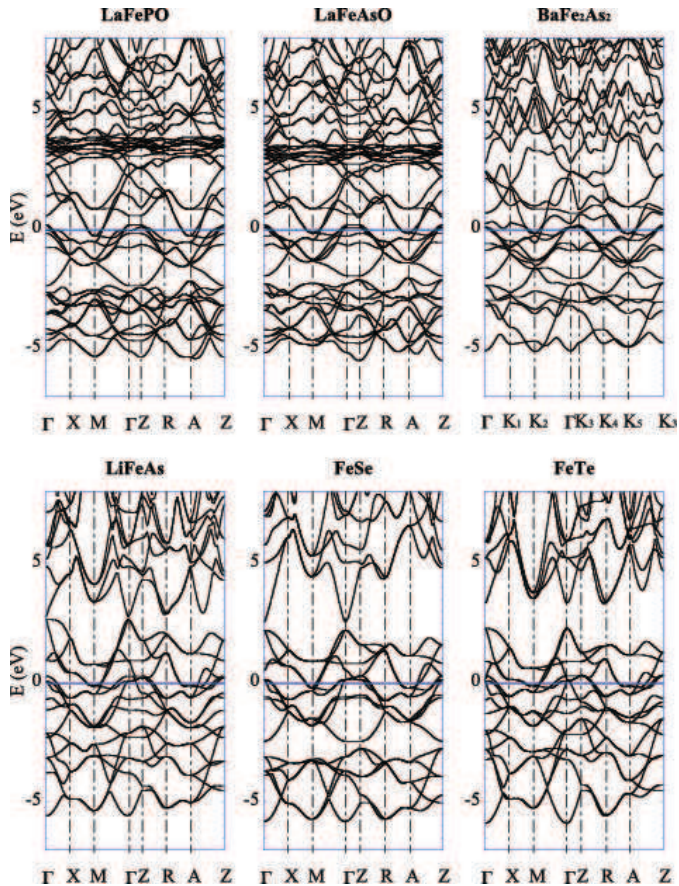}
\end{center} 
\caption{ 
Electronic band structures 
of six iron-based superconductors obtained by DFT-LDA.\cite{Miyake10} 
The $K_1-K_5$ points in BaFe$_2$As$_2$ are 
$
K_1 = \frac{2\pi}{a}(\frac{1}{2},0,0), 
K_2 = \frac{2\pi}{a}(\frac{1}{2},\frac{1}{2},0), 
K_3 = \frac{2\pi}{a}(0,0,\frac{a}{2c}),
K_4 = \frac{2\pi}{a}(\frac{1}{2},0,\frac{a}{2c}),
K_5 = \frac{2\pi}{a}(\frac{1}{2},\frac{1}{2},\frac{a}{2c}), 
$
respectively.
Energy is measured from the Fermi level. 
} 
\label{fig:Fe_Based_band} 
\end{figure} 

Roles of electron correlations are not fully understood so far and are under debate.\cite{Haule,Si,Nakamura,Anisimov,Craco,Liebsch10} 
Antiferromagnetic orders and fluctuations themselves revealed by the nuclear magnetic resonance and other probes imply some electron correlation effects.~\cite{Hosono-Nakai-Ishida,NMR}  
Small fraction of the Drude weight~\cite{Boris,Hu,Yang_Timusk,Basov} and bad metallic behaviors~\cite{Hosono,Chen}  
support substantial electron correlation effects. 
Recent fluctuation exchange calculation suggests a substantial self-energy effect, where  
the validity of weak coupling and nesting picture becomes questionable.\cite{AritaIkeda} 

Angle resolved photoemission spectroscopy~\cite{Lu_Shen_1111,Yi_Shen_122} 
 has shown some correspondence to the LDA result of Singh {\it et al}.\cite{Singh_122} 
Fe-2$p$ core-level spectra of X-ray photoemission suggest rather itinerant character.\cite{Malaeb,Shen_Xray} 
However, some role of moderate electron correlations has also been claimed.\cite{Takahashi,AnisimovPhysica} 
For FeSe, as we detail later, soft-Xray photoemission results~\cite{Shin,Yamasaki} appears to show a deviation from the LDA results and a crucial correlation effect~\cite{Aichhorn}.  

In the superconducting phase, even the pairing symmetry itself is highly controversial and no consensus has been reached. 
Although nodeless superconductivity is suggested~\cite{Kaminski,Ding,Hashimoto}, temperature dependence of nuclear-magnetic-relaxation  
time $T_1$ below $T_{\rm c}$ roughly scaled by $T^{-3}$ without the Hebel-Slichter peak
implies unconventional superconductivity driven by nontrivial electron-correlation  
effects.~\cite{NMR}  For example, orbital dependent gaps with sign-changing and fully-gapped $s\pm$ symmetry has been proposed.~\cite{Kuroki}
The gradual suppression of the superconducting transition temperature by Co doping into the Fe site was reported to be explained by
the $s$-wave singlet pairing without the sign change.\cite{Sato1,Sato2,Onari}   
Although overall experimental results suggest noticeable correlation  
effects, realistic roles on the pairing are not well established and controversial. 
  
To understand properties and mechanisms of magnetism and superconductivity in iron based superconductors, and to distinguish what are common and what are family dependent, 
effective low-energy models of these families have been derived\cite{Nakamura,Miyake10} from first principles
along the line of RMS based on the three-stage scheme.~\cite{aryasetiawan04,Imai1,Imai2}  
In the procedure, the LDA band structure was calculated as we see in Fig.\ref{fig:Fe_Based_band} and the maximally localized Wannier functions are constructed as we see examples in Fig.\ref{fig:mlwf_d}, from which the transfer and interaction parameters have been calculated by the {\it ab initio} downfolding and cRPA.  
So far, models for iron $3d$ orbitals ($d$ model) and models including additional pnictogen or chalcogen $p$ orbitals ($dp$ or $dpp$ models) have been derived.
For the $d$ model, the ratio of the averaged Hubbard diagonal onsite interaction $\bar{U}\sim 2.5$ eV to a typical nearest neighbor transfer $t\sim 0.3$ eV in the downfolded model for LaFeAsO
has been estimated to be $U/t\sim 8$-$10$ 
with the fivefold orbital degeneracy, 
indicating a moderately strong correlation. 
This moderately correlated nature has also been supported for the case of the 122-type, where $\bar{U}\sim 2.8$ eV.\cite{Anisimov_Vollhardt,Miyake10}
For the case of the 11 compounds, the effective interaction is even larger as $\bar{U}\sim 4.2$ eV for FeSe and 3.4 eV for FeTe. 
In Fig.\ref{fig:summary}, comparisons of the derived {\it ab initio} model parameters are shown.  
We note that the effective Coulomb interaction for the low-energy downfolded model estimated in these works is different from the interaction observed by experimental probes such as the X-ray photoemission.\cite{Malaeb,Shen_Xray}  
The measured interaction parameters are resulted from the further screening by the $3d$ electrons excluded in the model construction.
\begin{figure}[htb] 
\begin{center} 
\includegraphics[width=0.5\textwidth]{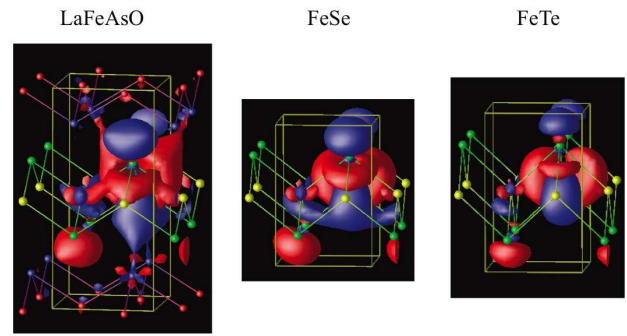}
\end{center} 
\caption{(Color online)
Isosurface of maximally localized Wannier function at $\pm$0.02 a.u.
for Fe ${x^2-y^2}$ orbital in $d$ model of 
LaFeAsO (left), FeSe (middle), and FeTe (right).\cite{Miyake10} 
This illustrates how the Wannier spread shrinks from LaFeAsO to FeTe.
 The dark shaded surfaces (color in blue) indicate the positive
isosurface at +0.02 and the light shaded surfaces (color in red)
indicate $-$0.02.
}
\label{fig:mlwf_d} 
\end{figure} 
\begin{figure*}[htb]
\begin{center} 
\includegraphics[width=0.6\textwidth]{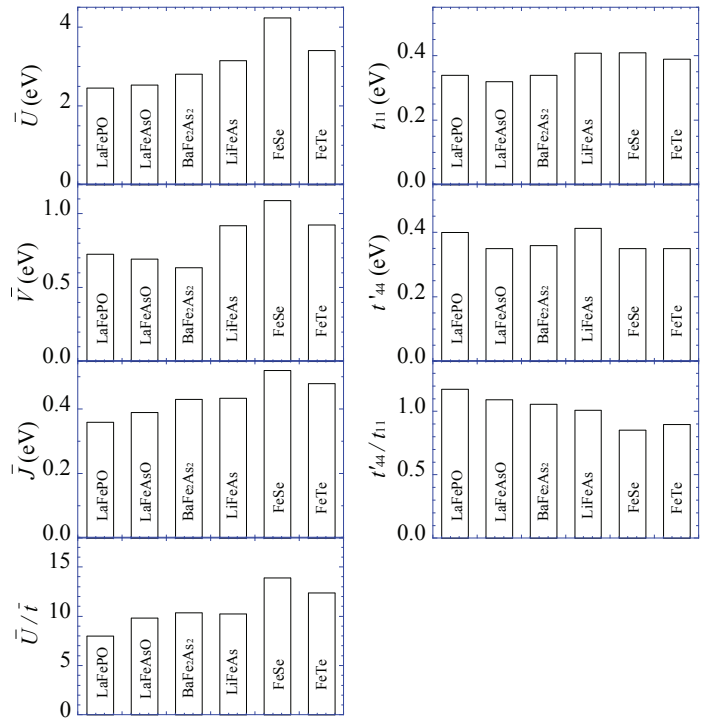}
\end{center} 
\caption{(Color online) Material dependence of parameters for models for iron $3d$ orbitals.\cite{Miyake10}   
The average of the onsite intraorbital effective Coulomb interactions ($\bar{U}$),  
the average of the offsite effective Coulomb interactions between  
the neighboring Fe sites at the same orbitals($\bar{V}$),  
the average of the onsite effective exchange interactions ($\bar{J}$),  
the maximum value of the transfer integrals between  
the neighboring Fe sites [$t_{11} = t_{11}(1/2,-1/2,0)$]
 and between the next-nearest neighbor [$t'_{44} = t_{44}(1,0,0)$],  
$\bar{U}/\bar{t}$, and $t'_{44}/t_{11}$ 
are compared.  
The subscripts of $t_{11}$ and $t'_{44}$ are orbital indices; 1 for $xy$ and 4 for $zx$.
$\bar{t}$ is the orbital average of the largest nearest $d$-$d$ transfer integrals. 
}
\label{fig:summary} 
\end{figure*} 

The systematic change in the model parameters is understood from the structural differences. 
A key quantity for understanding the systematic evolution from the 1111 to the 11 families is 
the height of the pnictogen/chalcogen layer $h$ measured from the iron plane.
The height $h$ increases from LaFePO (1.13 $\AA$), LaFeAsO (1.32 $\AA$), BaFe$_2$As$_2$ (1.36 $\AA$) FeSe (1.47$\AA$) to FeTe (1.77$\AA$).
It was pointed out in the early stage that the electronic band structure is altered significantly by changing $h$ \cite{Singh,vildosola08}. 
There is also a work claiming that the spin and charge susceptibilities are sensitive to $h$ \cite{KurokiPRB}. 
In terms of the correlation strength, $h$ controls the spatial extent of the 
Wannier orbitals and the strength of screening effect.  The smaller $h$ enhances the hybridization of the Fe $3d$ orbitals with the pnictogen/chalcogen $p$ orbitals leading to extended Wannier orbitals as is shown in Fig.\ref{fig:mlwf_d}.  This makes the bare onsite interaction small. The smaller $h$ also makes the pnictogen/chalcogen $p$ level closer to the Fermi level, which enhances the screening of the Coulomb interaction by the $p$ bands.  Furthermore, the number of $p$ bands contributing to the screening decreases in simpler compounds such as the 11 compounds, which reduces the screening channels and enhance the effective interaction for large $h$.
 The difference is similar in the effective models containing $p$ orbitals of As, 
Se or Te ($dp$ or $dpp$ model), where $U$ ranges from $\sim$ 4 eV for the 1111 family to 
$\sim 7$ eV for the 11 family. The exchange interaction $J$ has a similar tendency.
The family dependence of models indicates a wide variation ranging from 
weak correlation regime (LaFePO) to substantially strong correlation regime (FeSe). 
This variety of the electron correlation brings about the diversity of physical properties observed in different families of iron based superconductors in spite of similar band structures.

In fact it has been pointed out that FeSe may show the Hubbard splitting of the bands as a clear indication of the strong correlation effects.\cite{Aichhorn}  The available experimental results are consistently analyzed from this perspective.  
For the 1111 
family, the correlation effects have been analyzed in more detail as a moderately correlated system.\cite{Nakamura,miyake08b,aichhorn09}

The larger $h$ also explains why the ten-fold 3$d$ bands for the 11 family are more entangled with the smearing of the ``pseudogap" structure above the Fermi level observed in the 1111 family.\cite{Miyake10} 
While the family-dependent semimetallic splitting of the bands primarily consists of $d_{yz}/d_{zx}$ and $d_{x^2-y^2}$ orbitals, 
the size of the pseudogap structure is controlled by the hybridization between these orbitals and $d_{xy}/d_{3z^2-r^2}$: 
A large hybridization in the 1111 family generates a large ``band-insulating"-like pseudogap 
({\it hybridization gap}), whereas a large $h$ in the 11 family weakens them,
resulting in a ``half-filled" like bands of orbitals.
This enhances strong correlation effects in analogy with Mott physics and causes the orbital selective crossover in the three orbitals. 
On the other hand, the 
geometrical frustration $t'/t$, inferred from the ratio of the next-nearest transfer $t'$ to the nearest one $t$ of the $d$ model is
relatively larger for the 1111 family than FeTe. 

By using the {\it ab initio} model, magnetic properties have been analyzed by low-energy solvers.\cite{Misawa10} 
The magnetic transition with the correct pattern has been reproduced by the many-variable variational Monte Carlo calculations with a quantitative agreement of the ordered moment. In case of LaFeAsO, the unusually small moment has been naturally understood by the proximity to a quantum critical point between a paramagnetic metal and an antiferromagnetic metal. 
VMC results are shown in Fig.\ref{LaFeAsO_moment} for the ordered magnetic moment as a function of the scaled interaction strength to monitor the interaction effects.  The ordered moment shows systematic evolution from the quantum critical point near the interaction corresponding to LaFaAsO and is in quantitative agreement with the experimental results shown by crosses without an adjustable parameter.  Here, the {\it ab initio} model was constructed for LaFeAsO corresponding to $\tilde{\lambda}=1$ and the parameter $\tilde{\lambda}$ for other compounds is determined by the ratio of the averaged onsite intraorbital interaction between the compound and the reference system LaFeAsO.  This ratio was calculated from the {\it ab initio} model parameters obtained in Ref.\citen{Miyake10}.  The parameter $\tilde{\lambda}$ is obtained from the original scaling parameter $\lambda$ by considering the La $4f$ screening ignored in the model construction by Nakamura et al.\cite{Nakamura} and by considering the interlayer screening effects discussed in \S \ref{Dimensional downfolding}.\cite{Nakamura10} 
It is remarkable that the {\it ab initio} downfolded models for various different families of the iron superconductors are all within a few percent of errors given by a single effective Hamiltonian with a single parameter $\tilde{\lambda}$.
The robust metallic behavior in a large interaction region ($U/t\sim 10$) is also understood from the existence of two Dirac cones in the dispersion near the Fermi level.  The metal is protected as long as the Dirac cones are retained.  The Dirac cones can be annihilated in pair only by a large magnetic moment ($> 3 \mu_{\rm B}$).  
\begin{figure}[htb]
	\begin{center}
		\includegraphics[width=6cm]{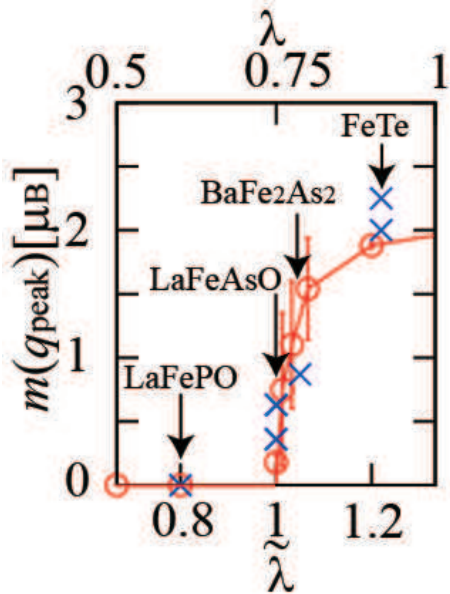}   
	\end{center}
\caption{(color online):Magnetic ordered moment $m(q_{\rm peak})$ as a function of the interaction parameter $\lambda$ 
scaling the overall interaction amplitude calculated by VMC and extrapolated to the thermodynamic limit for the {\it ab intitio} model of LaFeAsO (open circles).\cite{Misawa10} 　
The effective scaling factor $\tilde{\lambda}$ is obtained by further considering 
the interlayer screening effect (see the text). 
Experimentally observed materials dependence at corresponding $\tilde{\lambda}$ is also shown
by crosses for LaFePO,\cite{HosonoP} LaFeAsO\cite{Mook,Qureshi}, BaFe$_2$Se$_2$\cite{Huang,Matan} and FeTe.\cite{Bao,Li}.
%[$q_{\rm peak}=(0,\pi)$].
 Quantum critical point of the AF transition appears at slightly below $\lambda=0.75 (\tilde{\lambda}\sim 1)$.
}
\label{LaFeAsO_moment}
\end{figure}%

%%%%%%%%%%%%%%%%%%%%%%%%
\subsection{Organic conductor} \label{Organic conductor}
Families of organic conductors provide us with another type of strongly correlated electron systems.  Usually, a unit cell of molecular crystals contains many atoms.  However, in many cases, a unit cell contains only a small number of molecules and structures are simple in terms of molecular stackings.  Electrons in a single molecule occupy molecular orbitals, which has normally spread over the molecule.  Such molecular orbitals have a small overlap with those on neighboring molecules, when the molecules are stacked to form the bulk crystal.  Typical bands have simple structures near the Fermi level, where they consist only of lowest unoccupied molecular orbital (LUMO)) and highest occupied molecular orbital (HOMO).  The HOMO and LUMO bands are in many cases isolated from other bands.  This makes the effective Coulomb interaction between electrons on the HOMO and LUMO orbitals poorly screened by other bands.  The small overlap of the molecular orbitals between the neighboring molecules makes the HOMO and LUMO bandwidths small.  These two factors contribute to make the organic conductors mostly strongly correlated electron systems.  

%%%%%%%%%%%%%%%%%%%%%%%%
\subsubsection{$\kappa$-ET conductor} 
ET-type conductors are synthesized as a family of such organic conductors.
(ET)$_2X$ with a number of choices of anions $X$, offer a variety of prototypical behaviors of correlated electron systems with two-dimensional (2D) anisotropies.\cite{Kanoda} Examples range from correlated metals with superconductivity at low temperatures and competing charge orderings to Mott insulators either with a quantum spin liquid or with antiferromagnetic, charge-density or spin-Peierls orders.  Intriguing Mott transitions are also found. 
They have all been studied extensively at a front of research for unconventional quantum phases and quantum critical phenomena.  
Among them, $\kappa$-type ET conductors have stacking of dimerized pair of ET molecules.  The dimerization causes splittings of the HOMO and LUMO bands into bonding and antibonding bands. Since holes are quarter filled (electrons are three-quarter filled) at the HOMO band for (ET)$_2X$, after the dimerization splitting, the Fermi level is normally located at the antibonding HOMO band at half filled.   

In particular, an unconventional nonmagnetic Mott-insulating phase is found near the Mott transition in the $\kappa$-type structure of ET molecules, $\kappa$-(ET)$_2$Cu$_2$(CN)$_3$ referred to as $\kappa$-CN. Although this compound is a Mott insulator, no magnetic order is identified down to the temperature $T$=0.03 K, four orders of magnitude lower than the antiferromagnetic spin-exchange interaction $J$$\sim$250 K \cite{Shimizu}. The emergence of the quantum spin liquid near the Mott transition has been predicted in earlier numerical studies \cite{Kashima,Morita,Mizusaki}, while the full understanding of the spin liquid needs more thorough studies. It is also crucially important to elucidate the real relevance of the theoretical findings to the real $\kappa$-ET compounds.  Most of numerical \cite{Kyung,WataVMC01} and theoretical \cite{Lee} studies have also been performed for a simplified single-band 2D Hubbard model based on an empirical estimate of parameters combined with extended H\"uckel calculations \cite{Mori,Saito}.  A more realistic description of $\kappa$-ET compounds was certainly needed beyond the empirical model.  

Another fundamental finding achieved in this series of compound is the unconventional Mott transition found for $X$=Cu[N(CN)$_2$]Cl under pressure \cite{Kagawa}.   
The novel universality class of the Mott transition is in good agreement with the theoretically revealed marginal quantum criticality at the meeting point of the symmetry breaking and topological change \cite{Imada1,Imada2,Misawa1,Misawa2}.
 Because of its significance to the basic understanding on the physics of quantum criticality, the relevance of theoretical concept to the experimental observation needs to be further examined on realistic and first-principles grounds.  Furthermore, an unconventional superconductivity is found in the metallic sides of these compounds at low temperatures ($T$$<$$T_c$$\sim$10-13K), where the mechanism is not clearly understood yet \cite{Geiser,Cu2CN3}.  
These outstanding properties of $\kappa$-ET compounds have urged systematic studies based on realistic basis. As mentioned above, however, the first-principle studies are limited \cite{DFT_ET} and most of the studies so far were performed using the empirical models inferred from the H\"uckel studies.   

The three-stage RMS method has recently been applied to $\kappa$-ET conductors.\cite{NakamuraET}  
Effective low-energy models have been derived for two contrasting compounds, spin-liquid $\kappa$-CN and superconducting compound $X$=Cu(NCS)$_2$ abbreviated as $\kappa$-NCS \cite{CuSCN}, to get insights into the whole series of $\kappa$-(ET)$_2X$ compounds from metals to Mott insulators. 

The global band structures obtained by GGA are shown in Figs.\ref{fig:kappa-band}(a) and (b). They clearly and commonly show that the antibonding HOMO bands are isolated near the Fermi level as is anticipated.  Then the downfolding to an effective single-band model for the HOMO antibonding band has been performed after constructing the maximally localized Wannier orbital shown in Fig.\ref{fig:kappa-band}(c).\cite{NakamuraET}   
\begin{figure}[h]
\begin{center}
\includegraphics[width=0.46\textwidth]{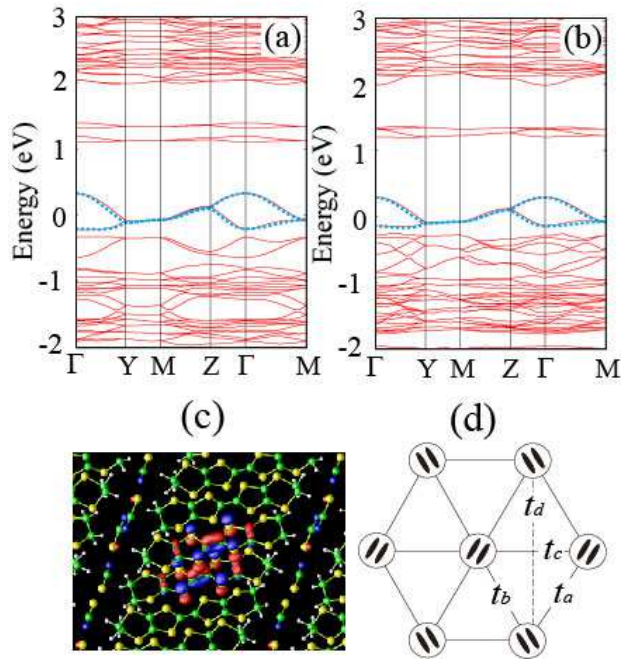}
\end{center}
\caption{(Color) {\it Ab initio} GGA band structures (red line) of $\kappa$-(BEDT-TTF)$_2$Cu(NCS)$_2$ (a) and $\kappa$-(BEDT-TTF)$_2$Cu$_2$(CN)$_3$ (b).\cite{NakamuraET} The zero of energy corresponds to the Fermi level. The blue dotted dispersions are obtained by the four transfer parameters listed in TABLE I. (c) Maximally localized Wannier function of $\kappa$-(BEDT-TTF)$_2$Cu(NCS)$_2$ constructed for effective low-energy model. The amplitudes of the contour surface are +1.5/$\sqrt{v}$ (blue) and $-$1.5/$\sqrt{v}$ (red), where $v$ is the volume of the primitive cell. S, C, H, N, and Cu nuclei are illustrated by green, yellow, silver, blue, and red spheres, respectively. (d) Schematic network of transfers in the triangular lattice.}
\label{fig:kappa-band}
\end{figure}

The parameters of the downfolded model are listed in Table~\ref{PARAM} for $\kappa$-NCS and $\kappa$-CN with the notation of the transfer in Fig.\ref{fig:kappa-band}(d). 
It contains dispersions of the highest occupied Wannier-type molecular orbitals with the nearest neighbor transfer $t$$\sim$0.067 eV for a metal $X$=Cu(NCS)$_2$ and 0.055 eV for a Mott insulator $X$=Cu$_2$(CN)$_3$, as well as the onsite screened Coulomb interactions. It shows a substantial difference from the previous simple extended H\"uckel results \cite{Shimizu,Saito,Mori}:  
%It shows much stronger onsite interaction $U$$\sim$0.8 eV ($U/t$$\sim$12-15) than the H\"uckel estimates ($U/t$$\sim$7-8) as well as an appreciable longer-ranged interaction. 
The derived parameters indicate that (i) the geometrical frustration parameter $|t'/t|$ is substantially smaller than the extended H\"uckel results and $\kappa$-CN estimated at $|t'/t|$$\sim$0.8 has turned out to be away from the right triangular structure\cite{Valenti} and (ii) the onsite Coulomb repulsion ($U$$\sim$0.8 eV characterized by $U/t$$\sim$12-15) is unexpectedly large compared to the H\"uckel estimate given by $U/t$$\sim$7-8, while the intersite Coulomb interaction was found to be also appreciable as we see in Figs.\ref{fig:Interaction}(a) and (b). 

%\begin{figure}[h]
%\includegraphics[width=0.45\textwidth]{kappa-ET_cuscn.w.eps} 
%\includegraphics[width=0.45\textwidth]{kappa-ET_fig1.eps} 
%\caption{(a) Calculated maximally localized Wannier functions of $\kappa$-(BEDT-TTF)$_2$Cu(NCS)$_2$. The amplitudes of the contour surface are +1.5/$\sqrt{v}$ (blue) and $-$1.5/$\sqrt{v}$ (red), where $v$ is the volume of the primitive cell. S, C, H, N, and Cu nuclei are illustrated by green, yellow, silver, blue, and red spheres, respectively. (b) Schematic diagram for transfer network in the triangular lattice.} 
%\label{fig:kappa-Wannier}
%\end{figure}
\begin{table}[h] 
\caption{List of the parameters in a form of the single-band extended Hubbard Hamiltonian in eq.~(\ref{ham}) for $\kappa$-(ET)$_2X$.} 
\centering 
\begin{tabular}{lr@{\ \ \ \ \ \ }r} \hline \hline
 & \multicolumn{1}{c}{$X$=Cu(NCS)$_2$} & \multicolumn{1}{c}{$X$=Cu$_2$(CN)$_3$} \\ \hline
   $t_a$ (meV)  & $-$64.8   & $-$54.5  \\ 
   $t_b$ (meV)  & $-$69.3   & $-$54.7  \\  
   $t_c$ (meV)  &    44.1   &    44.1  \\ 
   $t_d$ (meV)  & $-$11.5   & $-$ 6.8  \\ 
   $U  $ (eV)   &     0.83   &    0.85 \\ 
    \hline \hline
\end{tabular} 
\label{PARAM} 
\end{table}

\begin{figure}[t]
\begin{center}
\includegraphics[width=0.235\textwidth]{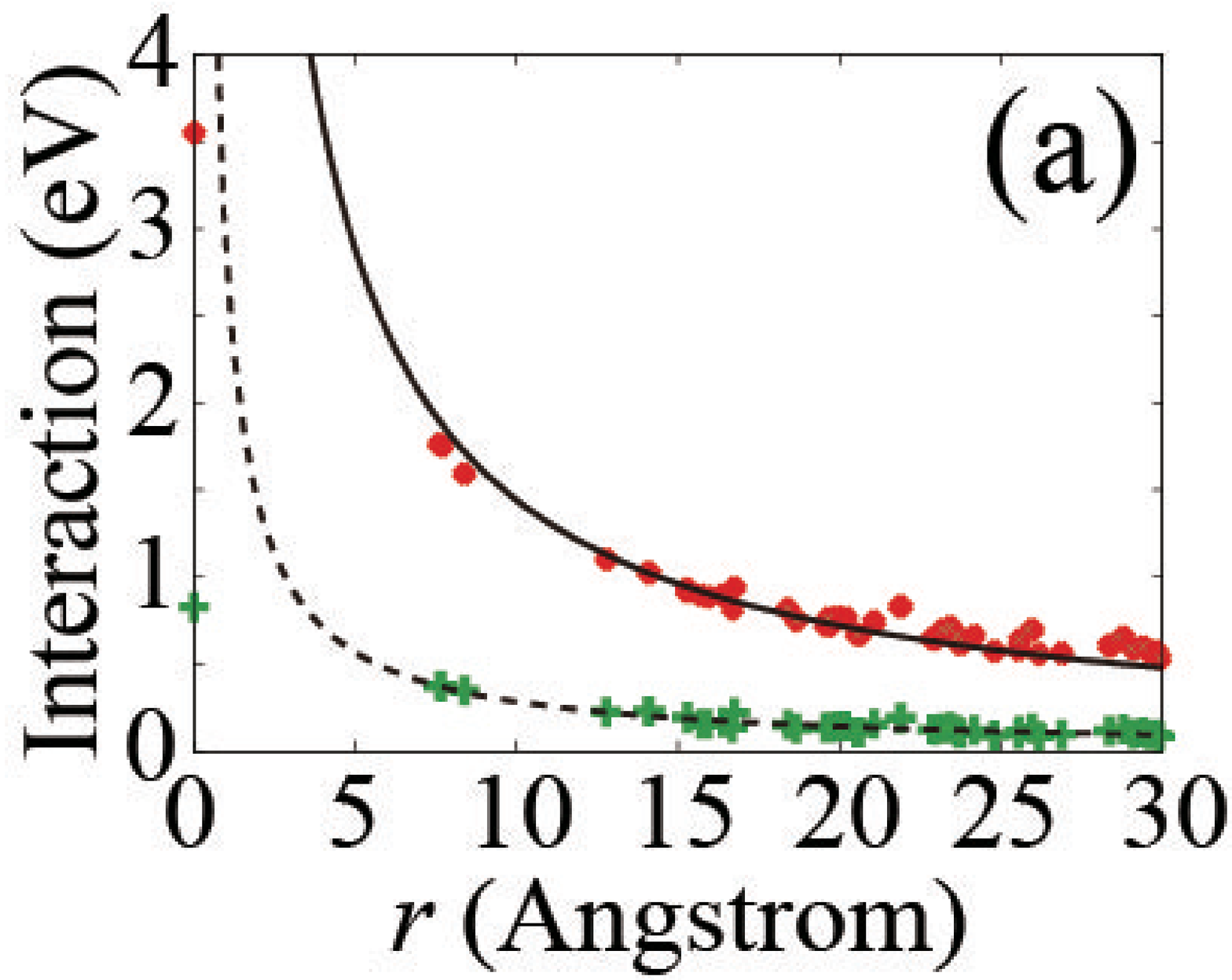} 
\includegraphics[width=0.235\textwidth]{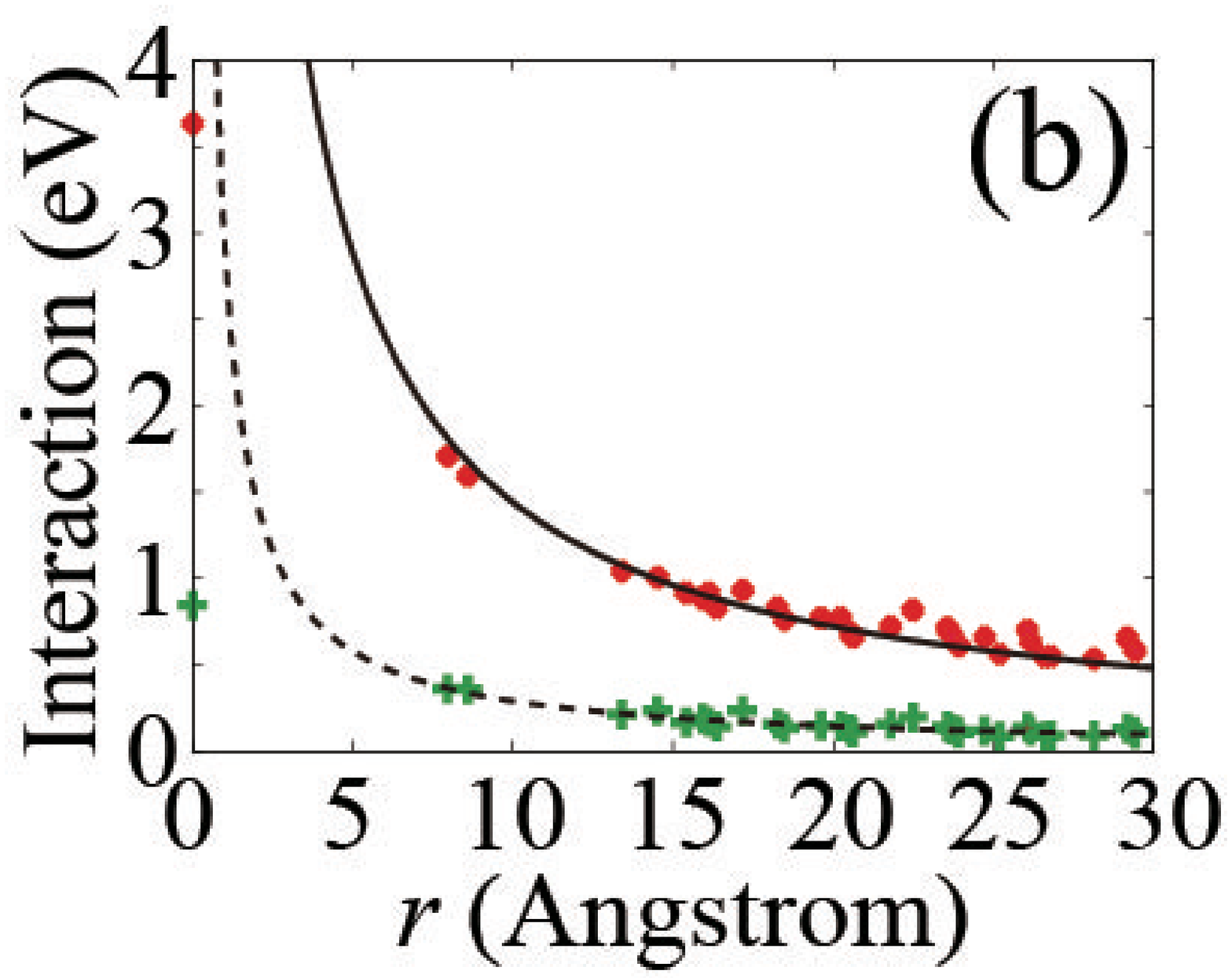} 
\end{center}
\caption{(Color online) Screened Coulomb interactions of $\kappa$-(BEDT-TTF)$_2$Cu(NCS)$_2$ (a) and $\kappa$-(BEDT-TTF)$_2$Cu$_2$(CN)$_3$ (b) as a function of distance between two centers of maximally localized Wannier orbitals, calculated by cRPA illustrated by circles (red). Crosses (green) show the bare interactions.\cite{NakamuraET} The solid and dotted curves denote $1/r$ and $1/(\lambda r)$, with a fitting parameter $\lambda \sim 5$ hartree$^{-1}$bohr$^{-1}$ in both of the compounds.}
\label{fig:Interaction}
\end{figure}

%%%%%%%%%%%%%%%%%%%%%%%%
\section{Concluding Remark and Outlook} \label{Concluding Remarks}
We have reviewed recent rapid advance in understanding electronic structures of strongly correlated electron systems by utilizing the electronic hierarchical structure.  The method starts from obtaining the global electronic structure by DFT or GW procedure.  Then low-energy effective models are derived by the downfolding eliminating the degrees of freedom away from the Fermi level. This is achieved first by extracting localized Wannier orbitals and constructing {\it ab initio} tight binding models in real space.  The interaction parameters are obtained by counting the screening with the constrained RPA. The resultant models are solved by highly accurate solvers such as the various Monte Carlo methods, path-integral renormalization group and dynamical mean field theory.  This scheme opens a way to understand electron correlations even when the single-particle picture breaks down. 

The applications to transition metal compounds including iron-based superconductors and perovskite oxides as well as to organic conductors have already proven its quantitative accuracy without any adjustable parameters and shown that a new powerful method has emerged.  Severely competing orders as well as quantum and many-body fluctuations are now under controlled treatment by this approach beyond the mean-field and one-body pictures, while it is still on the way toward further growth with diverse potentiality of improvement.  
Recent applications to alkali-cluster-loaded zeolites\cite{NakamuraSodalite,NoharaZeolite} show a wide feasibility and potential of the methods including the compounds with a large unit cell. Another direction of the application is surfaces and interfaces, which do not retain the bulk translational symmetry.  In such complex systems, the number of bands near the Fermi level can be very large because of large unit cells.        
%\textcolor{red}{ここで界面やゼオライトなど新たな相関系への展開に言及してもよいと思いました。}

When electron correlation effects are so large that the degrees of freedom far from the Fermi level are modified, one has either to include such a part under correlation effects into the low-energy effective model, or has to solve them selfconsistently with a feedback to a high-energy downfolded part as is illustrated as the broken arrow in Fig.\ref{fig:schematics}. Although preliminary results have been reported for LDA+DMFT and GW+DMFT, applications are so far limited because of computationally demanding iterations.  The total selfconsistency is certainly a future direction of challenge.

We have considered only the electronic degrees of freedom and the atomic structure is assumed to be given in this article. 
A future important subject is to combine with the structural optimization for the goal of the real first principles scheme.\cite{Leonov2010}  In addition, phonon degrees of freedom are in general coupled in a low-energy scale\cite{Merino00,Savrasov03,Sangiovanni06-2} and it determines a number of important properties including phonon mediated superconductivity, ferroelectricity and charge ordering.      

Another intriguing problem is dynamical processes far from equilibrium and relaxation phenomena.  Photoinduced transitions are typical examples of future issue. Experimentally, time resolved photoemission will open powerful probes and tools for new situation of nonwquilibrium phenomena.
Depending on the energy and time scales, we need to develop more involved but tractable framework with low-energy solvers for nonequilibrium. 

{\bf Acknowledgements}
The authors thank Ryotaro Arita, Kazuma Nakamura, Dieter Vollhardt and Giorgio Sangiovanni for useful and helpful comments.
The authors also thank Daisuke Tahara for providing benchmark results of VMC and useful discussions. This work is supported by a Grant-in-Aid for Scientific Research (No. 22104010) on ``First-principles effective models and frontiers in correlation science" from Ministry of Education, Culture, Sports, Science, and Technology, Japan. 

%\begin{thebibliography}{9}

\end{document}